\newcommand\papertitle{\textit{Planck} 2013 results. XXVI. Background geometry and topology of the Universe}

\pdfoutput=1
\documentclass[twocolumn,traditabstract,longauth]{aa} 
\usepackage{graphicx}

\usepackage{txfonts}
\usepackage{url}
\usepackage{natbib}
\usepackage{booktabs}
\usepackage{color}

\usepackage[labelfont=bf]{caption}
\usepackage{subcaption}   
\usepackage{fixltx2e}

\usepackage{hyperref}
\hypersetup{breaklinks,colorlinks,citecolor=blue}

\newcommand{\jdm}[1]{}

\bibpunct{(}{)}{;}{a}{}{,}  
\def\setsymbol#1#2{\expandafter\def\csname #1\endcsname{#2}}
\def\getsymbol#1{\csname #1\endcsname}

\def\Planck{\textit{Planck}}

\def\all2013resultspapers{\nocite{planck2013-p01, planck2013-p02, planck2013-p02a, planck2013-p02d, planck2013-p02b, planck2013-p03, planck2013-p03c, planck2013-p03f, planck2013-p03d, planck2013-p03e, planck2013-p06b, planck2013-p06, planck2013-p03a, planck2013-pip88, planck2013-p08, planck2013-p11, planck2013-p12, planck2013-p13, planck2013-p14, planck2013-p15, planck2013-p05b, planck2013-p17, planck2013-p09, planck2013-p09a, planck2013-p20, planck2013-p19, planck2013-pipaberration, planck2013-p05, planck2013-p05a, planck2013-pip56, planck2013-p01a}}

\newbox\tablebox    \newdimen\tablewidth
\def\leaderfil{\leaders\hbox to 5pt{\hss.\hss}\hfil}
\def\endPlancktable{\tablewidth=\columnwidth 
    $$\hss\copy\tablebox\hss$$
    \vskip-\lastskip\vskip -2pt}

\def\tablenote#1 #2\par{\begingroup \parindent=0.8em
    \abovedisplayshortskip=0pt\belowdisplayshortskip=0pt
    \noindent
    $$\hss\vbox{\hsize\tablewidth \hangindent=\parindent \hangafter=1 \noindent
    \hbox to \parindent{$^#1$\hss}\strut#2\strut\par}\hss$$
    \endgroup}
\def\doubleline{\vskip 3pt\hrule \vskip 1.5pt \hrule \vskip 5pt}

\def\L2{\ifmmode L_2\else $L_2$\fi}

\def\DeltaT{\ifmmode \Delta T\else $\Delta T$\fi}
\def\deltat{\ifmmode \Delta t\else $\Delta t$\fi}
\def\fknee{\ifmmode f_{\rm knee}\else $f_{\rm knee}$\fi}
\def\Fmax{\ifmmode F_{\rm max}\else $F_{\rm max}$\fi}
\def\solar{\ifmmode{\rm M}_{\mathord\odot}\else${\rm M}_{\mathord\odot}$\fi}
\def\Msolar{\ifmmode{\rm M}_{\mathord\odot}\else${\rm M}_{\mathord\odot}$\fi}
\def\Lsolar{\ifmmode{\rm L}_{\mathord\odot}\else${\rm L}_{\mathord\odot}$\fi}

\def\inv{\ifmmode^{-1}\else$^{-1}$\fi}
\def\mo{\ifmmode^{-1}\else$^{-1}$\fi}
\def\sup#1{\ifmmode ^{\rm #1}\else $^{\rm #1}$\fi}
\def\expo#1{\ifmmode \times 10^{#1}\else $\times 10^{#1}$\fi}
\def\,{\thinspace}
\def\lsim{\mathrel{\raise .4ex\hbox{\rlap{$<$}\lower 1.2ex\hbox{$\sim$}}}}
\def\gsim{\mathrel{\raise .4ex\hbox{\rlap{$>$}\lower 1.2ex\hbox{$\sim$}}}}

\def\simprop{\mathrel{\raise .4ex\hbox{\rlap{$\propto$}\lower 1.2ex\hbox{$\sim$}}}}
\def\deg{\ifmmode^\circ\else$^\circ$\fi}
\def\pdeg{\ifmmode $\setbox0=\hbox{$^{\circ}$}\rlap{\hskip.11\wd0 .}$^{\circ}
          \else \setbox0=\hbox{$^{\circ}$}\rlap{\hskip.11\wd0 .}$^{\circ}$\fi}
\def\arcs{\ifmmode {^{\scriptstyle\prime\prime}}
          \else $^{\scriptstyle\prime\prime}$\fi}
\def\arcm{\ifmmode {^{\scriptstyle\prime}}
          \else $^{\scriptstyle\prime}$\fi}
\newdimen\sa  \newdimen\sb
\def\parcs{\sa=.07em \sb=.03em
     \ifmmode \hbox{\rlap{.}}^{\scriptstyle\prime\kern -\sb\prime}\hbox{\kern -\sa}
     \else \rlap{.}$^{\scriptstyle\prime\kern -\sb\prime}$\kern -\sa\fi}
\def\parcm{\sa=.08em \sb=.03em
     \ifmmode \hbox{\rlap{.}\kern\sa}^{\scriptstyle\prime}\hbox{\kern-\sb}
     \else \rlap{.}\kern\sa$^{\scriptstyle\prime}$\kern-\sb\fi}
\def\ra[#1 #2 #3.#4]{#1\sup{h}#2\sup{m}#3\sup{s}\llap.#4}
\def\dec[#1 #2 #3.#4]{#1\deg#2\arcm#3\arcs\llap.#4}
\def\deco[#1 #2 #3]{#1\deg#2\arcm#3\arcs}
\def\rra[#1 #2]{#1\sup{h}#2\sup{m}}

\def\dots{\relax\ifmmode \ldots\else $\ldots$\fi}
\def\WHzsr{\ifmmode $W\,Hz\mo\,sr\mo$\else W\,Hz\mo\,sr\mo\fi}
\def\mHz{\ifmmode $\,mHz$\else \,mHz\fi}
\def\GHz{\ifmmode $\,GHz$\else \,GHz\fi}
\def\mKs{\ifmmode $\,mK\,s$^{1/2}\else \,mK\,s$^{1/2}$\fi}
\def\muKs{\ifmmode \,\mu$K\,s$^{1/2}\else \,$\mu$K\,s$^{1/2}$\fi}
\def\muKRJs{\ifmmode \,\mu$K$_{\rm RJ}$\,s$^{1/2}\else \,$\mu$K$_{\rm RJ}$\,s$^{1/2}$\fi}
\def\muKHz{\ifmmode \,\mu$K\,Hz$^{-1/2}\else \,$\mu$K\,Hz$^{-1/2}$\fi}
\def\MJysr{\ifmmode \,$MJy\,sr\mo$\else \,MJy\,sr\mo\fi}
\def\MJysrmK{\ifmmode \,$MJy\,sr\mo$\,mK$_{\rm CMB}\mo\else \,MJy\,sr\mo\,mK$_{\rm CMB}\mo$\fi}
\def\microns{\ifmmode \,\mu$m$\else \,$\mu$m\fi}

\def\muK{\ifmmode \,\mu$K$\else \,$\mu$\hbox{K}\fi}
\def\microK{\ifmmode \,\mu$K$\else \,$\mu$\hbox{K}\fi}
\def\muW{\ifmmode \,\mu$W$\else \,$\mu$\hbox{W}\fi}
\def\kms{\ifmmode $\,km\,s$^{-1}\else \,km\,s$^{-1}$\fi}
\def\kmsMpc{\ifmmode $\,\kms\,Mpc\mo$\else \,\kms\,Mpc\mo\fi}
\setsymbol{LFI:center:frequency:70GHz:units}{70.3\,GHz}
\setsymbol{LFI:center:frequency:44GHz:units}{44.1\,GHz}
\setsymbol{LFI:center:frequency:30GHz:units}{28.5\,GHz}

\setsymbol{LFI:center:frequency:70GHz}{70.3}
\setsymbol{LFI:center:frequency:44GHz}{44.1}
\setsymbol{LFI:center:frequency:30GHz}{28.5}

\setsymbol{LFI:center:frequency:LFI18:Rad:M:units}{71.7\GHz}
\setsymbol{LFI:center:frequency:LFI19:Rad:M:units}{67.5\GHz}
\setsymbol{LFI:center:frequency:LFI20:Rad:M:units}{69.2\GHz}
\setsymbol{LFI:center:frequency:LFI21:Rad:M:units}{70.4\GHz}
\setsymbol{LFI:center:frequency:LFI22:Rad:M:units}{71.5\GHz}
\setsymbol{LFI:center:frequency:LFI23:Rad:M:units}{70.8\GHz}
\setsymbol{LFI:center:frequency:LFI24:Rad:M:units}{44.4\GHz}
\setsymbol{LFI:center:frequency:LFI25:Rad:M:units}{44.0\GHz}
\setsymbol{LFI:center:frequency:LFI26:Rad:M:units}{43.9\GHz}
\setsymbol{LFI:center:frequency:LFI27:Rad:M:units}{28.3\GHz}
\setsymbol{LFI:center:frequency:LFI28:Rad:M:units}{28.8\GHz}
\setsymbol{LFI:center:frequency:LFI18:Rad:S:units}{70.1\GHz}
\setsymbol{LFI:center:frequency:LFI19:Rad:S:units}{69.6\GHz}
\setsymbol{LFI:center:frequency:LFI20:Rad:S:units}{69.5\GHz}
\setsymbol{LFI:center:frequency:LFI21:Rad:S:units}{69.5\GHz}
\setsymbol{LFI:center:frequency:LFI22:Rad:S:units}{72.8\GHz}
\setsymbol{LFI:center:frequency:LFI23:Rad:S:units}{71.3\GHz}
\setsymbol{LFI:center:frequency:LFI24:Rad:S:units}{44.1\GHz}
\setsymbol{LFI:center:frequency:LFI25:Rad:S:units}{44.1\GHz}
\setsymbol{LFI:center:frequency:LFI26:Rad:S:units}{44.1\GHz}
\setsymbol{LFI:center:frequency:LFI27:Rad:S:units}{28.5\GHz}
\setsymbol{LFI:center:frequency:LFI28:Rad:S:units}{28.2\GHz}

\setsymbol{LFI:center:frequency:LFI18:Rad:M}{71.7}
\setsymbol{LFI:center:frequency:LFI19:Rad:M}{67.5}
\setsymbol{LFI:center:frequency:LFI20:Rad:M}{69.2}
\setsymbol{LFI:center:frequency:LFI21:Rad:M}{70.4}
\setsymbol{LFI:center:frequency:LFI22:Rad:M}{71.5}
\setsymbol{LFI:center:frequency:LFI23:Rad:M}{70.8}
\setsymbol{LFI:center:frequency:LFI24:Rad:M}{44.4}
\setsymbol{LFI:center:frequency:LFI25:Rad:M}{44.0}
\setsymbol{LFI:center:frequency:LFI26:Rad:M}{43.9}
\setsymbol{LFI:center:frequency:LFI27:Rad:M}{28.3}
\setsymbol{LFI:center:frequency:LFI28:Rad:M}{28.8}
\setsymbol{LFI:center:frequency:LFI18:Rad:S}{70.1}
\setsymbol{LFI:center:frequency:LFI19:Rad:S}{69.6}
\setsymbol{LFI:center:frequency:LFI20:Rad:S}{69.5}
\setsymbol{LFI:center:frequency:LFI21:Rad:S}{69.5}
\setsymbol{LFI:center:frequency:LFI22:Rad:S}{72.8}
\setsymbol{LFI:center:frequency:LFI23:Rad:S}{71.3}
\setsymbol{LFI:center:frequency:LFI24:Rad:S}{44.1}
\setsymbol{LFI:center:frequency:LFI25:Rad:S}{44.1}
\setsymbol{LFI:center:frequency:LFI26:Rad:S}{44.1}
\setsymbol{LFI:center:frequency:LFI27:Rad:S}{28.5}
\setsymbol{LFI:center:frequency:LFI28:Rad:S}{28.2}

\setsymbol{LFI:white:noise:sensitivity:70GHz:units}{134.7\muKs}
\setsymbol{LFI:white:noise:sensitivity:44GHz:units}{164.7\muKs}
\setsymbol{LFI:white:noise:sensitivity:30GHz:units}{143.4\muKs}

\setsymbol{LFI:white:noise:sensitivity:70GHz}{134.7}
\setsymbol{LFI:white:noise:sensitivity:44GHz}{164.7}
\setsymbol{LFI:white:noise:sensitivity:30GHz}{143.4}

\setsymbol{LFI:white:noise:sensitivity:LFI18:Rad:M:units}{512.0\muKs}
\setsymbol{LFI:white:noise:sensitivity:LFI19:Rad:M:units}{581.4\muKs}
\setsymbol{LFI:white:noise:sensitivity:LFI20:Rad:M:units}{590.8\muKs}
\setsymbol{LFI:white:noise:sensitivity:LFI21:Rad:M:units}{455.2\muKs}
\setsymbol{LFI:white:noise:sensitivity:LFI22:Rad:M:units}{492.0\muKs}
\setsymbol{LFI:white:noise:sensitivity:LFI23:Rad:M:units}{507.7\muKs}
\setsymbol{LFI:white:noise:sensitivity:LFI24:Rad:M:units}{462.2\muKs}
\setsymbol{LFI:white:noise:sensitivity:LFI25:Rad:M:units}{413.6\muKs}
\setsymbol{LFI:white:noise:sensitivity:LFI26:Rad:M:units}{478.6\muKs}
\setsymbol{LFI:white:noise:sensitivity:LFI27:Rad:M:units}{277.7\muKs}
\setsymbol{LFI:white:noise:sensitivity:LFI28:Rad:M:units}{312.3\muKs}
\setsymbol{LFI:white:noise:sensitivity:LFI18:Rad:S:units}{465.7\muKs}
\setsymbol{LFI:white:noise:sensitivity:LFI19:Rad:S:units}{555.6\muKs}
\setsymbol{LFI:white:noise:sensitivity:LFI20:Rad:S:units}{623.2\muKs}
\setsymbol{LFI:white:noise:sensitivity:LFI21:Rad:S:units}{564.1\muKs}
\setsymbol{LFI:white:noise:sensitivity:LFI22:Rad:S:units}{534.4\muKs}
\setsymbol{LFI:white:noise:sensitivity:LFI23:Rad:S:units}{542.4\muKs}
\setsymbol{LFI:white:noise:sensitivity:LFI24:Rad:S:units}{399.2\muKs}
\setsymbol{LFI:white:noise:sensitivity:LFI25:Rad:S:units}{392.6\muKs}
\setsymbol{LFI:white:noise:sensitivity:LFI26:Rad:S:units}{418.6\muKs}
\setsymbol{LFI:white:noise:sensitivity:LFI27:Rad:S:units}{302.9\muKs}
\setsymbol{LFI:white:noise:sensitivity:LFI28:Rad:S:units}{285.3\muKs}

\setsymbol{LFI:white:noise:sensitivity:LFI18:Rad:M}{512.0}
\setsymbol{LFI:white:noise:sensitivity:LFI19:Rad:M}{581.4}
\setsymbol{LFI:white:noise:sensitivity:LFI20:Rad:M}{590.8}
\setsymbol{LFI:white:noise:sensitivity:LFI21:Rad:M}{455.2}
\setsymbol{LFI:white:noise:sensitivity:LFI22:Rad:M}{492.0}
\setsymbol{LFI:white:noise:sensitivity:LFI23:Rad:M}{507.7}
\setsymbol{LFI:white:noise:sensitivity:LFI24:Rad:M}{462.2}
\setsymbol{LFI:white:noise:sensitivity:LFI25:Rad:M}{413.6}
\setsymbol{LFI:white:noise:sensitivity:LFI26:Rad:M}{478.6}
\setsymbol{LFI:white:noise:sensitivity:LFI27:Rad:M}{277.7}
\setsymbol{LFI:white:noise:sensitivity:LFI28:Rad:M}{312.3}
\setsymbol{LFI:white:noise:sensitivity:LFI18:Rad:S}{465.7}
\setsymbol{LFI:white:noise:sensitivity:LFI19:Rad:S}{555.6}
\setsymbol{LFI:white:noise:sensitivity:LFI20:Rad:S}{623.2}
\setsymbol{LFI:white:noise:sensitivity:LFI21:Rad:S}{564.1}
\setsymbol{LFI:white:noise:sensitivity:LFI22:Rad:S}{534.4}
\setsymbol{LFI:white:noise:sensitivity:LFI23:Rad:S}{542.4}
\setsymbol{LFI:white:noise:sensitivity:LFI24:Rad:S}{399.2}
\setsymbol{LFI:white:noise:sensitivity:LFI25:Rad:S}{392.6}
\setsymbol{LFI:white:noise:sensitivity:LFI26:Rad:S}{418.6}
\setsymbol{LFI:white:noise:sensitivity:LFI27:Rad:S}{302.9}
\setsymbol{LFI:white:noise:sensitivity:LFI28:Rad:S}{285.3}

\setsymbol{LFI:knee:frequency:70GHz:units}{29.5\mHz}
\setsymbol{LFI:knee:frequency:44GHz:units}{56.2\mHz}
\setsymbol{LFI:knee:frequency:30GHz:units}{113.7\mHz}

\setsymbol{LFI:knee:frequency:70GHz}{29.5}
\setsymbol{LFI:knee:frequency:44GHz}{56.2}
\setsymbol{LFI:knee:frequency:30GHz}{113.7}

\setsymbol{LFI:knee:frequency:LFI18:Rad:M:units}{16.3\mHz}
\setsymbol{LFI:knee:frequency:LFI19:Rad:M:units}{15.1\mHz}
\setsymbol{LFI:knee:frequency:LFI20:Rad:M:units}{18.7\mHz}
\setsymbol{LFI:knee:frequency:LFI21:Rad:M:units}{37.2\mHz}
\setsymbol{LFI:knee:frequency:LFI22:Rad:M:units}{12.7\mHz}
\setsymbol{LFI:knee:frequency:LFI23:Rad:M:units}{34.6\mHz}
\setsymbol{LFI:knee:frequency:LFI24:Rad:M:units}{46.2\mHz}
\setsymbol{LFI:knee:frequency:LFI25:Rad:M:units}{24.9\mHz}
\setsymbol{LFI:knee:frequency:LFI26:Rad:M:units}{67.6\mHz}
\setsymbol{LFI:knee:frequency:LFI27:Rad:M:units}{187.4\mHz}
\setsymbol{LFI:knee:frequency:LFI28:Rad:M:units}{122.2\mHz}
\setsymbol{LFI:knee:frequency:LFI18:Rad:S:units}{17.7\mHz}
\setsymbol{LFI:knee:frequency:LFI19:Rad:S:units}{22.0\mHz}
\setsymbol{LFI:knee:frequency:LFI20:Rad:S:units}{8.7\mHz}
\setsymbol{LFI:knee:frequency:LFI21:Rad:S:units}{25.9\mHz}
\setsymbol{LFI:knee:frequency:LFI22:Rad:S:units}{15.8\mHz}
\setsymbol{LFI:knee:frequency:LFI23:Rad:S:units}{129.8\mHz}
\setsymbol{LFI:knee:frequency:LFI24:Rad:S:units}{100.9\mHz}
\setsymbol{LFI:knee:frequency:LFI25:Rad:S:units}{38.9\mHz}
\setsymbol{LFI:knee:frequency:LFI26:Rad:S:units}{58.9\mHz}
\setsymbol{LFI:knee:frequency:LFI27:Rad:S:units}{104.4\mHz}
\setsymbol{LFI:knee:frequency:LFI28:Rad:S:units}{40.7\mHz}

\setsymbol{LFI:knee:frequency:LFI18:Rad:M}{16.3}
\setsymbol{LFI:knee:frequency:LFI19:Rad:M}{15.1}
\setsymbol{LFI:knee:frequency:LFI20:Rad:M}{18.7}
\setsymbol{LFI:knee:frequency:LFI21:Rad:M}{37.2}
\setsymbol{LFI:knee:frequency:LFI22:Rad:M}{12.7}
\setsymbol{LFI:knee:frequency:LFI23:Rad:M}{34.6}
\setsymbol{LFI:knee:frequency:LFI24:Rad:M}{46.2}
\setsymbol{LFI:knee:frequency:LFI25:Rad:M}{24.9}
\setsymbol{LFI:knee:frequency:LFI26:Rad:M}{67.6}
\setsymbol{LFI:knee:frequency:LFI27:Rad:M}{187.4}
\setsymbol{LFI:knee:frequency:LFI28:Rad:M}{122.2}
\setsymbol{LFI:knee:frequency:LFI18:Rad:S}{17.7}
\setsymbol{LFI:knee:frequency:LFI19:Rad:S}{22.0}
\setsymbol{LFI:knee:frequency:LFI20:Rad:S}{8.7}
\setsymbol{LFI:knee:frequency:LFI21:Rad:S}{25.9}
\setsymbol{LFI:knee:frequency:LFI22:Rad:S}{15.8}
\setsymbol{LFI:knee:frequency:LFI23:Rad:S}{129.8}
\setsymbol{LFI:knee:frequency:LFI24:Rad:S}{100.9}
\setsymbol{LFI:knee:frequency:LFI25:Rad:S}{38.9}
\setsymbol{LFI:knee:frequency:LFI26:Rad:S}{58.9}
\setsymbol{LFI:knee:frequency:LFI27:Rad:S}{104.4}
\setsymbol{LFI:knee:frequency:LFI28:Rad:S}{40.7}

\setsymbol{LFI:slope:70GHz:units}{$-1.03$\mHz}
\setsymbol{LFI:slope:44GHz:units}{$-0.89$\mHz}
\setsymbol{LFI:slope:30GHz:units}{$-0.87$\mHz}

\setsymbol{LFI:slope:70GHz}{$-1.03$}
\setsymbol{LFI:slope:44GHz}{$-0.89$}
\setsymbol{LFI:slope:30GHz}{$-0.87$}

\setsymbol{LFI:slope:LFI18:Rad:M:units}{$-1.04$\mHz}
\setsymbol{LFI:slope:LFI19:Rad:M:units}{$-1.09$\mHz}
\setsymbol{LFI:slope:LFI20:Rad:M:units}{$-0.69$\mHz}
\setsymbol{LFI:slope:LFI21:Rad:M:units}{$-1.56$\mHz}
\setsymbol{LFI:slope:LFI22:Rad:M:units}{$-1.01$\mHz}
\setsymbol{LFI:slope:LFI23:Rad:M:units}{$-0.96$\mHz}
\setsymbol{LFI:slope:LFI24:Rad:M:units}{$-0.83$\mHz}
\setsymbol{LFI:slope:LFI25:Rad:M:units}{$-0.91$\mHz}
\setsymbol{LFI:slope:LFI26:Rad:M:units}{$-0.95$\mHz}
\setsymbol{LFI:slope:LFI27:Rad:M:units}{$-0.87$\mHz}
\setsymbol{LFI:slope:LFI28:Rad:M:units}{$-0.88$\mHz}
\setsymbol{LFI:slope:LFI18:Rad:S:units}{$-1.15$\mHz}
\setsymbol{LFI:slope:LFI19:Rad:S:units}{$-1.00$\mHz}
\setsymbol{LFI:slope:LFI20:Rad:S:units}{$-0.95$\mHz}
\setsymbol{LFI:slope:LFI21:Rad:S:units}{$-0.92$\mHz}
\setsymbol{LFI:slope:LFI22:Rad:S:units}{$-1.01$\mHz}
\setsymbol{LFI:slope:LFI23:Rad:S:units}{$-0.95$\mHz}
\setsymbol{LFI:slope:LFI24:Rad:S:units}{$-0.73$\mHz}
\setsymbol{LFI:slope:LFI25:Rad:S:units}{$-1.16$\mHz}
\setsymbol{LFI:slope:LFI26:Rad:S:units}{$-0.79$\mHz}
\setsymbol{LFI:slope:LFI27:Rad:S:units}{$-0.82$\mHz}
\setsymbol{LFI:slope:LFI28:Rad:S:units}{$-0.91$\mHz}

\setsymbol{LFI:slope:LFI18:Rad:M}{$-1.04$}
\setsymbol{LFI:slope:LFI19:Rad:M}{$-1.09$}
\setsymbol{LFI:slope:LFI20:Rad:M}{$-0.69$}
\setsymbol{LFI:slope:LFI21:Rad:M}{$-1.56$}
\setsymbol{LFI:slope:LFI22:Rad:M}{$-1.01$}
\setsymbol{LFI:slope:LFI23:Rad:M}{$-0.96$}
\setsymbol{LFI:slope:LFI24:Rad:M}{$-0.83$}
\setsymbol{LFI:slope:LFI25:Rad:M}{$-0.91$}
\setsymbol{LFI:slope:LFI26:Rad:M}{$-0.95$}
\setsymbol{LFI:slope:LFI27:Rad:M}{$-0.87$}
\setsymbol{LFI:slope:LFI28:Rad:M}{$-0.88$}
\setsymbol{LFI:slope:LFI18:Rad:S}{$-1.15$}
\setsymbol{LFI:slope:LFI19:Rad:S}{$-1.00$}
\setsymbol{LFI:slope:LFI20:Rad:S}{$-0.95$}
\setsymbol{LFI:slope:LFI21:Rad:S}{$-0.92$}
\setsymbol{LFI:slope:LFI22:Rad:S}{$-1.01$}
\setsymbol{LFI:slope:LFI23:Rad:S}{$-0.95$}
\setsymbol{LFI:slope:LFI24:Rad:S}{$-0.73$}
\setsymbol{LFI:slope:LFI25:Rad:S}{$-1.16$}
\setsymbol{LFI:slope:LFI26:Rad:S}{$-0.79$}
\setsymbol{LFI:slope:LFI27:Rad:S}{$-0.82$}
\setsymbol{LFI:slope:LFI28:Rad:S}{$-0.91$}

\setsymbol{LFI:FWHM:70GHz:units}{13\parcm01}
\setsymbol{LFI:FWHM:44GHz:units}{27\parcm92}
\setsymbol{LFI:FWHM:30GHz:units}{32\parcm65}

\setsymbol{LFI:FWHM:70GHz}{13.01}
\setsymbol{LFI:FWHM:44GHz}{27.92}
\setsymbol{LFI:FWHM:30GHz}{32.65}

\setsymbol{LFI:FWHM:LFI18:units}{13\parcm39}
\setsymbol{LFI:FWHM:LFI19:units}{13\parcm01}
\setsymbol{LFI:FWHM:LFI20:units}{12\parcm75}
\setsymbol{LFI:FWHM:LFI21:units}{12\parcm74}
\setsymbol{LFI:FWHM:LFI22:units}{12\parcm87}
\setsymbol{LFI:FWHM:LFI23:units}{13\parcm27}
\setsymbol{LFI:FWHM:LFI24:units}{22\parcm98}
\setsymbol{LFI:FWHM:LFI25:units}{30\parcm46}
\setsymbol{LFI:FWHM:LFI26:units}{30\parcm31}
\setsymbol{LFI:FWHM:LFI27:units}{32\parcm65}
\setsymbol{LFI:FWHM:LFI28:units}{32\parcm66}

\setsymbol{LFI:FWHM:LFI18}{13.39}
\setsymbol{LFI:FWHM:LFI19}{13.01}
\setsymbol{LFI:FWHM:LFI20}{12.75}
\setsymbol{LFI:FWHM:LFI21}{12.74}
\setsymbol{LFI:FWHM:LFI22}{12.87}
\setsymbol{LFI:FWHM:LFI23}{13.27}
\setsymbol{LFI:FWHM:LFI24}{22.98}
\setsymbol{LFI:FWHM:LFI25}{30.46}
\setsymbol{LFI:FWHM:LFI26}{30.31}
\setsymbol{LFI:FWHM:LFI27}{32.65}
\setsymbol{LFI:FWHM:LFI28}{32.66}

\setsymbol{LFI:FWHM:uncertainty:LFI18:units}{0.170\arcm}
\setsymbol{LFI:FWHM:uncertainty:LFI19:units}{0.174\arcm}
\setsymbol{LFI:FWHM:uncertainty:LFI20:units}{0.170\arcm}
\setsymbol{LFI:FWHM:uncertainty:LFI21:units}{0.156\arcm}
\setsymbol{LFI:FWHM:uncertainty:LFI22:units}{0.164\arcm}
\setsymbol{LFI:FWHM:uncertainty:LFI23:units}{0.171\arcm}
\setsymbol{LFI:FWHM:uncertainty:LFI24:units}{0.652\arcm}
\setsymbol{LFI:FWHM:uncertainty:LFI25:units}{1.075\arcm}
\setsymbol{LFI:FWHM:uncertainty:LFI26:units}{1.131\arcm}
\setsymbol{LFI:FWHM:uncertainty:LFI27:units}{1.266\arcm}
\setsymbol{LFI:FWHM:uncertainty:LFI28:units}{1.287\arcm}

\setsymbol{LFI:FWHM:uncertainty:LFI18}{0.170}
\setsymbol{LFI:FWHM:uncertainty:LFI19}{0.174}
\setsymbol{LFI:FWHM:uncertainty:LFI20}{0.170}
\setsymbol{LFI:FWHM:uncertainty:LFI21}{0.156}
\setsymbol{LFI:FWHM:uncertainty:LFI22}{0.164}
\setsymbol{LFI:FWHM:uncertainty:LFI23}{0.171}
\setsymbol{LFI:FWHM:uncertainty:LFI24}{0.652}
\setsymbol{LFI:FWHM:uncertainty:LFI25}{1.075}
\setsymbol{LFI:FWHM:uncertainty:LFI26}{1.131}
\setsymbol{LFI:FWHM:uncertainty:LFI27}{1.266}
\setsymbol{LFI:FWHM:uncertainty:LFI28}{1.287}

\setsymbol{HFI:center:frequency:100GHz:units}{100\,GHz}
\setsymbol{HFI:center:frequency:143GHz:units}{143\,GHz}
\setsymbol{HFI:center:frequency:217GHz:units}{217\,GHz}
\setsymbol{HFI:center:frequency:353GHz:units}{353\,GHz}
\setsymbol{HFI:center:frequency:545GHz:units}{545\,GHz}
\setsymbol{HFI:center:frequency:857GHz:units}{857\,GHz}

\setsymbol{HFI:center:frequency:100GHz}{100}
\setsymbol{HFI:center:frequency:143GHz}{143}
\setsymbol{HFI:center:frequency:217GHz}{217}
\setsymbol{HFI:center:frequency:353GHz}{353}
\setsymbol{HFI:center:frequency:545GHz}{545}
\setsymbol{HFI:center:frequency:857GHz}{857}

\setsymbol{HFI:Ndetectors:100GHz}{8}
\setsymbol{HFI:Ndetectors:143GHz}{11}
\setsymbol{HFI:Ndetectors:217GHz}{12}
\setsymbol{HFI:Ndetectors:353GHz}{12}
\setsymbol{HFI:Ndetectors:545GHz}{3}
\setsymbol{HFI:Ndetectors:857GHz}{4}

\setsymbol{HFI:FWHM:Maps:100GHz:units}{9\parcm88}
\setsymbol{HFI:FWHM:Maps:143GHz:units}{7\parcm18}
\setsymbol{HFI:FWHM:Maps:217GHz:units}{4\parcm87}
\setsymbol{HFI:FWHM:Maps:353GHz:units}{4\parcm65}
\setsymbol{HFI:FWHM:Maps:545GHz:units}{4\parcm72}
\setsymbol{HFI:FWHM:Maps:857GHz:units}{4\parcm39}
\setsymbol{HFI:FWHM:Maps:100GHz}{9.88}
\setsymbol{HFI:FWHM:Maps:143GHz}{7.18}
\setsymbol{HFI:FWHM:Maps:217GHz}{4.87}
\setsymbol{HFI:FWHM:Maps:353GHz}{4.65}
\setsymbol{HFI:FWHM:Maps:545GHz}{4.72}
\setsymbol{HFI:FWHM:Maps:857GHz}{4.39}

\setsymbol{HFI:beam:ellipticity:Maps:100GHz}{1.15}
\setsymbol{HFI:beam:ellipticity:Maps:143GHz}{1.01}
\setsymbol{HFI:beam:ellipticity:Maps:217GHz}{1.06}
\setsymbol{HFI:beam:ellipticity:Maps:353GHz}{1.05}
\setsymbol{HFI:beam:ellipticity:Maps:545GHz}{1.14}
\setsymbol{HFI:beam:ellipticity:Maps:857GHz}{1.19}

\setsymbol{HFI:FWHM:Mars:100GHz:units}{9\parcm37}
\setsymbol{HFI:FWHM:Mars:143GHz:units}{7\parcm04}
\setsymbol{HFI:FWHM:Mars:217GHz:units}{4\parcm68}
\setsymbol{HFI:FWHM:Mars:353GHz:units}{4\parcm43}
\setsymbol{HFI:FWHM:Mars:545GHz:units}{3\parcm80}
\setsymbol{HFI:FWHM:Mars:857GHz:units}{3\parcm67}

\setsymbol{HFI:FWHM:Mars:100GHz}{9.37}
\setsymbol{HFI:FWHM:Mars:143GHz}{7.04}
\setsymbol{HFI:FWHM:Mars:217GHz}{4.68}
\setsymbol{HFI:FWHM:Mars:353GHz}{4.43}
\setsymbol{HFI:FWHM:Mars:545GHz}{3.80}
\setsymbol{HFI:FWHM:Mars:857GHz}{3.67}

\setsymbol{HFI:beam:ellipticity:Mars:100GHz}{1.18}
\setsymbol{HFI:beam:ellipticity:Mars:143GHz}{1.03}
\setsymbol{HFI:beam:ellipticity:Mars:217GHz}{1.14}
\setsymbol{HFI:beam:ellipticity:Mars:353GHz}{1.09}
\setsymbol{HFI:beam:ellipticity:Mars:545GHz}{1.25}
\setsymbol{HFI:beam:ellipticity:Mars:857GHz}{1.03}

\setsymbol{HFI:CMB:relative:calibration:100GHz}{$\lsim 1\%$}
\setsymbol{HFI:CMB:relative:calibration:143GHz}{$\lsim 1\%$}
\setsymbol{HFI:CMB:relative:calibration:217GHz}{$\lsim 1\%$}
\setsymbol{HFI:CMB:relative:calibration:353GHz}{$\lsim 1\%$}
\setsymbol{HFI:CMB:relative:calibration:545GHz}{}
\setsymbol{HFI:CMB:relative:calibration:857GHz}{}

\setsymbol{HFI:CMB:absolute:calibration:100GHz}{$\lsim 2\%$}
\setsymbol{HFI:CMB:absolute:calibration:143GHz}{$\lsim 2\%$}
\setsymbol{HFI:CMB:absolute:calibration:217GHz}{$\lsim 2\%$}
\setsymbol{HFI:CMB:absolute:calibration:353GHz}{$\lsim 2\%$}
\setsymbol{HFI:CMB:absolute:calibration:545GHz}{}
\setsymbol{HFI:CMB:absolute:calibration:857GHz}{}

\setsymbol{HFI:FIRAS:gain:calibration:accuracy:statistical:100GHz}{}
\setsymbol{HFI:FIRAS:gain:calibration:accuracy:statistical:143GHz}{}
\setsymbol{HFI:FIRAS:gain:calibration:accuracy:statistical:217GHz}{}
\setsymbol{HFI:FIRAS:gain:calibration:accuracy:statistical:353GHz}{2.5\%}
\setsymbol{HFI:FIRAS:gain:calibration:accuracy:statistical:545GHz}{1\%}
\setsymbol{HFI:FIRAS:gain:calibration:accuracy:statistical:857GHz}{0.5\%}

\setsymbol{HFI:FIRAS:gain:calibration:accuracy:systematic:100GHz}{}
\setsymbol{HFI:FIRAS:gain:calibration:accuracy:systematic:143GHz}{}
\setsymbol{HFI:FIRAS:gain:calibration:accuracy:systematic:217GHz}{}
\setsymbol{HFI:FIRAS:gain:calibration:accuracy:systematic:353GHz}{}
\setsymbol{HFI:FIRAS:gain:calibration:accuracy:systematic:545GHz}{7\%}
\setsymbol{HFI:FIRAS:gain:calibration:accuracy:systematic:857GHz}{7\%}

\setsymbol{HFI:FIRAS:zero:point:accuracy:100GHz:units}{0.8\MJysr}
\setsymbol{HFI:FIRAS:zero:point:accuracy:143GHz:units}{}
\setsymbol{HFI:FIRAS:zero:point:accuracy:217GHz:units}{}
\setsymbol{HFI:FIRAS:zero:point:accuracy:353GHz:units}{1.4\MJysr}
\setsymbol{HFI:FIRAS:zero:point:accuracy:545GHz:units}{2.2\MJysr}
\setsymbol{HFI:FIRAS:zero:point:accuracy:857GHz:units}{1.7\MJysr}

\setsymbol{HFI:FIRAS:zero:point:accuracy:100GHz}{0.8}
\setsymbol{HFI:FIRAS:zero:point:accuracy:143GHz}{}
\setsymbol{HFI:FIRAS:zero:point:accuracy:217GHz}{}
\setsymbol{HFI:FIRAS:zero:point:accuracy:353GHz}{1.4}
\setsymbol{HFI:FIRAS:zero:point:accuracy:545GHz}{2.2}
\setsymbol{HFI:FIRAS:zero:point:accuracy:857GHz}{1.7}

\setsymbol{HFI:unit:conversion:100GHz:units}{0.2415\MJysrmK}
\setsymbol{HFI:unit:conversion:143GHz:units}{0.3694\MJysrmK}
\setsymbol{HFI:unit:conversion:217GHz:units}{0.4811\MJysrmK}
\setsymbol{HFI:unit:conversion:353GHz:units}{0.2883\MJysrmK}
\setsymbol{HFI:unit:conversion:545GHz:units}{0.05826\MJysrmK}
\setsymbol{HFI:unit:conversion:857GHz:units}{0.002238\MJysrmK}

\setsymbol{HFI:unit:conversion:100GHz}{0.2415}
\setsymbol{HFI:unit:conversion:143GHz}{0.3694}
\setsymbol{HFI:unit:conversion:217GHz}{0.4811}
\setsymbol{HFI:unit:conversion:353GHz}{0.2883}
\setsymbol{HFI:unit:conversion:545GHz}{0.05826}
\setsymbol{HFI:unit:conversion:857GHz}{0.002238}

\setsymbol{HFI:colour:correction:alpha=-2:V1.01:100GHz}{0.9893}
\setsymbol{HFI:colour:correction:alpha=-2:V1.01:143GHz}{0.9759}
\setsymbol{HFI:colour:correction:alpha=-2:V1.01:217GHz}{1.0007}
\setsymbol{HFI:colour:correction:alpha=-2:V1.01:353GHz}{1.0028}
\setsymbol{HFI:colour:correction:alpha=-2:V1.01:545GHz}{1.0019}
\setsymbol{HFI:colour:correction:alpha=-2:V1.01:857GHz}{0.9889}

\setsymbol{HFI:colour:correction:alpha=0:V1.01:100GHz}{1.0008}
\setsymbol{HFI:colour:correction:alpha=0:V1.01:143GHz}{1.0148}
\setsymbol{HFI:colour:correction:alpha=0:V1.01:217GHz}{0.9909}
\setsymbol{HFI:colour:correction:alpha=0:V1.01:353GHz}{0.9888}
\setsymbol{HFI:colour:correction:alpha=0:V1.01:545GHz}{0.9878}
\setsymbol{HFI:colour:correction:alpha=0:V1.01:857GHz}{1.0014}

\newcommand{\fig}[1]{Fig.~#1}

\newcommand{\eg}{\mbox{e.g.}}
\newcommand{\ie}{\mbox{i.e.}}
\newcommand{\cmb}{{CMB}}
\newcommand{\bianchi}{{Bianchi}}
\newcommand{\bianchiviih}{{Bianchi VII{$_{\lowercase{h}}$}}}
\newcommand{\spcend}{\ensuremath{\:}}

\newcommand{\dx}{\ensuremath{\mathrm{\,d}}}
\newcommand{\hub}{\ensuremath{H}}
\newcommand{\Den}{\ensuremath{\Omega}}
\newcommand{\Dentot}{\ensuremath{\Den_\mathrm{tot}}}

\newcommand{\euls}{\ensuremath{\eula, \eulb, \eulc}}
\newcommand{\eula}{\ensuremath{\alpha}}
\newcommand{\eulb}{\ensuremath{\beta}}
\newcommand{\eulc}{\ensuremath{\gamma}}
\newcommand{\el}{\ensuremath{\ell}}
\newcommand{\m}{\ensuremath{m}}
\newcommand{\shc}[3]{\ensuremath{{#1}_{{#2}{#3}}}}
\newcommand{\bx}{\ensuremath{x}}

\newcommand{\bh}{\ensuremath{h}}

\newcommand{\bvort}{\ensuremath{\left(\frac{\omega}{H}\right)_0}}

\newcommand{\bshearijinline}{\ensuremath{(\sigma_{ij}/H)_0}}
\newcommand{\bvortinline}{\ensuremath{(\omega/H)_0}}
\newcommand{\prob}{\ensuremath{P}}
\newcommand{\bparam}{\ensuremath{\Theta_{\rm B}}}
\newcommand{\cosmoparam}{\ensuremath{\Theta_{\rm C}}}
\newcommand{\fitdata}{\ensuremath{\vec{d}}}
\newcommand{\fitmodelsel}{\ensuremath{M}}
\newcommand{\fitdataalm}{\ensuremath{\shc{d}{\el}{\m}}}
\newcommand{\fittmplalm}{\ensuremath{\shc{b}{\el}{\m}}}

\newcommand{\elmax}{\ensuremath{{\ell_{\rm max}}}}
\newcommand{\fittmpl}{\ensuremath{\vec{b}}}

\newcommand{\mtrx}[1]{\ensuremath{\tens{{#1}}}}
\newcommand{\nilc}{{\tt NILC}}
\newcommand{\smica}{{\tt SMICA}}
\newcommand{\sevem}{{\tt SEVEM}}
\newcommand{\commanderruler}{{\tt Commander-Ruler}}
\def\all2103resultspapers{\nocite{planck2013-p01, planck2013-p02, planck2013-p02a, planck2013-p02d, planck2013-p02b, planck2013-p03, planck2013-p03c, planck2013-p03f, planck2013-p03d, planck2013-p03e, planck2013-p01a, planck2013-p06, planck2013-p03a, planck2013-pip88, planck2013-p08, planck2013-p11, planck2013-p12, planck2013-p13, planck2013-p14, planck2013-p15, planck2013-p05b, planck2013-p17, planck2013-p09, planck2013-p09a, planck2013-p20, planck2013-p19, planck2013-pipaberration, planck2013-p05, planck2013-p05a, planck2013-pip56, planck2013-p06b}}

\begin{document}
\author{\small
Planck Collaboration:
P.~A.~R.~Ade\inst{86}
\and
N.~Aghanim\inst{58}
\and
C.~Armitage-Caplan\inst{91}
\and
M.~Arnaud\inst{71}
\and
M.~Ashdown\inst{68, 6}
\and
F.~Atrio-Barandela\inst{18}
\and
J.~Aumont\inst{58}
\and
C.~Baccigalupi\inst{85}
\and
A.~J.~Banday\inst{94, 9}
\and
R.~B.~Barreiro\inst{65}
% ...
}
\institute{\small
APC, AstroParticule et Cosmologie, Universit\'{e} Paris Diderot, CNRS/IN2P3, CEA/lrfu, Observatoire de Paris, Sorbonne Paris Cit\'{e}, 10, rue Alice Domon et L\'{e}onie Duquet, 75205 Paris Cedex 13, France\\
\and
Aalto University Mets\"{a}hovi Radio Observatory and Dept of Radio Science and Engineering, P.O. Box 13000, FI-00076 AALTO, Finland\\
\and
African Institute for Mathematical Sciences, 6-8 Melrose Road, Muizenberg, Cape Town, South Africa\\
\and
Agenzia Spaziale Italiana Science Data Center, Via del Politecnico snc, 00133, Roma, Italy\\
\and
Agenzia Spaziale Italiana, Viale Liegi 26, Roma, Italy\\
\and
Astrophysics Group, Cavendish Laboratory, University of Cambridge, J J Thomson Avenue, Cambridge CB3 0HE, U.K.\\
\and
Astrophysics \& Cosmology Research Unit, School of Mathematics, Statistics \& Computer Science, University of KwaZulu-Natal, Westville Campus, Private Bag X54001, Durban 4000, South Africa\\
\and
CITA, University of Toronto, 60 St. George St., Toronto, ON M5S 3H8, Canada\\
\and
CNRS, IRAP, 9 Av. colonel Roche, BP 44346, F-31028 Toulouse cedex 4, France\\
\and
California Institute of Technology, Pasadena, California, U.S.A.\\
\and
Centre for Theoretical Cosmology, DAMTP, University of Cambridge, Wilberforce Road, Cambridge CB3 0WA, U.K.\\
\and
Centro de Estudios de F\'{i}sica del Cosmos de Arag\'{o}n (CEFCA), Plaza San Juan, 1, planta 2, E-44001, Teruel, Spain\\
\and
Computational Cosmology Center, Lawrence Berkeley National Laboratory, Berkeley, California, U.S.A.\\
\and
Consejo Superior de Investigaciones Cient\'{\i}ficas (CSIC), Madrid, Spain\\
\and
DSM/Irfu/SPP, CEA-Saclay, F-91191 Gif-sur-Yvette Cedex, France\\
\and
DTU Space, National Space Institute, Technical University of Denmark, Elektrovej 327, DK-2800 Kgs. Lyngby, Denmark\\
\and
D\'{e}partement de Physique Th\'{e}orique, Universit\'{e} de Gen\`{e}ve, 24, Quai E. Ansermet,1211 Gen\`{e}ve 4, Switzerland\\
\and
Departamento de F\'{\i}sica Fundamental, Facultad de Ciencias, Universidad de Salamanca, 37008 Salamanca, Spain\\
\and
Departamento de F\'{\i}sica, Universidad de Oviedo, Avda. Calvo Sotelo s/n, Oviedo, Spain\\
\and
Department of Astronomy and Astrophysics, University of Toronto, 50 Saint George Street, Toronto, Ontario, Canada\\
\and
Department of Astrophysics/IMAPP, Radboud University Nijmegen, P.O. Box 9010, 6500 GL Nijmegen, The Netherlands\\
\and
Department of Electrical Engineering and Computer Sciences, University of California, Berkeley, California, U.S.A.\\
\and
Department of Physics \& Astronomy, University of British Columbia, 6224 Agricultural Road, Vancouver, British Columbia, Canada\\
\and
Department of Physics and Astronomy, Dana and David Dornsife College of Letter, Arts and Sciences, University of Southern California, Los Angeles, CA 90089, U.S.A.\\
\and
Department of Physics and Astronomy, University College London, London WC1E 6BT, U.K.\\
\and
Department of Physics, Florida State University, Keen Physics Building, 77 Chieftan Way, Tallahassee, Florida, U.S.A.\\
\and
Department of Physics, Gustaf H\"{a}llstr\"{o}min katu 2a, University of Helsinki, Helsinki, Finland\\
\and
Department of Physics, Princeton University, Princeton, New Jersey, U.S.A.\\
\and
Department of Physics, University of Alberta, 11322-89 Avenue, Edmonton, Alberta, T6G 2G7, Canada\\
\and
Department of Physics, University of California, One Shields Avenue, Davis, California, U.S.A.\\
\and
Department of Physics, University of California, Santa Barbara, California, U.S.A.\\
\and
Department of Physics, University of Illinois at Urbana-Champaign, 1110 West Green Street, Urbana, Illinois, U.S.A.\\
\and
Dipartimento di Fisica e Astronomia G. Galilei, Universit\`{a} degli Studi di Padova, via Marzolo 8, 35131 Padova, Italy\\
\and
Dipartimento di Fisica e Scienze della Terra, Universit\`{a} di Ferrara, Via Saragat 1, 44122 Ferrara, Italy\\
\and
Dipartimento di Fisica, Universit\`{a} La Sapienza, P. le A. Moro 2, Roma, Italy\\
\and
Dipartimento di Fisica, Universit\`{a} degli Studi di Milano, Via Celoria, 16, Milano, Italy\\
\and
Dipartimento di Fisica, Universit\`{a} degli Studi di Trieste, via A. Valerio 2, Trieste, Italy\\
\and
Dipartimento di Fisica, Universit\`{a} di Roma Tor Vergata, Via della Ricerca Scientifica, 1, Roma, Italy\\
\and
Discovery Center, Niels Bohr Institute, Blegdamsvej 17, Copenhagen, Denmark\\
\and
Dpto. Astrof\'{i}sica, Universidad de La Laguna (ULL), E-38206 La Laguna, Tenerife, Spain\\
\and
European Space Agency, ESAC, Planck Science Office, Camino bajo del Castillo, s/n, Urbanizaci\'{o}n Villafranca del Castillo, Villanueva de la Ca\~{n}ada, Madrid, Spain\\
\and
European Space Agency, ESTEC, Keplerlaan 1, 2201 AZ Noordwijk, The Netherlands\\
\and
Helsinki Institute of Physics, Gustaf H\"{a}llstr\"{o}min katu 2, University of Helsinki, Helsinki, Finland\\
\and
INAF - Osservatorio Astronomico di Padova, Vicolo dell'Osservatorio 5, Padova, Italy\\
\and
INAF - Osservatorio Astronomico di Roma, via di Frascati 33, Monte Porzio Catone, Italy\\
\and
INAF - Osservatorio Astronomico di Trieste, Via G.B. Tiepolo 11, Trieste, Italy\\
\and
INAF Istituto di Radioastronomia, Via P. Gobetti 101, 40129 Bologna, Italy\\
\and
INAF/IASF Bologna, Via Gobetti 101, Bologna, Italy\\
\and
INAF/IASF Milano, Via E. Bassini 15, Milano, Italy\\
\and
INFN, Sezione di Bologna, Via Irnerio 46, I-40126, Bologna, Italy\\
\and
INFN, Sezione di Roma 1, Universit\`{a} di Roma Sapienza, Piazzale Aldo Moro 2, 00185, Roma, Italy\\
\and
IPAG: Institut de Plan\'{e}tologie et d'Astrophysique de Grenoble, Universit\'{e} Joseph Fourier, Grenoble 1 / CNRS-INSU, UMR 5274, Grenoble, F-38041, France\\
\and
IUCAA, Post Bag 4, Ganeshkhind, Pune University Campus, Pune 411 007, India\\
\and
Imperial College London, Astrophysics group, Blackett Laboratory, Prince Consort Road, London, SW7 2AZ, U.K.\\
\and
Infrared Processing and Analysis Center, California Institute of Technology, Pasadena, CA 91125, U.S.A.\\
\and
Institut N\'{e}el, CNRS, Universit\'{e} Joseph Fourier Grenoble I, 25 rue des Martyrs, Grenoble, France\\
\and
Institut Universitaire de France, 103, bd Saint-Michel, 75005, Paris, France\\
\and
Institut d'Astrophysique Spatiale, CNRS (UMR8617) Universit\'{e} Paris-Sud 11, B\^{a}timent 121, Orsay, France\\
\and
Institut d'Astrophysique de Paris, CNRS (UMR7095), 98 bis Boulevard Arago, F-75014, Paris, France\\
\and
Institute for Space Sciences, Bucharest-Magurale, Romania\\
\and
Institute of Astronomy and Astrophysics, Academia Sinica, Taipei, Taiwan\\
\and
Institute of Astronomy, University of Cambridge, Madingley Road, Cambridge CB3 0HA, U.K.\\
\and
Institute of Theoretical Astrophysics, University of Oslo, Blindern, Oslo, Norway\\
\and
Instituto de Astrof\'{\i}sica de Canarias, C/V\'{\i}a L\'{a}ctea s/n, La Laguna, Tenerife, Spain\\
\and
Instituto de F\'{\i}sica de Cantabria (CSIC-Universidad de Cantabria), Avda. de los Castros s/n, Santander, Spain\\
\and
Jet Propulsion Laboratory, California Institute of Technology, 4800 Oak Grove Drive, Pasadena, California, U.S.A.\\
\and
Jodrell Bank Centre for Astrophysics, Alan Turing Building, School of Physics and Astronomy, The University of Manchester, Oxford Road, Manchester, M13 9PL, U.K.\\
\and
Kavli Institute for Cosmology Cambridge, Madingley Road, Cambridge, CB3 0HA, U.K.\\
\and
LAL, Universit\'{e} Paris-Sud, CNRS/IN2P3, Orsay, France\\
\and
LERMA, CNRS, Observatoire de Paris, 61 Avenue de l'Observatoire, Paris, France\\
\and
Laboratoire AIM, IRFU/Service d'Astrophysique - CEA/DSM - CNRS - Universit\'{e} Paris Diderot, B\^{a}t. 709, CEA-Saclay, F-91191 Gif-sur-Yvette Cedex, France\\
\and
Laboratoire Traitement et Communication de l'Information, CNRS (UMR 5141) and T\'{e}l\'{e}com ParisTech, 46 rue Barrault F-75634 Paris Cedex 13, France\\
\and
Laboratoire de Physique Subatomique et de Cosmologie, Universit\'{e} Joseph Fourier Grenoble I, CNRS/IN2P3, Institut National Polytechnique de Grenoble, 53 rue des Martyrs, 38026 Grenoble cedex, France\\
\and
Laboratoire de Physique Th\'{e}orique, Universit\'{e} Paris-Sud 11 \& CNRS, B\^{a}timent 210, 91405 Orsay, France\\
\and
Lawrence Berkeley National Laboratory, Berkeley, California, U.S.A.\\
\and
Max-Planck-Institut f\"{u}r Astrophysik, Karl-Schwarzschild-Str. 1, 85741 Garching, Germany\\
\and
McGill Physics, Ernest Rutherford Physics Building, McGill University, 3600 rue University, Montr\'{e}al, QC, H3A 2T8, Canada\\
\and
MilliLab, VTT Technical Research Centre of Finland, Tietotie 3, Espoo, Finland\\
\and
Mullard Space Science Laboratory, University College London, Surrey RH5 6NT, U.K.\\
\and
National University of Ireland, Department of Experimental Physics, Maynooth, Co. Kildare, Ireland\\
\and
Niels Bohr Institute, Blegdamsvej 17, Copenhagen, Denmark\\
\and
Observational Cosmology, Mail Stop 367-17, California Institute of Technology, Pasadena, CA, 91125, U.S.A.\\
\and
Optical Science Laboratory, University College London, Gower Street, London, U.K.\\
\and
SB-ITP-LPPC, EPFL, CH-1015, Lausanne, Switzerland\\
\and
SISSA, Astrophysics Sector, via Bonomea 265, 34136, Trieste, Italy\\
\and
School of Physics and Astronomy, Cardiff University, Queens Buildings, The Parade, Cardiff, CF24 3AA, U.K.\\
\and
School of Physics and Astronomy, University of Nottingham, Nottingham NG7 2RD, U.K.\\
\and
Space Sciences Laboratory, University of California, Berkeley, California, U.S.A.\\
\and
Special Astrophysical Observatory, Russian Academy of Sciences, Nizhnij Arkhyz, Zelenchukskiy region, Karachai-Cherkessian Republic, 369167, Russia\\
\and
Stanford University, Dept of Physics, Varian Physics Bldg, 382 Via Pueblo Mall, Stanford, California, U.S.A.\\
\and
Sub-Department of Astrophysics, University of Oxford, Keble Road, Oxford OX1 3RH, U.K.\\
\and
Theory Division, PH-TH, CERN, CH-1211, Geneva 23, Switzerland\\
\and
UPMC Univ Paris 06, UMR7095, 98 bis Boulevard Arago, F-75014, Paris, France\\
\and
Universit\'{e} de Toulouse, UPS-OMP, IRAP, F-31028 Toulouse cedex 4, France\\
% ...
}

\title{\papertitle}
%
%

\abstract{

  The new cosmic microwave background (CMB) temperature maps from \Planck\ provide the
  highest-quality full-sky view of the surface of last scattering available to date. This allows us
  to detect possible departures from the standard model of a globally homogeneous and isotropic
  cosmology on the largest scales. We search for correlations induced by a possible non-trivial
  topology with a fundamental domain intersecting, or nearly intersecting, the last scattering
  surface (at comoving distance ${\chi_\mathrm{rec}}$), both via a direct search for matched circular
  patterns at the intersections and by an optimal likelihood search
  for specific topologies. For the latter we
  consider flat spaces with cubic toroidal (T3), equal-sided chimney (T2) and slab (T1) topologies,
  three multi-connected spaces of constant positive curvature (dodecahedral, truncated cube and
  octahedral) and two compact negative-curvature spaces. These searches yield no detection of the
  compact topology with the scale below the diameter of the last scattering surface. For most compact
  topologies studied the likelihood maximized over the orientation of the space relative to the
  observed map shows some preference for multi-connected models just larger than the diameter of the
  last scattering surface. Since this effect is also present in simulated realizations of isotropic
  maps, we interpret it as the inevitable alignment of mild anisotropic correlations with chance
  features in a single sky realization; such a feature can also be present, in milder form, when the
  likelihood is marginalized over orientations. Thus marginalized, the limits on the radius 
  ${\cal R}_{\rm i}$ of the largest sphere inscribed in topological domain (at log-likelihood-ratio
  $\Delta\textrm{ln}{\cal L}> -5$ relative to a simply-connected flat \Planck\ best-fit model) are: 
  in a flat Universe, 
  ${\cal R}_\mathrm{i}>0.92\chi_\mathrm{rec}$ for the T3 cubic torus; 
  ${\cal R}_\mathrm{i}>0.71\chi_\mathrm{rec}$ for the T2 chimney; 
  ${\cal R}_\mathrm{i}>0.50\chi_\mathrm{rec}$ for the T1 slab; 
  and in a positively curved Universe, 
  ${\cal R}_\mathrm{i}>1.03 \chi_\mathrm{rec}$ for the dodecahedral space; 
  ${\cal R}_\mathrm{i}>1.0\chi_\mathrm{rec}$ for the truncated cube; and 
  ${\cal R}_\mathrm{i}>0.89\chi_\mathrm{rec}$ for the octahedral
  space. The limit for a wider class of topologies, i.e., those predicting matching
    pairs of back-to-back circles, among them tori and the three spherical cases listed above, coming from the
    matched-circles search is ${\cal R}_\mathrm{i}>0.94\chi_\mathrm{rec}$ at
    99\,\% confidence level. Similar limits apply to a wide, although not exhaustive, range of topologies.
  
  We also perform a Bayesian search for an anisotropic global \bianchiviih\ geometry. In the
  non-physical setting where the Bianchi cosmology is decoupled from the standard cosmology, \Planck\
  data favour the inclusion of a Bianchi component with a Bayes factor of at least 1.5 units of
  log-evidence. Indeed, the Bianchi pattern is quite efficient at accounting for some of the
  large-scale anomalies found in \Planck\ data. However, the cosmological parameters that generate
  this pattern are in strong disagreement with those found from CMB anisotropy data alone. In the
  physically motivated setting where the Bianchi parameters are coupled and fitted simultaneously
  with the standard cosmological parameters, we find no evidence for a \bianchiviih\ cosmology and
  constrain the vorticity of such models to $(\omega/H)_0 < 8.1 \times 10^{-10}$ (95\,\% confidence
  level).}


\titlerunning{\papertitle}
\authorrunning{Planck Collaboration}

\maketitle
\all2103resultspapers


\section{Introduction} 
\label{sec:Introduction}

This paper, one of a set of papers associated with
the 2013 release of data from the \Planck\footnote{\Planck\
  (\url{http://www.esa.int/Planck}) is a project of the European Space
  Agency (ESA) with instruments provided by two scientific consortia
  funded by ESA member states (in particular the lead countries France
  and Italy), with contributions from NASA (USA) and telescope
  reflectors provided by a collaboration between ESA and a scientific
  consortium led and funded by Denmark.}  mission
\citep{planck2013-p01}, describes the use of \Planck\ data to limit
departures from the global isotropy and homogeneity of spacetime. We
will use \Planck's measurements of the cosmic microwave background
(CMB) to assess the properties of anisotropic geometries (i.e.,
Bianchi models) and non-trivial topologies (e.g., the torus). The
simplest models of spacetime are globally isotropic and simply
connected. Although both are supported by both local observations and
previous CMB observations, without a fundamental theory of the birth
of the Universe, observational constraints on departures from global
isotropy are necessary. General Relativity itself places no
restrictions upon the topology of the Universe, as was recognised very
early on \citep[e.g.,][]{deSitter1917}; most proposed theories of
quantum gravity predict topology-change in the early Universe which
could be visible at large scales today.

The Einstein field equations relate local properties of the curvature
to the matter content in spacetime.  By themselves they do not
restrict the global properties of the space, allowing a universe with
a given local geometry to have various global
topologies. Friedmann--Robertson-–Walker (FRW) models of the universe
observed to have the same average local properties everywhere still
have freedom to describe quite different spaces at large
scales. Perhaps the most remarkable possibility is that a vanishing or
negative local curvature ($\Omega_{K}\equiv 1-\Omega_\mathrm{tot} \ge 0$) does not
necessarily mean that our Universe is infinite.  Indeed we can still
be living in a universe of finite volume due to the global topological
multi-connectivity of space, even if described by the flat or hyperbolic FRW
solutions. In particular, quantum fluctuations can produce compact spaces of 
constant curvature, both flat  \citep[e.g.,][]{Zeldovich1984} and curved 
\citep[e.g.,][]{Coule:2000dh,Linde:2004gg}, within the inflationary scenario.

The primary CMB anisotropy alone is incapable of constraining curvature
due to the well-known geometrical degeneracy which produces identical small-scale fluctuations
when the recombination sound speed, initial fluctuations, and comoving distance to the last scattering surface
are kept constant \citep[e.g.,][]{BET97,ZalSel97,Stompor1999}.
The present results from \Planck\ \citep{planck2013-p11} can therefore place restrictive constraints
on the curvature of the Universe only when considering secondary anisotropies or non-CMB data:
$\Omega_{K} = -K(R_0H_0)^{-2} = -0.0010^{+0.0018}_{-0.0019}$ at 95\,\%,
considering CMB primary anisotropy and lensing from \Planck\ (in the natural units with $c=1$ we use throughout).
This is equivalent to constraints on the radius of curvature $R_0H_0>19$
for positive curvature ($K=+1$) and $R_0H_0>33$ for negative curvature ($K=-1$).
CMB primary anisotropy alone gives limits on $R_0H_0$ roughly a factor of two less restrictive
(and strongly dependent on priors).

Thus, the global nature of the Universe we live in is still an open
question and studying the observational effects of a possible finite
universe is one way to address it.  With topology not affecting local mean properties that are found to be
well described by FRW parameters, its main observational effect is
in setting boundary conditions on perturbation modes that can be
excited and developed into the structure that we observe. Studying
structure on the last scattering surface is the best-known way to probe
the global organisation of our Universe and the CMB provides the most
detailed and best understood dataset for this purpose.

We can also relax assumptions about the global structure of spacetime
by allowing anisotropy about each point in the Universe. This yields
more general solutions to Einstein's field equations, leading to the
so-called Bianchi cosmologies.  For small anisotropy, as demanded by
current observations, linear perturbation about the standard FRW model
may be applied.  A universal shear and rotation induce a
characteristic subdominant, deterministic signature in the \cmb, which
is embedded in the usual stochastic anisotropies.  The deterministic
\cmb\ temperature fluctuations that result in the homogenous Bianchi
models were first examined by \cite{collins:1973} and
\cite{barrow:1985} (and subsequently \citealt{barrow:1986}),
however no dark energy component was included as it was not considered
plausible at the time.  More recently, \cite{jaffe:2006b}, and
independently \cite{bridges:2006b}, extended these solutions for the
open and flat \bianchiviih\ models to include cosmologies with dark
energy.  It is these solutions to \bianchiviih\ models that we
  study in the current article.  More accurate solutions were since
  derived by \cite{pontzen:2007}, \cite{pontzen:2009} and
  \cite{pontzen:2011}, where recombination is treated in a more
  sophisticated manner and reionisation is supported.  Furthermore, we
  note that in these works
  \citep{pontzen:2007,pontzen:2009,pontzen:2011} the induced \cmb\
  polarisation contributions that arise in Bianchi models have also
  been derived, although here focus is given to temperature
  contributions.

In this paper, we will explicitly consider models of global topology and anisotropy. 
In a chaotic inflation scenario, however, our post-inflationary patch might exhibit
large-scale \emph{local} topological features (``handles'' and ``holes'')
the can mimic a \emph{global} multiply-connected topology in our observable volume. 
Similarly, it might also have residual shear or rotation which could mimic the properties of a global Bianchi spacetime.

\Planck's ability to discriminate and remove large-scale
astrophysical foregrounds \citep{planck2013-p06} reduces the
systematic error budget associated with measurements of the CMB sky significantly.
\Planck\ data therefore allow refined limits on the scale of the
topology and the presence of anisotropy. Moreover, previous work in
this field has been done by a wide variety of authors using a wide
variety of data (e.g., \textit{COBE}, \textit{WMAP} 1-year, 3-year, 5-year, etc.) and in this work we perform a coherent analysis. 

In Sect.~\ref{sec:Previous results}, we discuss previous attempts to
limit the topology and global isotropy of the Universe. In
Sect.~\ref{sec:Correlations} we discuss the signals induced in
topologically non-trivial and Bianchi universes. In
Sect.~\ref{sec:Data} the \Planck\ data we use in the analysis are presented, and in
Sect.~\ref{sec:Methods} the methods we have developed to detect those
signals are discussed. We apply those methods in Sect.~\ref{sec:Results} and discuss
the results in Sect.~\ref{sec:Discussion}.


\section{Previous results} 
\label{sec:Previous results}

The first searches for non-trivial topology on cosmic scales looked for
repeated patterns or individual objects in the distribution of
galaxies
\citep{Shvartsman1974,FangSato1983,Fagundes:1987dn,1996A&A...313..339L,1996MNRAS.283.1147R,Weatherley:2003gw,2011A&A...529A.121F}.
The last scattering surface from which the CMB is released represents
the most distant source of photons in the Universe, and hence the
largest scales with which we could probe the topology of the Universe.
This first became possible with the DMR instrument on the
\textit{COBE} satellite \citep{bennett:1996}: various searches found no
evidence for non-trivial topologies
(e.g., \citealt{Starobinskii1993,Sokolov1993,StevensScottSilk1993,deoliveira-costa1995,Levin:1998bw,
BPS1998,BPS2000b,Rocha2004}; but see also \citealt{2000MNRAS.312..712R,2000CQGra..17.3951R}),
but sparked the creation of robust statistical tools, along with
greater care in the enumeration of the possible topologies for a given
geometry (see, for example, \citealt{LachiezeRey:1995wi} and
\citealt{Levin:2002co} for reviews). With data from the \textit{WMAP}
satellite \citep{jarosik2010}, these theoretical and observational
tools were applied to a high-quality dataset for the first time.
\citet{Luminet:2003bp} and \citet{Caillerie:2007jv} claimed the low
value of the low multipoles (compared to standard $\Lambda$CDM
cosmology) as evidence for missing large-scale power as predicted in a
closed universe with a small fundamental domain (see also
\citealt{1999ApJ...524..497A,aurich2004,aurich2005,aurich2006,aurich2008,aurich2013,lew2008,roukema2008}). However,
searches in pixel space \citep{cornish2004,key2007,Niarchou:2003gi,bielewicz2009,Dineen2005}
and in harmonic space \citep{Kunz:2005wh}
determined that this was an unlikely explanation for the low power.
\cite{BPS1998,BPS2000a} and \cite{riazuelo2004a,riazuelo2004b}
presented some of the mathematical formalism for the computation of
the correlations induced by topology in a form suitable for use in
cosmological calculations. \citet{phillips2006} presented efficient
algorithms for the computation of the correlation structure of the
flat torus and applied it via a Bayesian formalism to the
\textit{WMAP} data; similar computations for a wider range of
geometries were performed by \cite{Niarchou:2007fe}.

These calculations used a variety of different vintages of the
\emph{COBE} and \emph{WMAP} data, as well as a variety of different
sky cuts (including the unmasked internal linear combination (ILC)
map, not originally intended for cosmological studies). Nonetheless,
none of the pixel-space calculations which took advantage of the full
correlation structure induced by the topology found evidence for a
multiply-connected topology with a fundamental domain within or
intersecting the last scattering surface.
Hence in this paper we will attempt to corroborate this earlier work
and put the calculations on a consistent footing.

The open and flat Bianchi type VII$_{h}$ models have been compared
previously to both the \textit{COBE} \citep{bunn:1996,kogut:1997} and
\textit{WMAP} \citep{jaffe:2005,jaffe:2006a} data, albeit ignoring
dark energy, in order to place limits on the global rotation and shear
of the Universe.
A statistically significant correlation between one of the
\bianchiviih\ models and the \textit{WMAP} ILC map
(\citealt{bennett:2003}) was first detected by
\cite{jaffe:2005}. However, it was noted that the parameters of this
model are inconsistent with standard constraints. Nevertheless, when
the \textit{WMAP} ILC map was ``corrected'' for the best-fit Bianchi
template, some of the so-called ``anomalies'' reported in
\textit{WMAP} data disappear
\citep{jaffe:2005,jaffe:2006a,cayon:2006,mcewen:2006:bianchi}. A
modified template fitting technique was performed by \cite{lm:2006}
and, although a statistically significant template fit was not
reported, the corresponding ``corrected'' \textit{WMAP} data were
again free of many large scale ``anomalies''. Subsequently,
  \citet{ghosh:2007} used the bipolar power spectrum of \textit{WMAP}
  data to constrain the amplitude of any Bianchi component in the
  \cmb.
Due to the renewed interest in \bianchi\ models, solutions to the CMB
temperature fluctuations induced in \bianchiviih\ models when
incorporating dark energy were since derived by \cite{jaffe:2006b} and
\cite{bridges:2006b}.  Nevertheless, the cosmological parameters of
the Bianchi template embedded in \textit{WMAP} data in this setting
remain inconsistent with constraints from the CMB alone
\citep{jaffe:2006c,jaffe:2006b}.  Furthermore,
  \cite{pontzen:2007} compared the polarisation power spectra of the
  best-fit \bianchiviih\ model found by \cite{jaffe:2006c} with the
  \textit{WMAP} 3-year data \citep{page2007} and also concluded that
  the model could be ruled out since it produced greater polarization
  than observed in the \textit{WMAP} data. A Bayesian analysis of
\bianchiviih\ models was performed by \cite{bridges:2006b} using
\textit{WMAP} ILC data to explore the joint cosmological and Bianchi
parameter space via Markov chain Monte Carlo sampling, where it was
again determined that the parameters of the resulting Bianchi
cosmology were inconsistent with standard constraints.  In a following
study by \cite{bridges:2008} it was suggested that the CMB ``cold
spot'' \citep{vielva:2004,cruz:2006a,vielva:2010} could be driving
evidence for a Bianchi component. Recently, this Bayesian analysis has
been revisited by \cite{mcewen:bianchi} to handle partial-sky
observations and to use nested sampling methods
\citep{skilling:2004,feroz:multinest1,feroz:multinest2}. \cite{mcewen:bianchi}
conclude that \textit{WMAP} 9-year temperature data do not favour
\bianchiviih\ cosmologies over $\Lambda$CDM.





\section{CMB correlations in anisotropic and multiply-connected universes} 
\label{sec:Correlations}

\subsection{Topology} 
\label{sub:Topology}

All FRW models can describe multi-connected universes.  In the case of flat
space, there are a finite number of compactifications, the simplest of which are those of the torus. All of them have continuous
parameters that describe the length of periodicity in some or all
directions \citep[e.g.,][]{riazuelo2004b}.  In a space of constant non-zero curvature
the situation is notably different --- the
presence of a length scale (the curvature radius $R_0$) precludes
topological compactification at an arbitrary scale. The size of the
space must now reflect its curvature, linking topological properties
to $\Omega_\mathrm{tot}=1-\Omega_{K}$.  In the case of hyperbolic spacetimes,
the list of possible compact spaces of constant negative curvature is
still infinite, but discrete \citep{Thurston:1982}, while in the positive
curvature spherical space there is only a finite set of well-proportioned possibilities (i.e., those with roughly comparable sizes in all directions; there are also the {countably} infinite lens and prism topologies)
for a multi-connected space (e.g., \citealt{Gausmann2001, riazuelo2004a}).

The effect of topology is equivalent to considering
the full simply-connected three-dimensional spatial slice of the
spacetime (known as the \emph{covering space}) as being filled with
repetitions of a shape which is finite in some or all directions (the
\emph{fundamental domain}) --- by analogy with the two-dimensional
case, we say that the fundamental domain \emph{tiles} the covering
space. For the flat and hyperbolic geometries, there are infinite copies of
the fundamental domain; for the spherical geometry, with a finite volume,
there is a finite number of tiles. Physical fields repeat their configuration
in every tile, and thus can be viewed as defined on the covering space but 
subject to periodic boundary conditions.
Topological compactification always break isotropy,
and for some topologies also the global homogeneity of physical 
fields. Positively curved
and flat  spaces studied in this paper are homogeneous, however
hyperbolic multi-connected spaces are never homogeneous.

\begin{table*}[tmb]
\centering
\caption{Parameters of analysed curved spaces.}
\label{tbl:spaces}
	\begin{tabular}{llllllcc}
           \noalign{\doubleline}
        {\bf Size} && \multicolumn{3}{c}{\bf Spherical} && \multicolumn{2}{c}{\bf Hyperbolic} \\
                   && Dodecahedral & Truncated Cube &  Octahedral &&  m004($-5$,1) & v3543(2,3)  \\ 
\noalign{\vskip 3pt\hrule\vskip 5pt}
${\cal V}/R_0^3$  && 0.16  & 0.41 & 0.82  && 0.98 & 6.45 \\
${\cal R}_{\rm i}/R_0$  && 0.31 ($\pi/10$) & 0.39 ($\pi/8$) & 0.45 && 0.54  & 0.89 \\
${\cal R}_{\rm m}/R_0$  && 0.37  & 0.56 & 0.56 && 0.64 & 1.22 \\
${\cal R}_{\rm u}/R_0$  && 0.40  & 0.58 & 0.79 ($\pi/4$) && 0.75 & 1.33 \\
\noalign{\vskip 5pt\hrule\vskip 3pt}
\end{tabular}
\end{table*}

The primary observable effect of a multi-connected universe is the
existence of directions in which light could circumnavigate the space
in cosmological time more than once, i.e., the radial distance
$\chi_\mathrm{rec}$ to the surface of last scattering exceeds the size of the
universe.  In these cases, the surface of last scattering can
intersect the (notional) edge of a fundamental domain. At this
intersection, we can view the same spacetime event from multiple
directions --- conversely, it appears in different directions when
observed from a single point. 

Thus, temperature perturbations in one direction, $T(\vec{\hat n})$,
become correlated with those in another direction $T(\vec{\hat m})$ by an
amount that differs from the usual isotropic correlation function
$C(\theta)$, where $\theta$ denotes the angle between $\vec{\hat n}$
  and $\vec{\hat m}$. Considering a pixelized map, this induces a correlation
matrix $C_{pp'}$ which depends on quantities other than the angular
distance between pixels $p$ and $p'$. This break from statistical
isotropy can therefore be used to constrain topological models. Hence,
we need to calculate the pixel-space correlation matrix or its
equivalent in harmonic space.

In this paper we consider the following topologies {using the
  likelihood method}: a) toroidal flat
models with equal-length compactification size $L$ in three
directions, denoted $T[L,L,L]$;\footnote{In a slight abuse of notation, the lengths $L_i$ will be given in units of $H_0^{-1}$ in $T[L_1,L_2,L_3]$, but in physical units elsewhere.}
 b) toroidal flat models with
different compactification lengths, parametrized by $L_x, L_y, L_z$,
denoted $T[L_x,L_y,L_z]$; c) three major types of single-action
positively curved spherical manifolds with dodecahedral, truncated
cubical and octahedral fundamental domains ($I^*$, $O^*$, $T^*$
compactification groups correspondingly, see \citealt{Gausmann2001});
and d) two sample negative curvature hyperbolic spaces, m004($-5$,1) being one of the smallest
known compact hyperbolic spaces as well as the relatively large v3543(2,3).\footnote{The nomenclature for
hyperbolic spaces follows J.~Weeks' census, as incorporated in the freely available \texttt{SnapPea} software, 
\url{http://www.geometrygames.org/SnapPea}; see also \cite{thurston1997three}.}
Scales of
fundamental domains of compactified curved spaces are fixed in the
units of curvature and are summarised in Table~\ref{tbl:spaces}, where
we quote the volume ${\cal V}$, radius of the largest sphere that can
be inscribed in the domain ${\cal R}_\mathrm{i}$ (equal to the distance to the
nearest face from the origin of the domain), the smallest sphere in
which the domain can be inscribed ${\cal R}_\mathrm{u}$ (equal to the distance
to the farthest vertex), and the intermediate scale ${\cal R}_\mathrm{m}$ that
is taken to be the distance to the edges for spherical spaces and
the ``spine'' distance for hyperbolic topologies. 
For the cubic torus with edge length $L$, these lengths are 
${\cal R}_\mathrm{i}=L/2$, ${\cal R}_\mathrm{m}=\sqrt{2}L/2$ and ${\cal R}_\mathrm{u}=\sqrt{3}L/2$.  
The ratio ${\cal R}_\mathrm{u}/{\cal R}_i$ is a good indicator of the shape of the
fundamental domain. Note that when $\chi_\mathrm{rec}$ is less than ${\cal
  R}_\mathrm{i}$, multiple images on large scales are not present, although the
$C_{pp^\prime}$ correlation matrix is still modified versus the
singly-connected limit. The effects of topology usually become strong
when $\chi_\mathrm{rec}$ exceeds the intermediate ${\cal R}_\mathrm{m}$; {conversely, for flat and nearly-flat geometries, there are limits to the allowed topologies \citep{PhysRevD.84.083507}.}

{A much wider class of topologies is explicitly constrained using the matched
  circles method. As discussed in Sect.~\ref{sec:method_circles},
  because of computational limitations we restrict our analysis to
  pairs of circles centered around antipodal points, so called
  back-to-back circles. Thus, we can constrain all topologies
  predicting pairs of such circles. }
The strongest constraints are imposed
on topologies predicting back-to-back circles in all directions
i.e., all the single action manifolds, among them tori of any shape and
the three spherical cases {considered explicitly in the likelihood analysis}. Weaker constraints
are imposed on topologies with all back-to-back circles centred on a great
circle of the celestial sphere such as half-turn, quarter-turn, third-turn and sixth-turn spaces, as well as Klein and chimney spaces. The
statistic can also constrain the multi-connected spaces predicting one pair of antipodal
matching circles such as Klein or chimney spaces with horizontal flip, vertical
flip or half-turn and slab space translated without screw motion.
Other topologies catalogued in \cite{riazuelo2004b} are not constrained by this analysis: 
the Hantzsche-Wendt space; the chimney space with half-turn and
flip; the generic slab space; the slab space with flip; spherical manifolds
with double and linked action; and all the hyperbolic topologies
{including those two cases considered using the likelihood method.}

\subsubsection{Computing correlation matrices}

The CMB temperature pixel-pixel correlation matrix is {defined as the ensemble-average product of the temperature at two different pixels:}
\begin{equation}
	C_{pp^\prime} = \left< T_p \; T_{p^\prime} \right> \;.
\end{equation}
It can be calculated as a double radial
integral of the ensemble average of the product of the source
functions that describe the transport of photons through the universe
from the last scattering surface to the observer:
\begin{equation}
C_{pp^\prime} = \int_0^{\chi_\mathrm{rec}} d\chi \int_0^{\chi_\mathrm{rec}} d \chi^\prime \langle S (\chi \vec{\hat q}_p) 
S (\chi^\prime \vec{\hat q}_{p^\prime}) \rangle ~,
\end{equation}
where $\vec{\hat q}_p$ and $\vec{\hat q}_{p^\prime}$ are unit vectors that point
at pixels $p$ and $p^\prime$ on the sky, and $\chi$ and $\chi^\prime$
are proper distances along radial rays pointing towards the last
scattering surface.

Two techniques have been developed to compute the CMB correlation
function for multiply-connected universes.  In one approach, one
constructs the orthonormal set of basis functions that satisfy the
boundary conditions imposed by compactification (eigenfunctions of the
Laplacian operator furnish such a basis), and assembles the {\it
  spatial} correlation function of the source $ \langle S (\chi \hat
q_p ) S ( \chi^\prime \vec{\hat q}_{p^\prime}) \rangle$ from such a basis
\citep{Cornish1999,Lehoucq2002}.  In the other approach, one applies the method
of images to create the compactified version of $ \left\langle S ( \chi
  \vec{\hat q}_p ) S ( \chi^\prime \vec{\hat q}_{p^\prime})
\right\rangle^{\rm c}$ from
the one computed on the universal covering space by resumming the
latter over the images of the 3D spatial positions $\chi \vec{\hat q}_p$
\citep{BPS1998,BPS2000a,BPS2000b}:
\begin{equation}
\left\langle S ( \chi\vec{\hat q}_p ) S ( \chi^\prime \vec{\hat
    q}_{p^\prime}) \right\rangle^{\rm c}
= \widetilde{\sum_{\gamma\in\Gamma}} \langle {\cal S} ( \chi\vec{\hat q}_p)
\gamma[{\cal S} (\gamma[\chi^\prime \vec{\hat q}_{p^\prime} ])]
\rangle^{\rm u} ,
\label{moi_source}
\end{equation}
where the superscripts ${\rm c}$ and ${\rm u}$ refer to the quantity in the
multiply-connected space and its universal cover, respectively.  The tilde
refers to the need for sum regularization in the models with an infinite set of images, 
e.g., hyperbolic and flat toroidal ones.
$\Gamma$ is the discrete subgroup of motions which defines the
multiply-connected space and $\gamma[\vec{x}]$ is the spatial point on
the universal cover obtained by the action of the motion
$\gamma\in\Gamma$ on the point $\vec{x}$. {Note that we can consider the location of one of the pixels as fixed and consider the action of $\gamma$ on the other due to symmetry.} This equation defines the action of $\gamma$ on the source function itself, needed unless all the terms in the source function are
scalar quantities (which is the case if one limits consideration to Sachs-Wolfe terms)
when the action is trivial.

Both methods are general, but have
practical considerations to take into account when one increases the
pixel resolution.  For computing $C_{pp^\prime}$ up to the resolution
corresponding to harmonic mode $\ell \approx 40$ both methods have
been tested and were found to work equally well.  In this paper we
employ both approaches.

The main effect of the compactification is that $C_{pp^\prime}$ is no
longer a function of the angular separation between the pixels $p$ and
$p^\prime$ only, due to the lack of global isotropy.  In harmonic
space the two-point correlation function of the CMB is given by
\begin{equation}\label{eq:corrmat}
C_{\ell \ell'}^{m m'} = \langle a_{\ell m} a^*_{\ell' m'} \rangle \ne C_\ell \delta_{\ell \ell'} \delta_{m m'}  \, ,
\end{equation}
where $\delta_{\ell \ell'}$ is the Kronecker delta symbol and $a_{\ell
  m}$ are the spherical harmonic coefficients of the temperature on
the sky when decomposed into the spherical harmonics $Y_{\ell
  m}({\vec{\hat q}})$ by
\begin{equation}
	T({\vec{\hat q}}) = \sum_{\ell m} a_{\ell m} Y_{\ell m}({\vec{\hat q}}) \ .
\end{equation}
Note that the two-point correlation function $C_{\ell \ell'}^{m m'}$
is no longer diagonal, nor is it $m$-independent, as in an isotropic
universe.

A flat universe provides an example when the eigenfunctions of the
Laplacian are readily available in a set of plane waves.  The
topological compactification in the flat space discretizes the
spectrum of the wavevector magnitudes $\vec{k}^2$ and selects the
subset of allowed directions.  For example, for a toroidal universe
the length of the fundamental cell needs to be an integer multiple of
the wavelength of the modes. We therefore recover a discrete sum
over modes $\vec{k}_{\bf n} = (2 \pi / L) {\bf n}$ for ${\bf n} =
(n_x,n_y,n_z)$ a triplet of integers, instead of an integral over
$\vec{k}$,
\begin{eqnarray}\label{topocorr}
C_{\ell \ell'}^{m m'}  &\propto&\int d^3 k \Delta_{\ell}(k, \Delta\eta) \Delta_{\ell'}(k, \Delta\eta) P(k)
\, \rightarrow  \nonumber \\
&&\sum_{\bf n}  \Delta_{\ell}(k_n, \Delta\eta) \Delta_{\ell'}(k_n, \Delta\eta) P(k_n) Y_{\ell m}({\bf \hat{n}})
Y_{\ell' m'}^* ({\bf \hat{n}}) \, ,\nonumber\\
\end{eqnarray}
where $\Delta_\ell (k, \Delta\eta)$ is the radiation transfer
function \citep[e.g.,][]{Bond:1987uc,Seljak1996}. We refer to the cubic torus with three equal sides as the T3 topology; it is also possible for the fundamental domain
to be compact in only two spatial dimensions (e.g., the so-called T2 ``chimney'' space) or one (the T1 ``slab'', similar to the ``lens'' spaces available in manifolds with constant positive curvature)
in which case the sum is replaced by an integral in those directions. These models serve as approximations
to modifications to the local topology of the global manifold (albeit on cosmological scales):
for example, the chimney space can mimic a ``handle'' connecting different regions of an approximately flat manifold.

\begin{figure}
	\centering
\includegraphics[width=0.32\columnwidth]{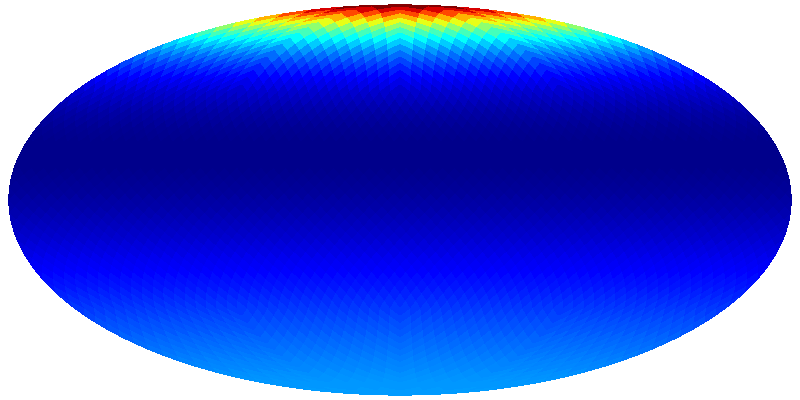}
\\
\includegraphics[width=0.32\columnwidth]{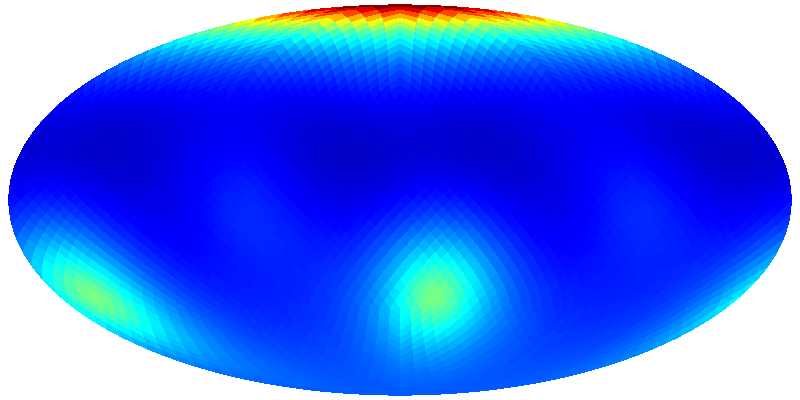}
\includegraphics[width=0.32\columnwidth]{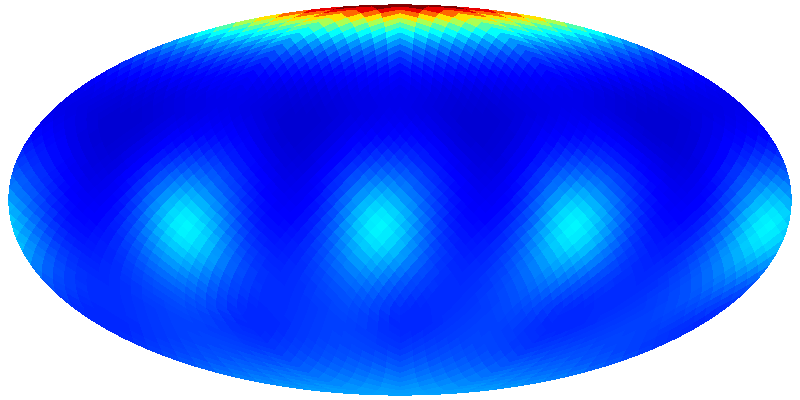}
\includegraphics[width=0.32\columnwidth]{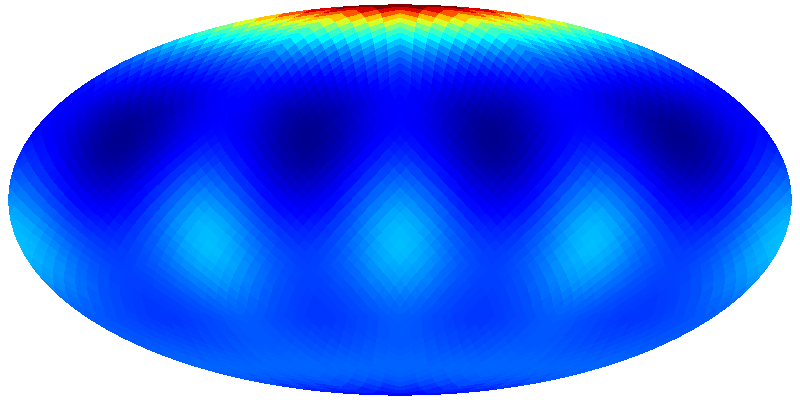}
\\
\includegraphics[width=0.32\columnwidth]{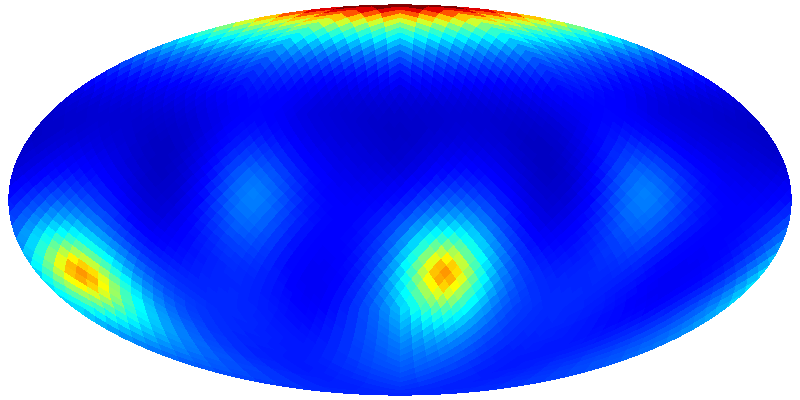}
\includegraphics[width=0.32\columnwidth]{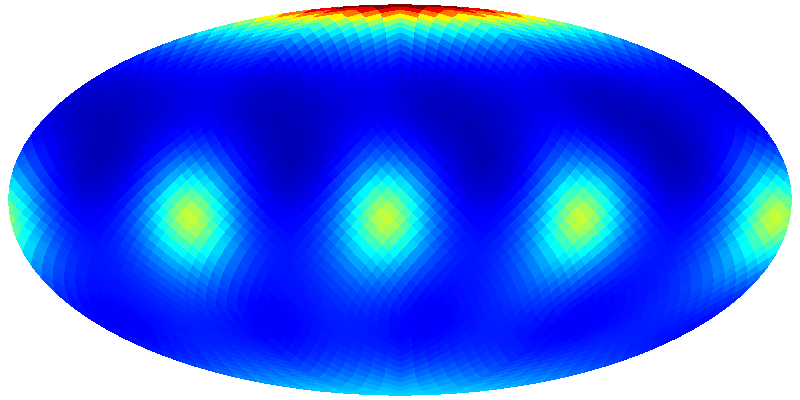}
\includegraphics[width=0.32\columnwidth]{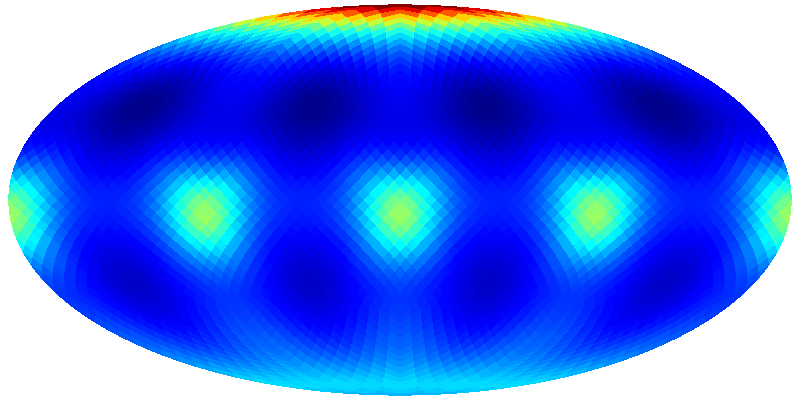}
\\
\includegraphics[width=0.32\columnwidth]{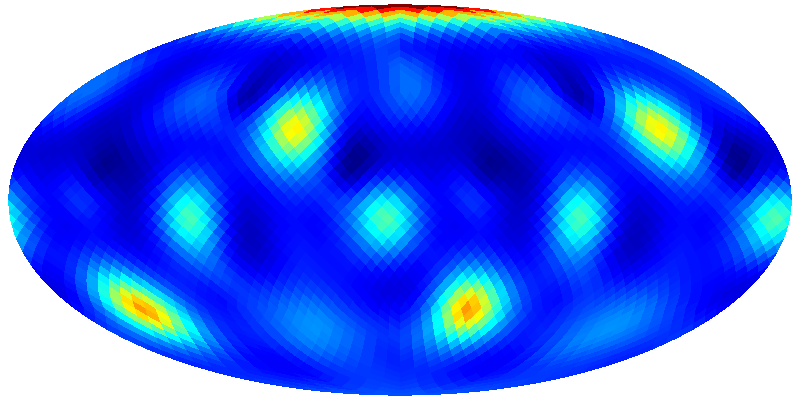}
\includegraphics[width=0.32\columnwidth]{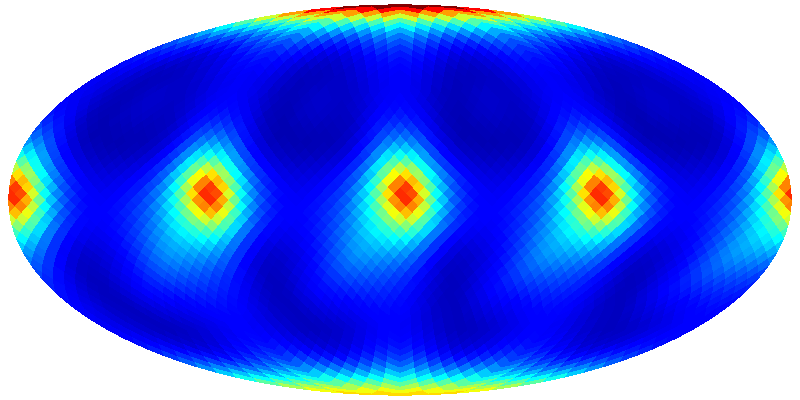}
\includegraphics[width=0.32\columnwidth]{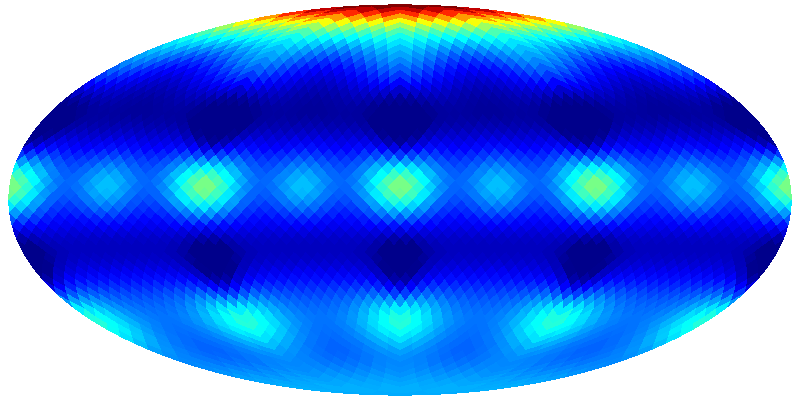}
\caption{The top row shows the correlation structure (i.e., a single row of the correlation matrix) of a simply-connected universe with isotropic correlations. For subsequent rows, the left and middle column show positively curved multiply-connected spaces (left: dedocahedral, middle: octahedral) and the right column shows equal sided tori.
The upper row of three maps corresponds to the case when the size of the fundamental domain
is of the size of the diameter to the last scattering surface and hence the first 
evidence for large angle excess correlation appears. Subsequent rows correspond
to decreasing fundamental domain size with respect to the last scattering diameter, with
parameters roughly chosen to maintain the same ratio between the models.
}
\label{fig:pixcorr}
\end{figure}

\begin{figure}[htbp]
	\centering
\includegraphics[width=0.33\columnwidth]{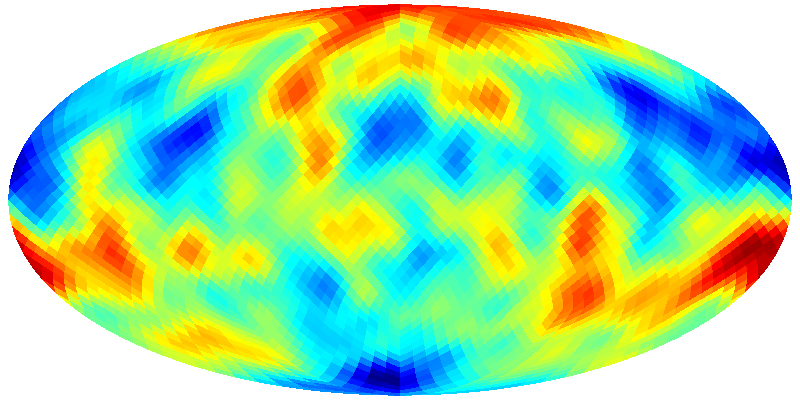}
\\
\includegraphics[width=0.32\columnwidth]{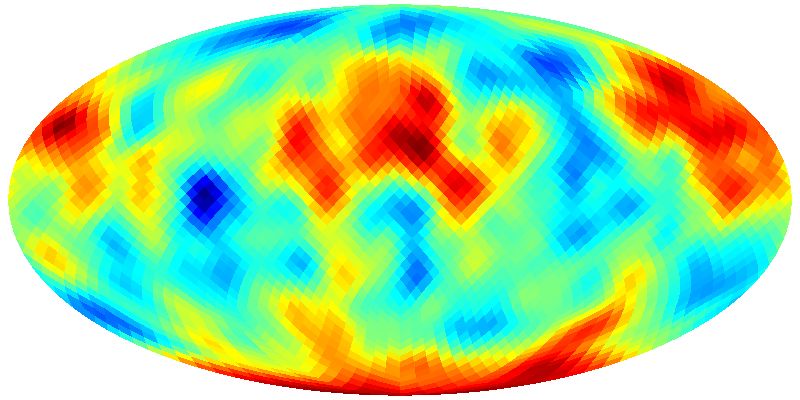}
\includegraphics[width=0.32\columnwidth]{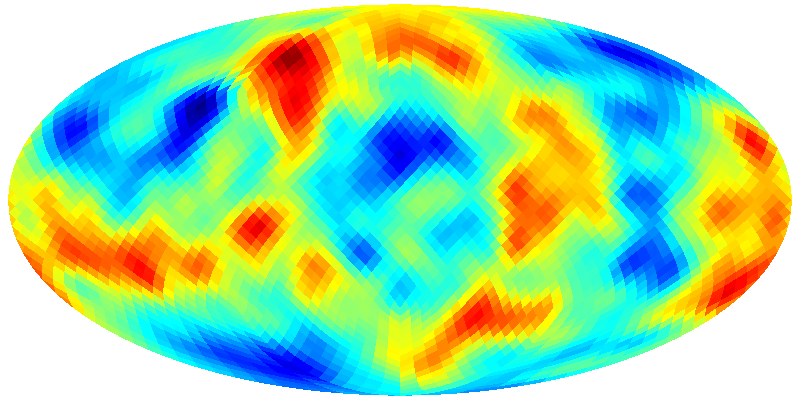}
\includegraphics[width=0.32\columnwidth]{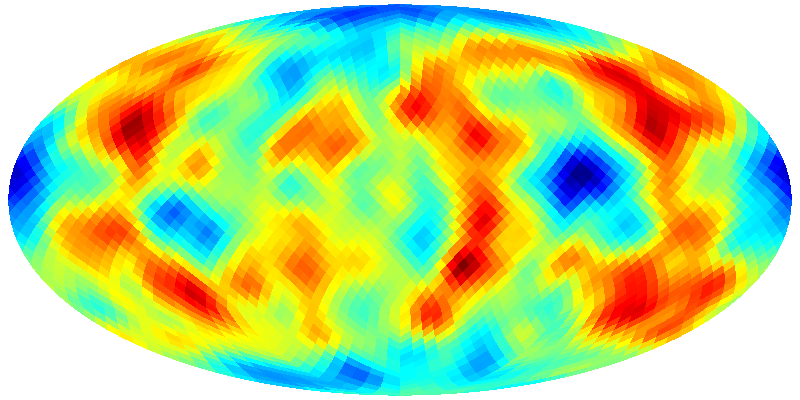}
\\
\includegraphics[width=0.32\columnwidth]{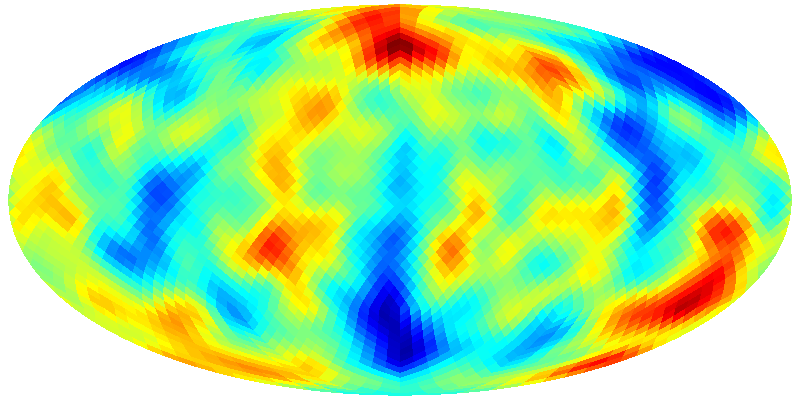}
\includegraphics[width=0.32\columnwidth]{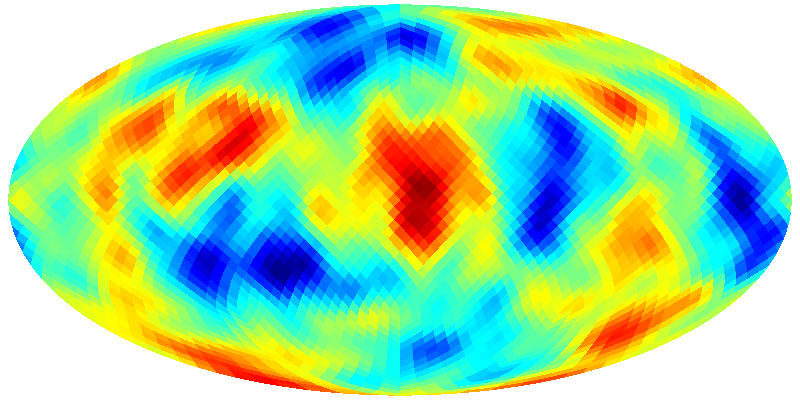}
\includegraphics[width=0.32\columnwidth]{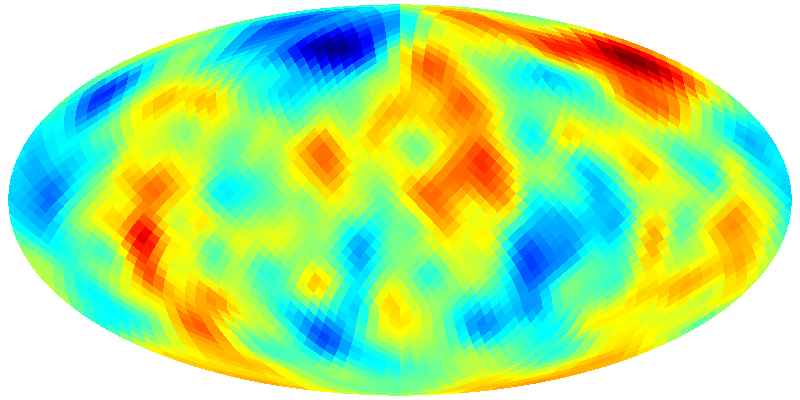}
\\
\includegraphics[width=0.32\columnwidth]{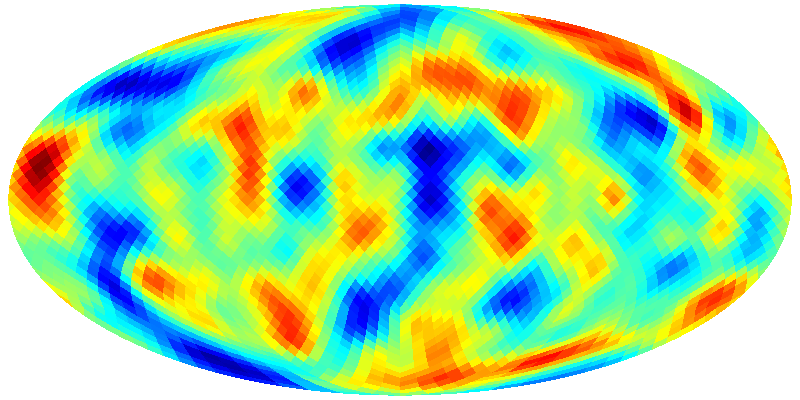}
\includegraphics[width=0.32\columnwidth]{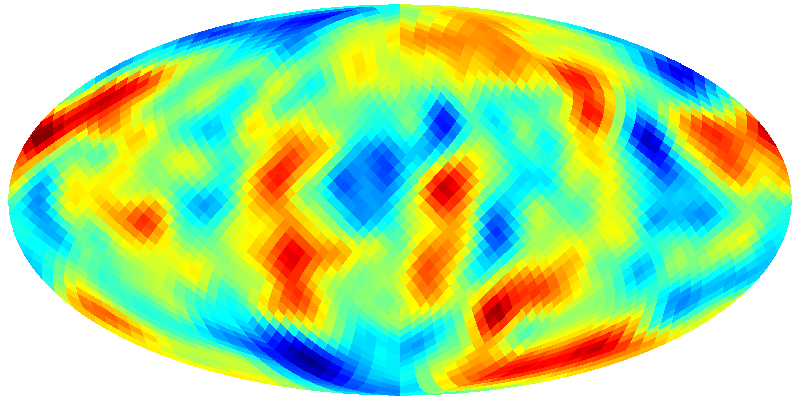}
\includegraphics[width=0.32\columnwidth]{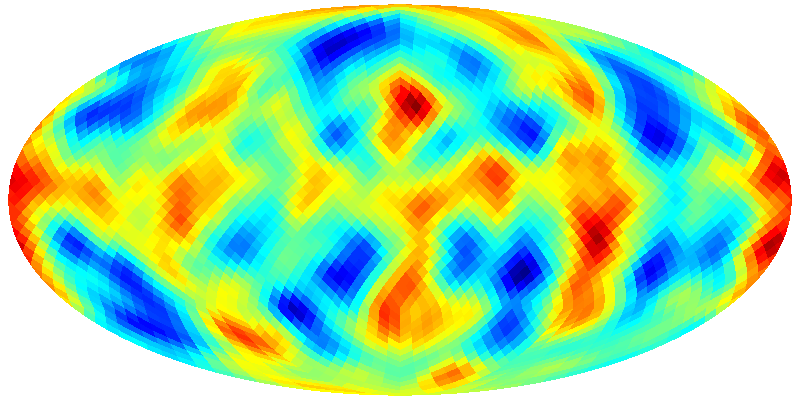}
\caption{Random realisations of temperature maps for the models in Fig.~\ref{fig:pixcorr}.
The maps are smoothed with a Gaussian filter with
full-width-half-maximum ${\rm FWHM}=640\,\arcm$.}
\label{fig:pixcorr_realization}
\end{figure}

In Fig.~\ref{fig:pixcorr} we show rows of the pixel-space
correlation matrix for a number of multiply-connected topologies as a
map, showing the magnitude of the correlation within a particular pixel.
For the simply-connected case, the map simply shows the same information
as the correlation function $C(\theta)$; for the topologically
non-trivial cases, we see the correlations depend on distance and
direction and differ from pixel to pixel (i.e., from row to row of the
matrix).  In Fig.~\ref{fig:pixcorr_realization} we show example maps
of CMB anisotropies in universes with these topologies, created by
direct realisations of Gaussian fields with the correlation matrices
of Fig.~\ref{fig:pixcorr}.

\subsection{Bianchi} 
\label{sub:Bianchi}

Bianchi cosmologies include the class of homogeneous but anisotropic
cosmologies, where the assumption of isotropy about each point in the
Universe is relaxed. For small anisotropy, as demanded by current
observations, linear perturbation about the standard FRW model may be
applied, leading to a subdominant, deterministic contribution to the
\cmb\ fluctuations.  In this setting \cmb\ fluctuations may be viewed
as the sum of a deterministic Bianchi contribution and the usual
stochastic contribution that arises in the $\Lambda$CDM model.  The
deterministic \cmb\ temperature fluctuations that result in the
Bianchi models were derived by \cite{barrow:1985}, although
no dark energy component was included.  More recently,
\cite{jaffe:2006b}, and independently \cite{bridges:2006b}, extended
these solutions for the open and flat \bianchiviih\ models to include
cosmologies with dark energy.  We defer the details of the CMB
temperature fluctuations induced in Bianchi models to these works and 
give only a brief description here.

\bianchiviih\ models describe a universe with overall rotation,
parameterized by an angular velocity, $\omega$, and a
three-dimensional rate of shear, parameterized by the tensor
$\sigma_{ij}$; we take these to be relative to the $z$ axis.  The
model has a free parameter, first identified by \cite{collins:1973},
describing the comoving length-scale over which the principal axes of
shear and rotation change orientation.  The ratio of this length scale
to the present Hubble radius is typically denoted $\bx$, which defines
the \bh\ parameter of type VII${}_\bh$ models through
\citep{barrow:1985}
\begin{equation}
\bx = \sqrt{\frac{\bh}{1-\Omega_\mathrm{tot}}}
\spcend ,
\end{equation}
where the total energy density $\Omega_\mathrm{tot} = \Omega_{\rm m} +
\Omega_{\Lambda}$.  The parameter \bx\ acts to change the ``tightness''
of the spiral-type \cmb\ temperature contributions that arise due to
the geodesic focusing of \bianchiviih\ cosmologies.  
The shear modes $\sigma_{ij}$ of combinations of orthogonal coordinate
axes are also required to describe a Bianchi cosmology.  
The present dimensionless vorticity \bvortinline\ may be related to the
dimensionless shear modes \bshearijinline\ by \citep{barrow:1985}
\begin{equation}
\label{eqn:vort}
\bvort= \frac{(1+\bh)^{1/2} (1+9\bh)^{1/2}}{6 \bx^2\Dentot} 
\sqrt{\left(\frac{\sigma_{12}}{H}\right)_0^2 + \left(\frac{\sigma_{13}}{H}\right)_0^2}
\spcend ,
\end{equation}
where \hub\ is the Hubble parameter.  
{The spherical harmonic coefficients of the \bianchiviih\ induced
  temperature component are proportional to \mbox{$[(\sigma_{12} \pm i
    \sigma_{13})/H]_0 $} and are non-zero for azimuthal modes $m=\mp
  1$ only \citep{barrow:1985, mcewen:2006:bianchi,
    pontzen:2007}. Hence, varying the phase of
  $\sigma_{12}+i\sigma_{13}$ corresponds to an azimuthal rotation,
  i.e.\ a change of coordinates, while the rotationally invariant part
  depends on $\sigma_{12}^2+\sigma_{13}^2$, and we are thus free to
  choose equality of shear modes $\sigma=\sigma_{12}=\sigma_{13}$
  \citep{pontzen:2007}, which we do for consistency with previous
  studies (e.g.\ \citealt{jaffe:2005}).}
The amplitude of the deterministic CMB
temperature fluctuations induced in \bianchiviih\ cosmologies may be
characterised by either $(\sigma/H)_0$ or $(\omega/H)_0$ since these
parameters influence the amplitude of the induced temperature
contribution only and not its morphology.  The handedness of the
coordinate system is also free in \bianchiviih\ models, hence both
left- and right-handed models arise.
Since the Bianchi-induced temperature fluctuations are anisotropic on
the sky the orientation of the resulting map may vary also,
introducing three additional degrees-of-freedom.  The orientation of
the map is described by the Euler angles\footnote{The active $zyz$
  Euler convention is adopted, corresponding to the rotation of a
  physical body in a {\it{fixed}} coordinate system about the $z$, $y$
  and $z$ axes by $\gamma$, $\beta$ and $\alpha$ respectively.}
$(\euls)$, where for $(\euls)=(0^\circ,0^\circ,0^\circ)$ the swirl
pattern typical of Bianchi templates is centred on the South pole.

Examples of simulated \bianchiviih\ CMB temperature maps are
illustrated in \fig{\ref{fig:bianchi_sims}} for a range of parameters.
In the analysis performed herein the {\tt
  BIANCHI2}\footnote{\url{http://www.jasonmcewen.org/}} 
  \citep{mcewen:bianchi} code is used
to simulate the temperature fluctuations induced in \bianchiviih\
models.  \bianchiviih\ models induce only large scale temperature
fluctuations in the \cmb\ and consequently Bianchi maps have a
particularly low band-limit, both globally and azimuthally (\ie, in
both \el\ and \m\ in spherical harmonic space; indeed, {as mentioned} only those
harmonic coefficients with $\m=\pm1$ are non-zero).

\newlength{\plotwidth}
\setlength{\plotwidth}{0.3\columnwidth}

\begin{figure}
\centering

\mbox{
  \includegraphics[bb= 0 40 800 440,clip=,width=\plotwidth]{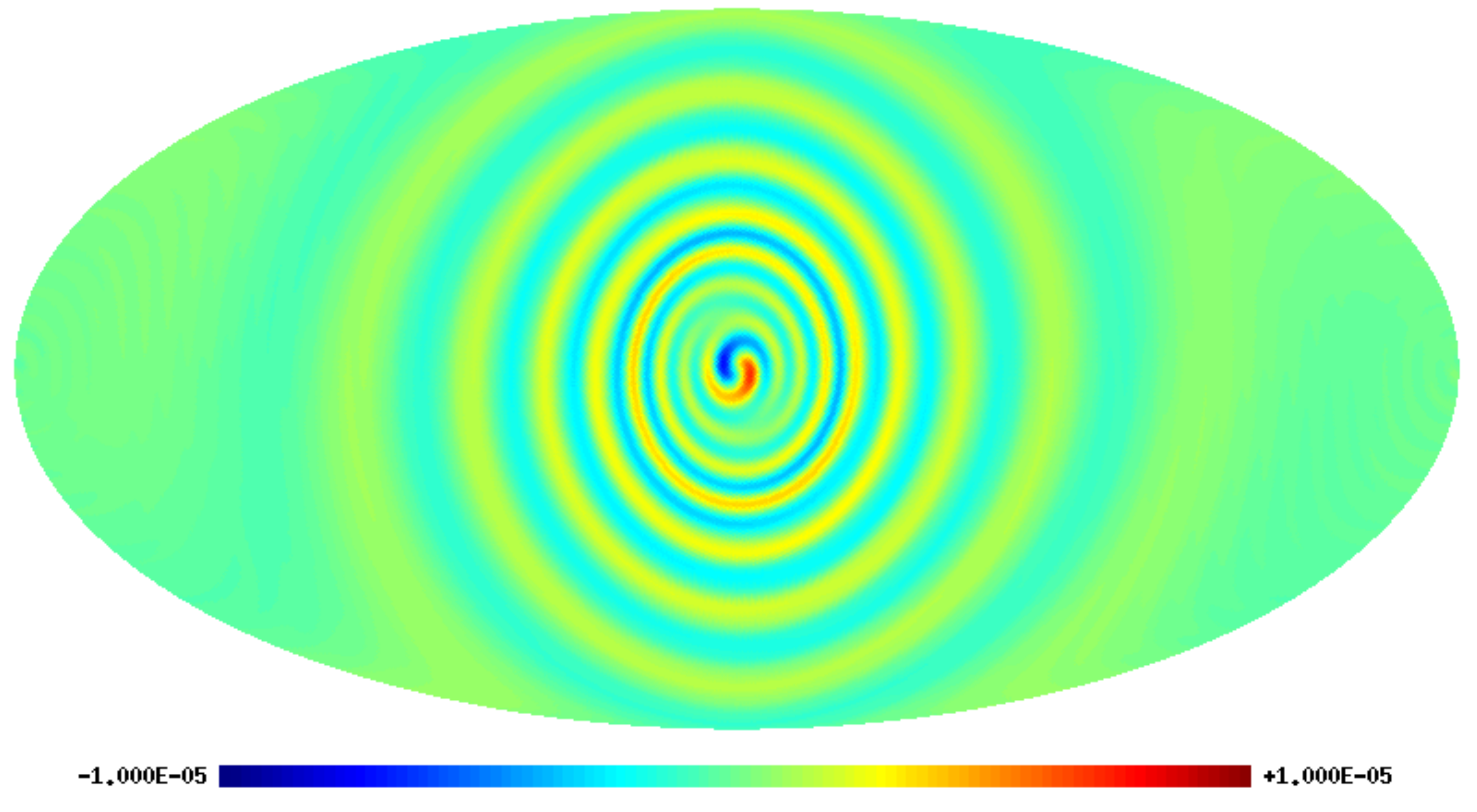}
  \includegraphics[bb= 0 40 800 440,clip=,width=\plotwidth]{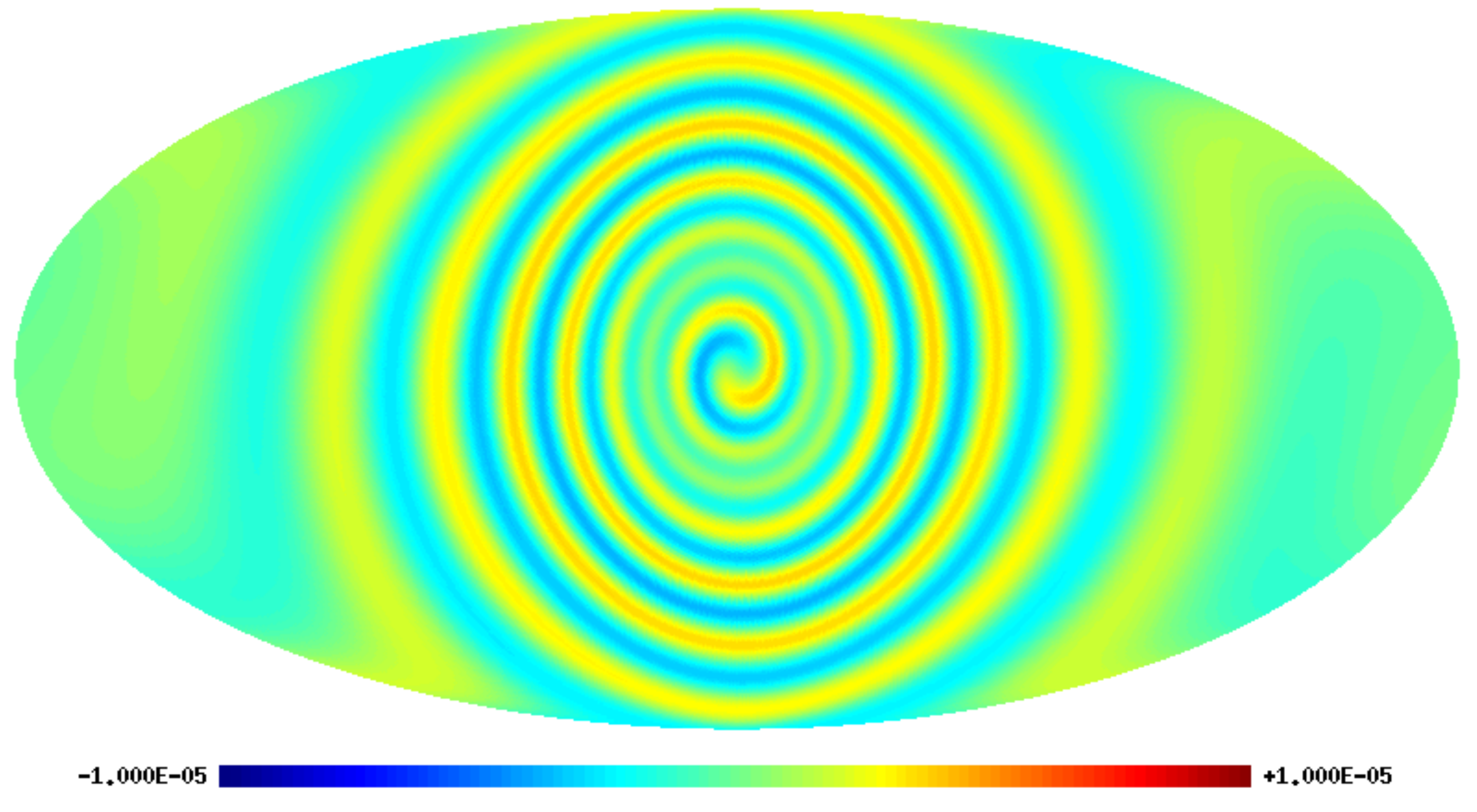} 
  \includegraphics[bb= 0 40 800 440,clip=,width=\plotwidth]{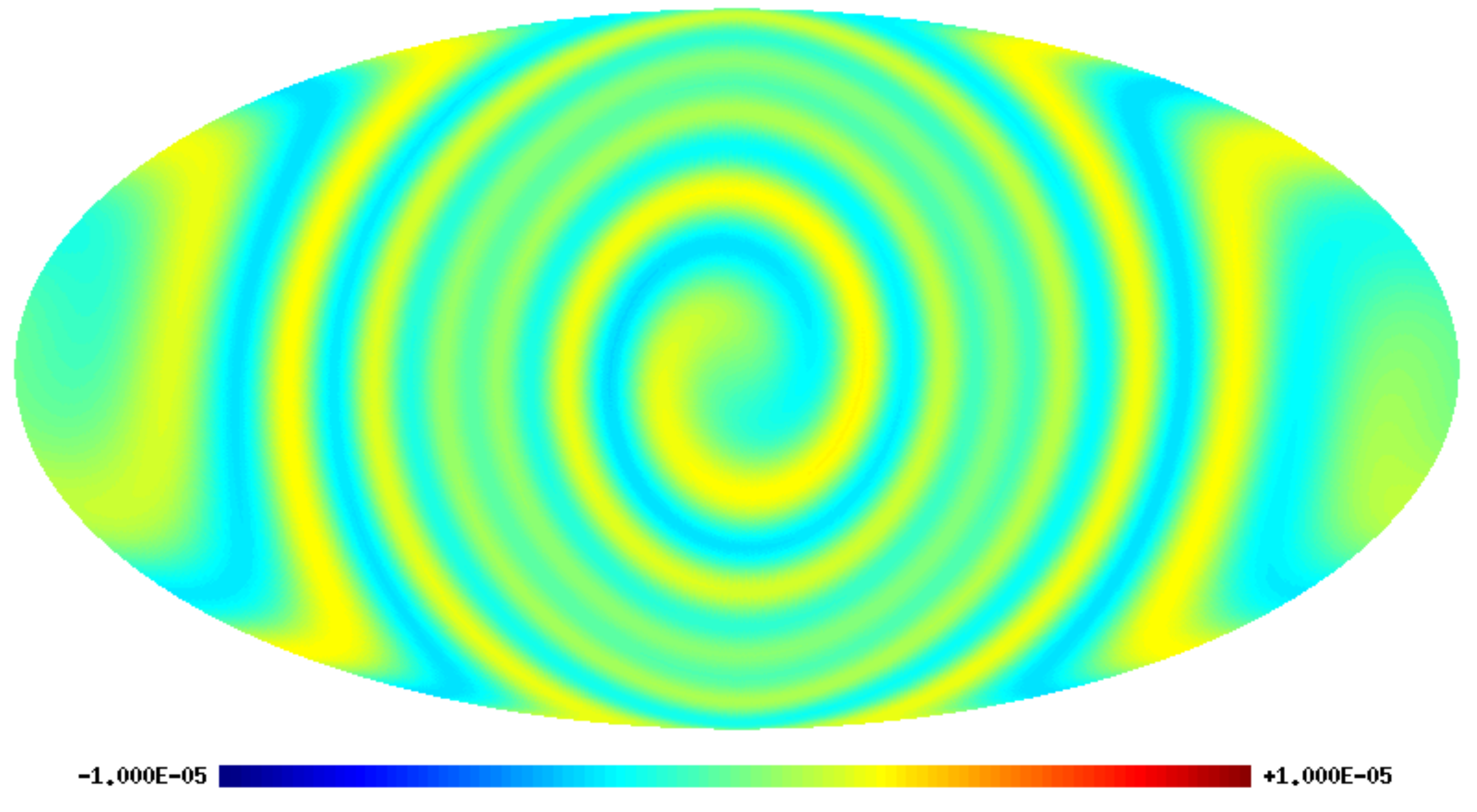}
}
\mbox{
  \includegraphics[bb= 0 40 800 440,clip=,width=\plotwidth]{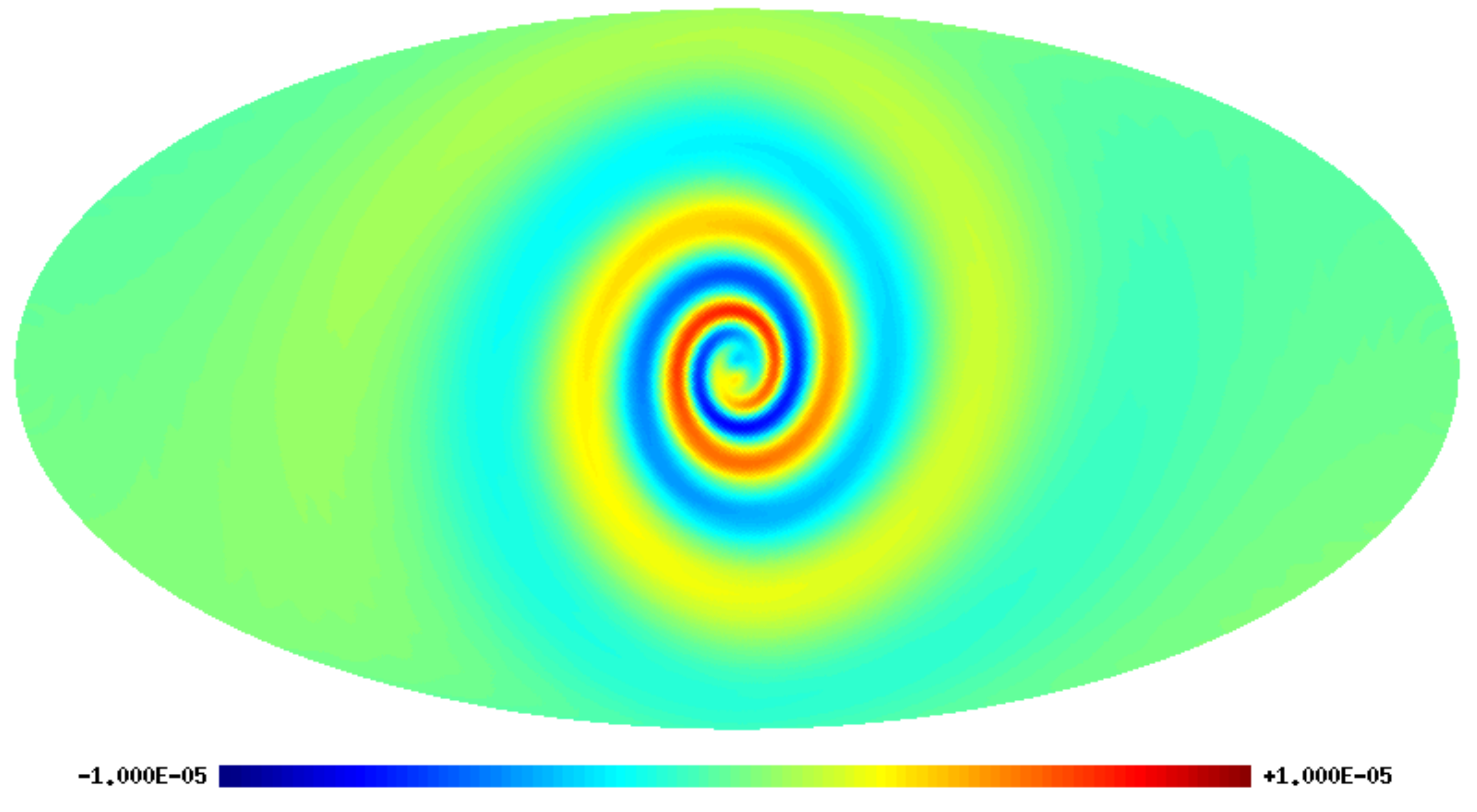} 
  \includegraphics[bb= 0 40 800 440,clip=,width=\plotwidth]{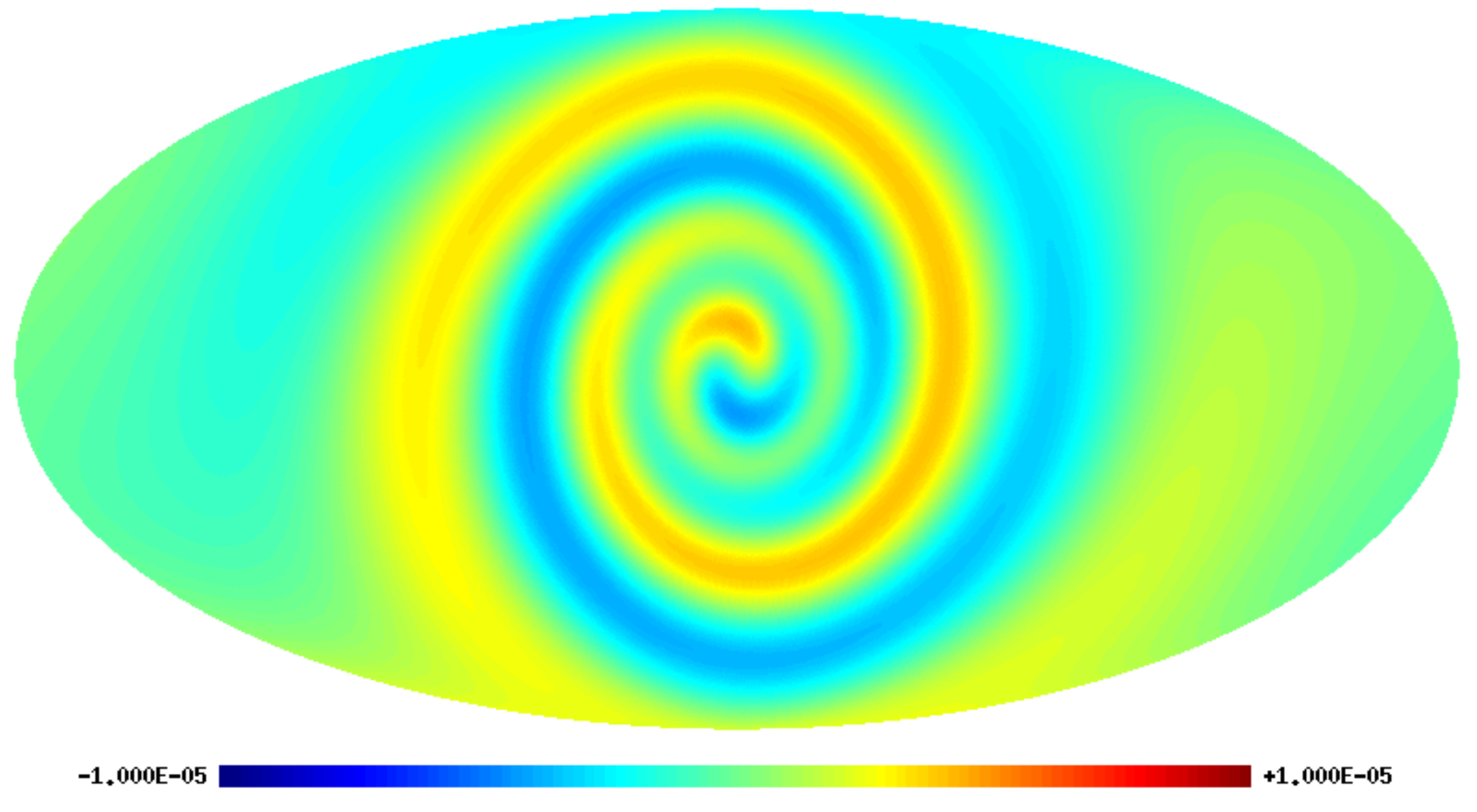} 
  \includegraphics[bb= 0 40 800 440,clip=,width=\plotwidth]{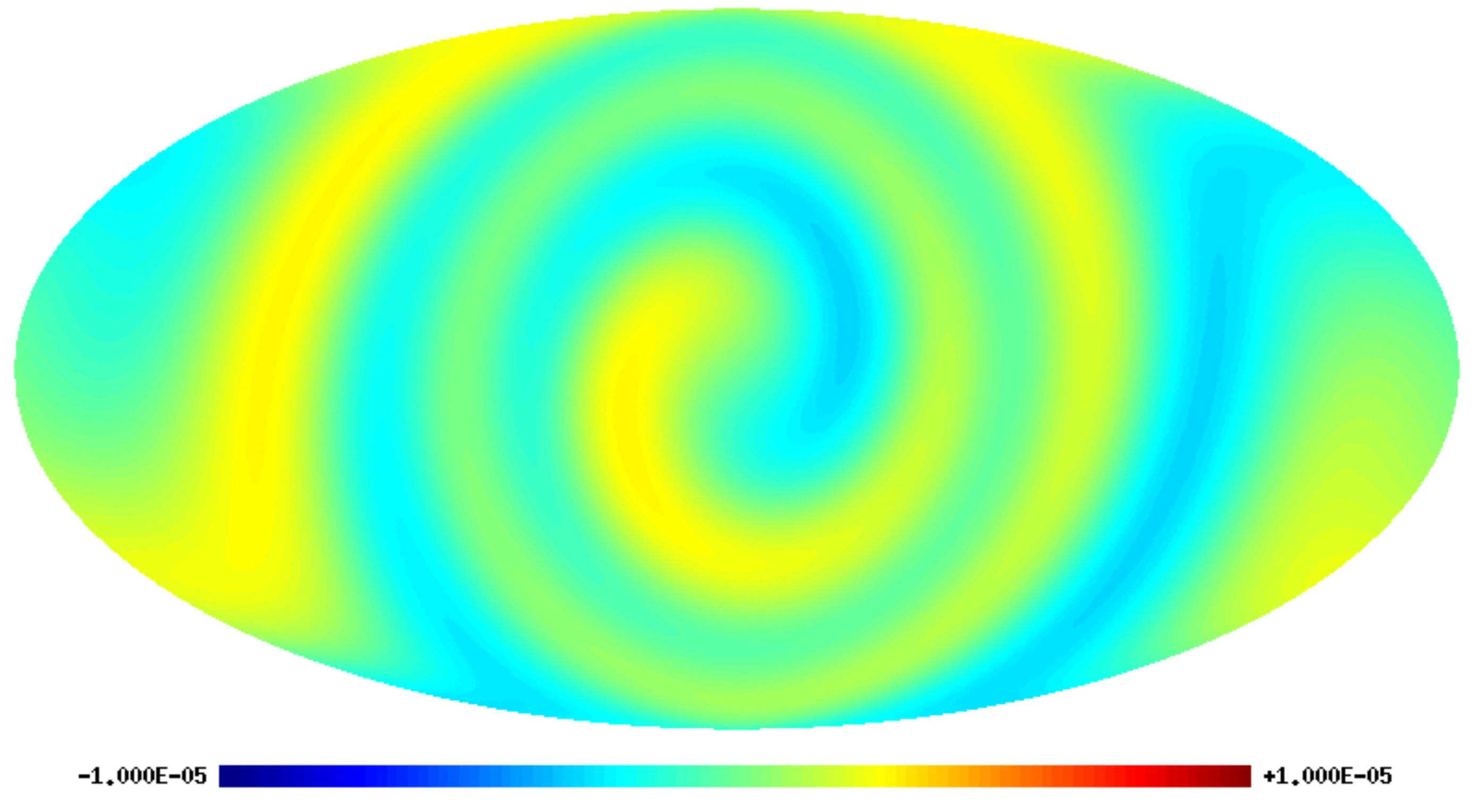}
}
\mbox{
  \includegraphics[bb= 0 40 800 440,clip=,width=\plotwidth]{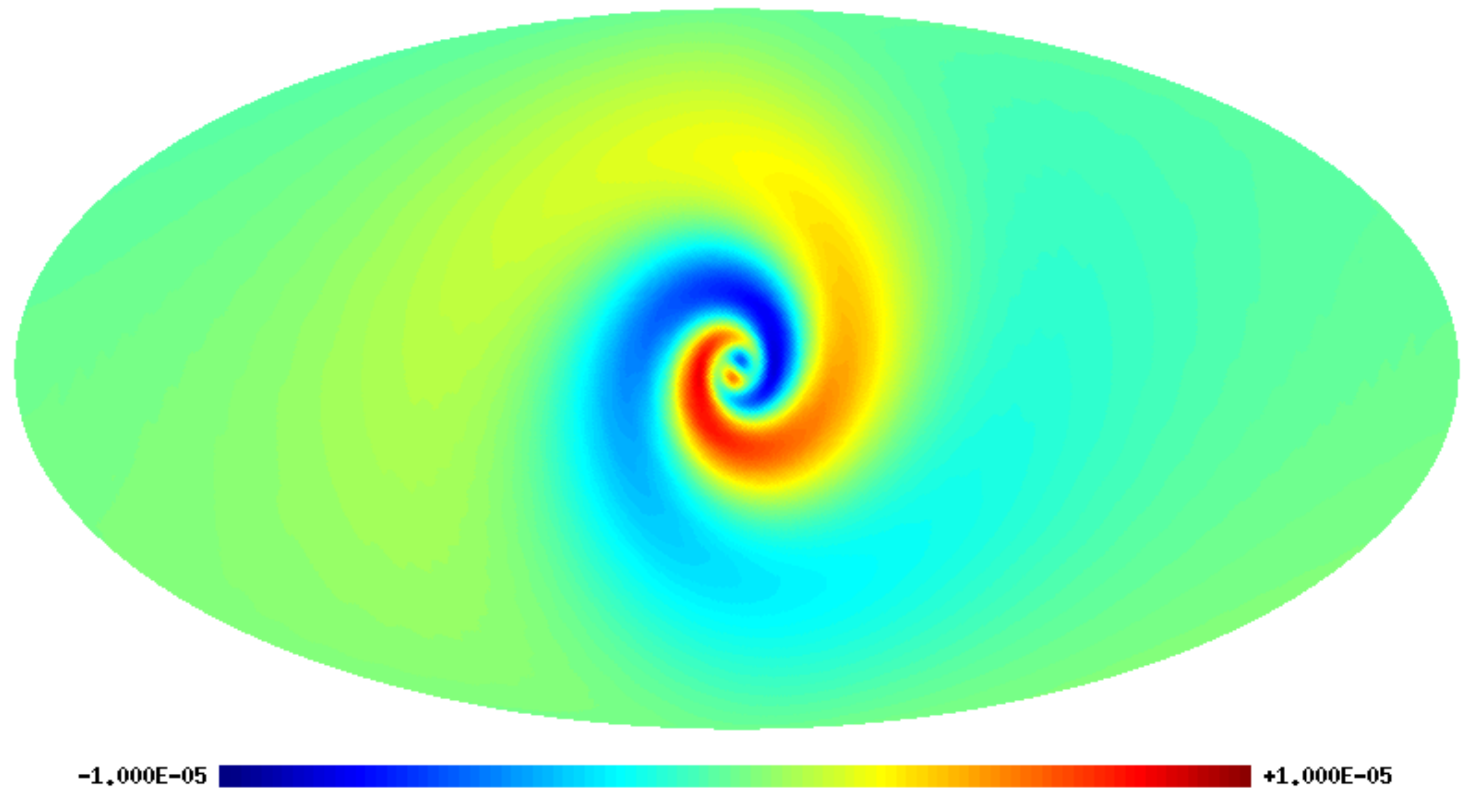} 
  \includegraphics[bb= 0 40 800 440,clip=,width=\plotwidth]{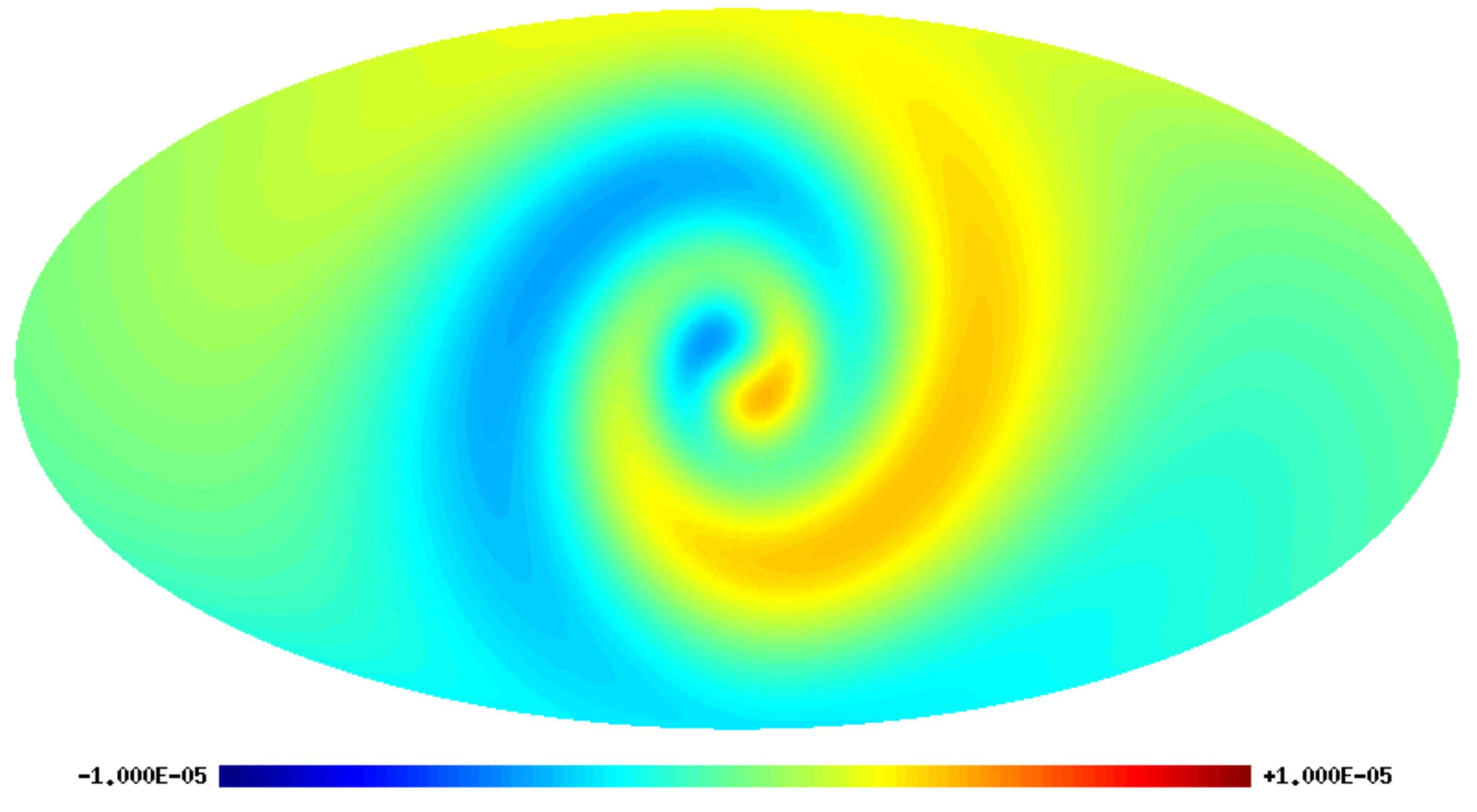} 
  \includegraphics[bb= 0 40 800 440,clip=,width=\plotwidth]{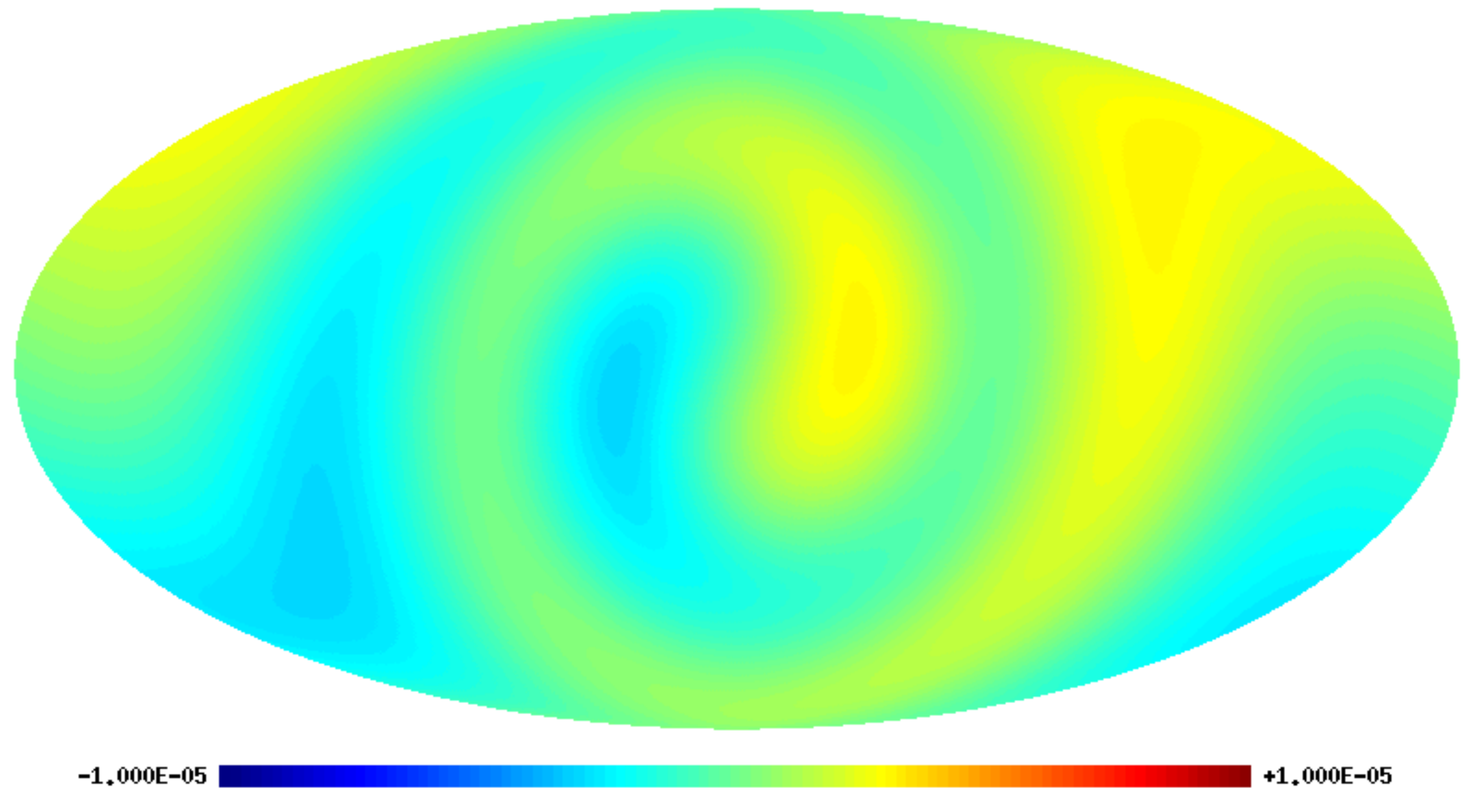}
}
\mbox{
  \includegraphics[bb= 0 40 800 440,clip=,width=\plotwidth]{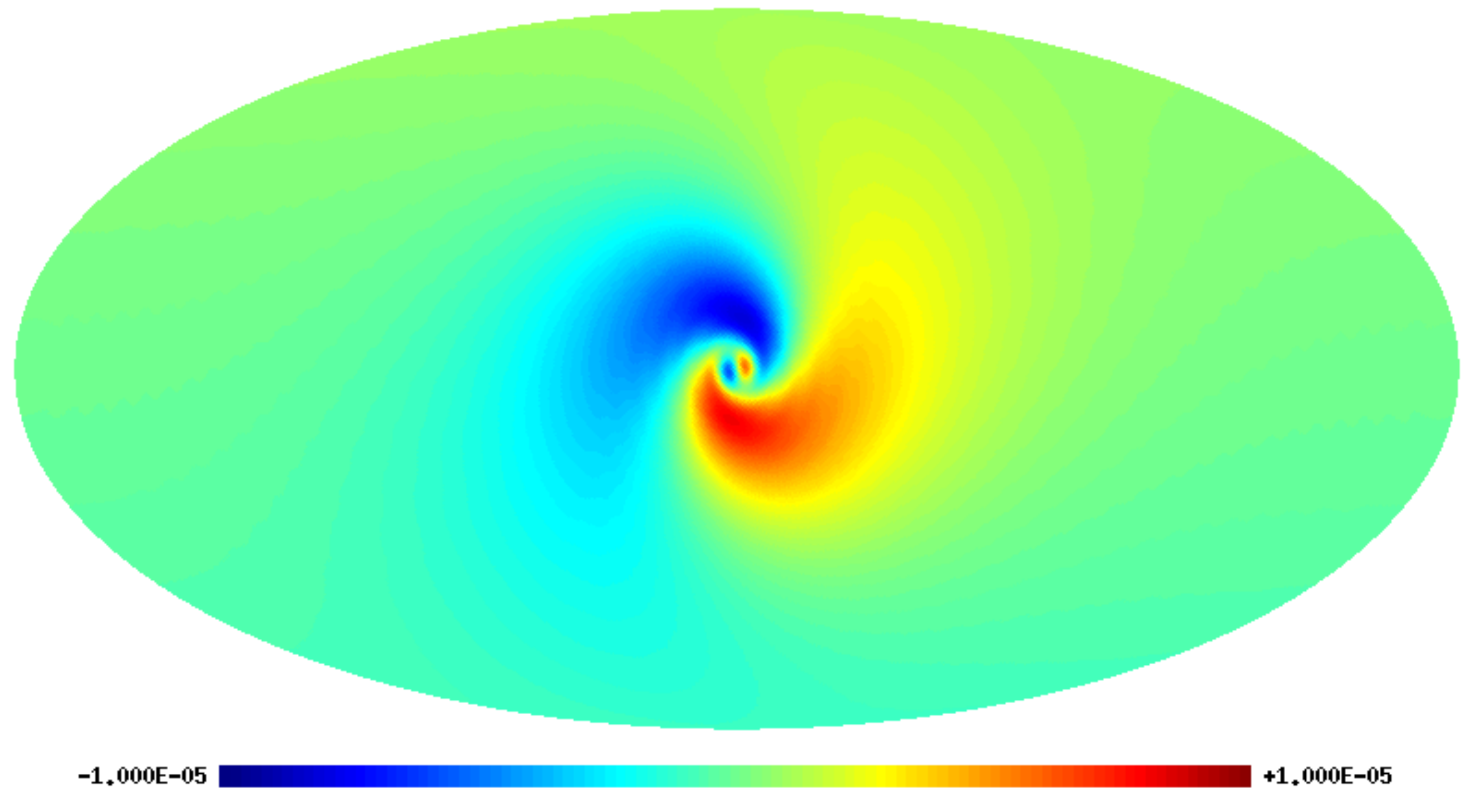} 
  \includegraphics[bb= 0 40 800 440,clip=,width=\plotwidth]{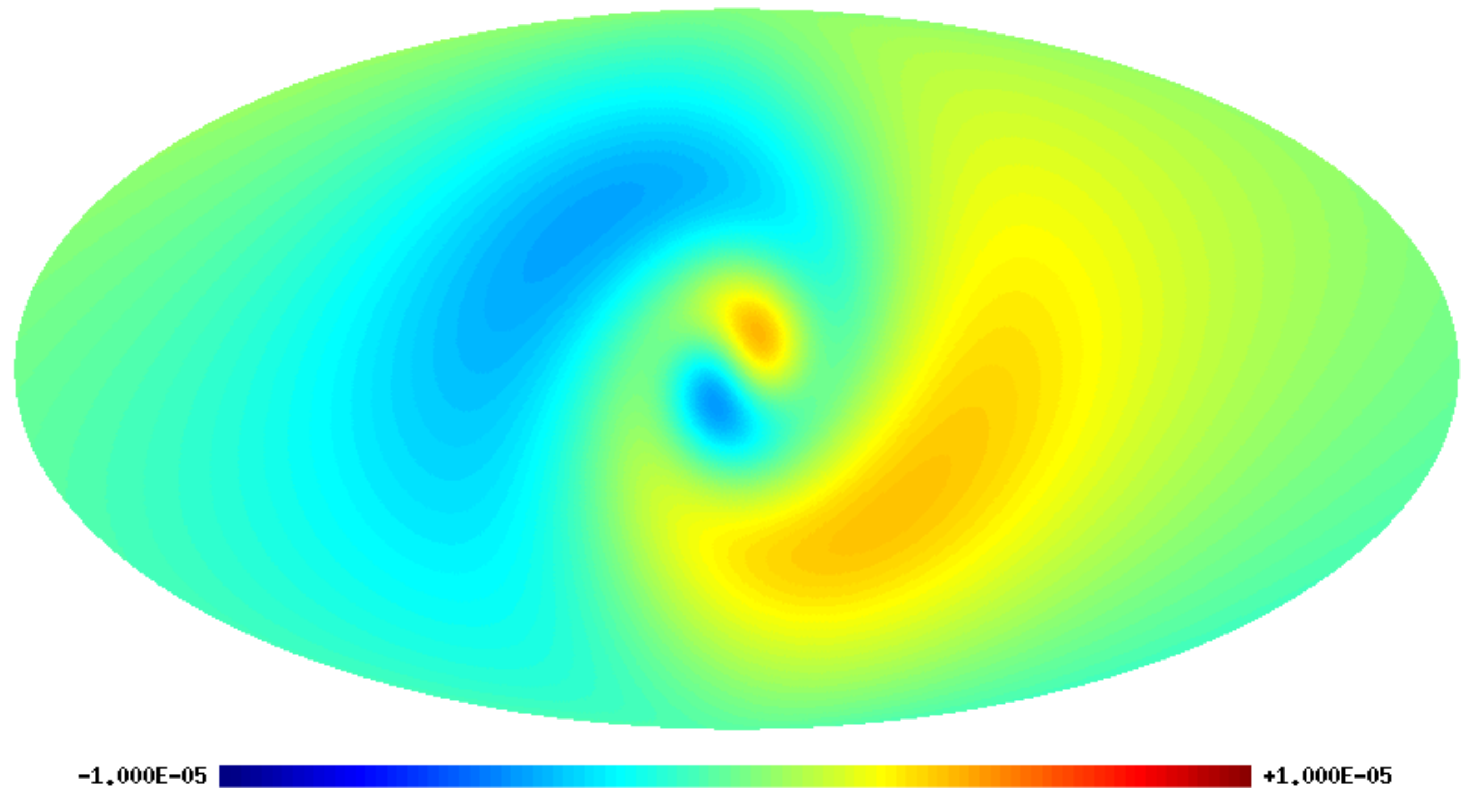} 
  \includegraphics[bb= 0 40 800 440,clip=,width=\plotwidth]{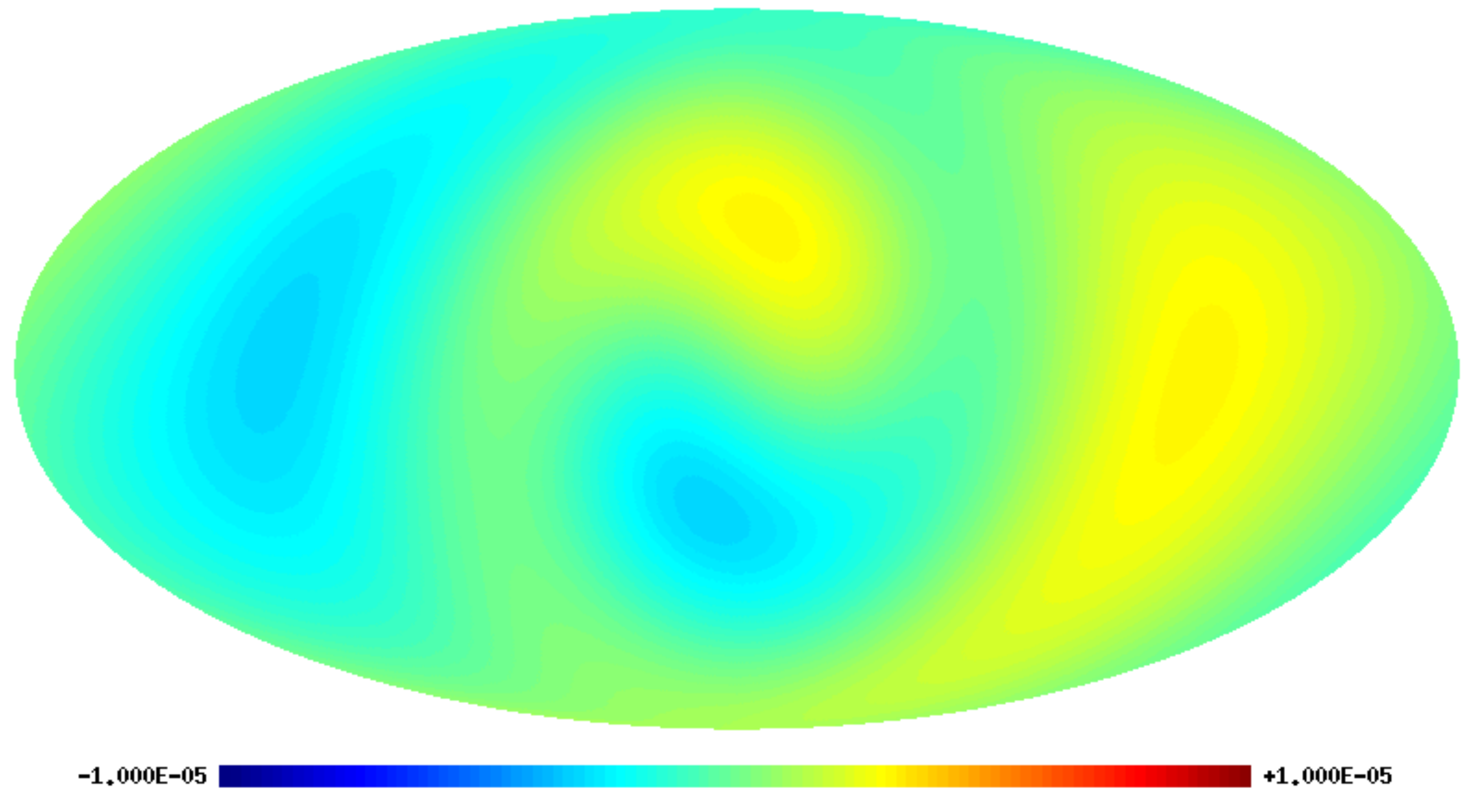}
}
\mbox{
  \includegraphics[bb= 0 40 800 440,clip=,width=\plotwidth]{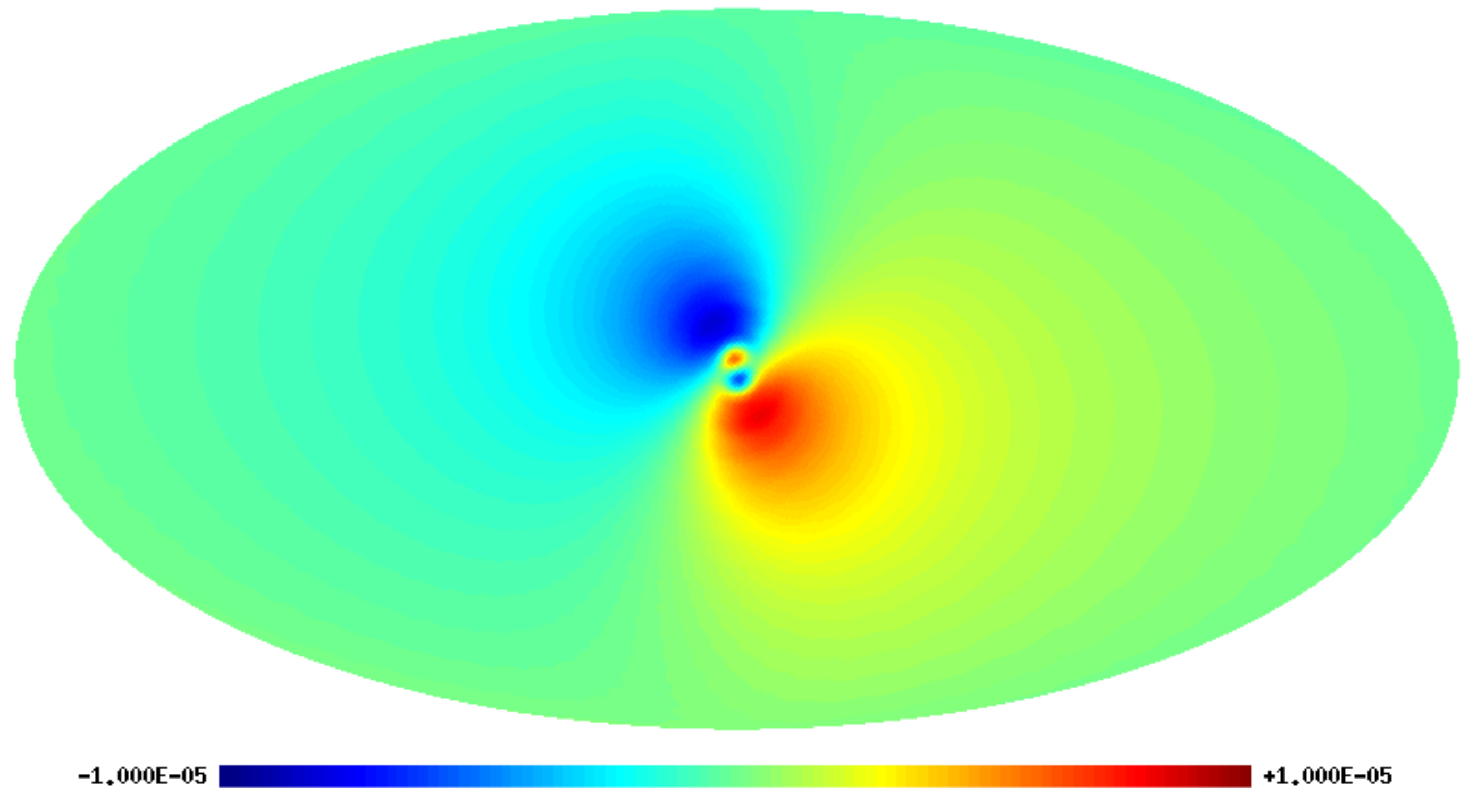} 
  \includegraphics[bb= 0 40 800 440,clip=,width=\plotwidth]{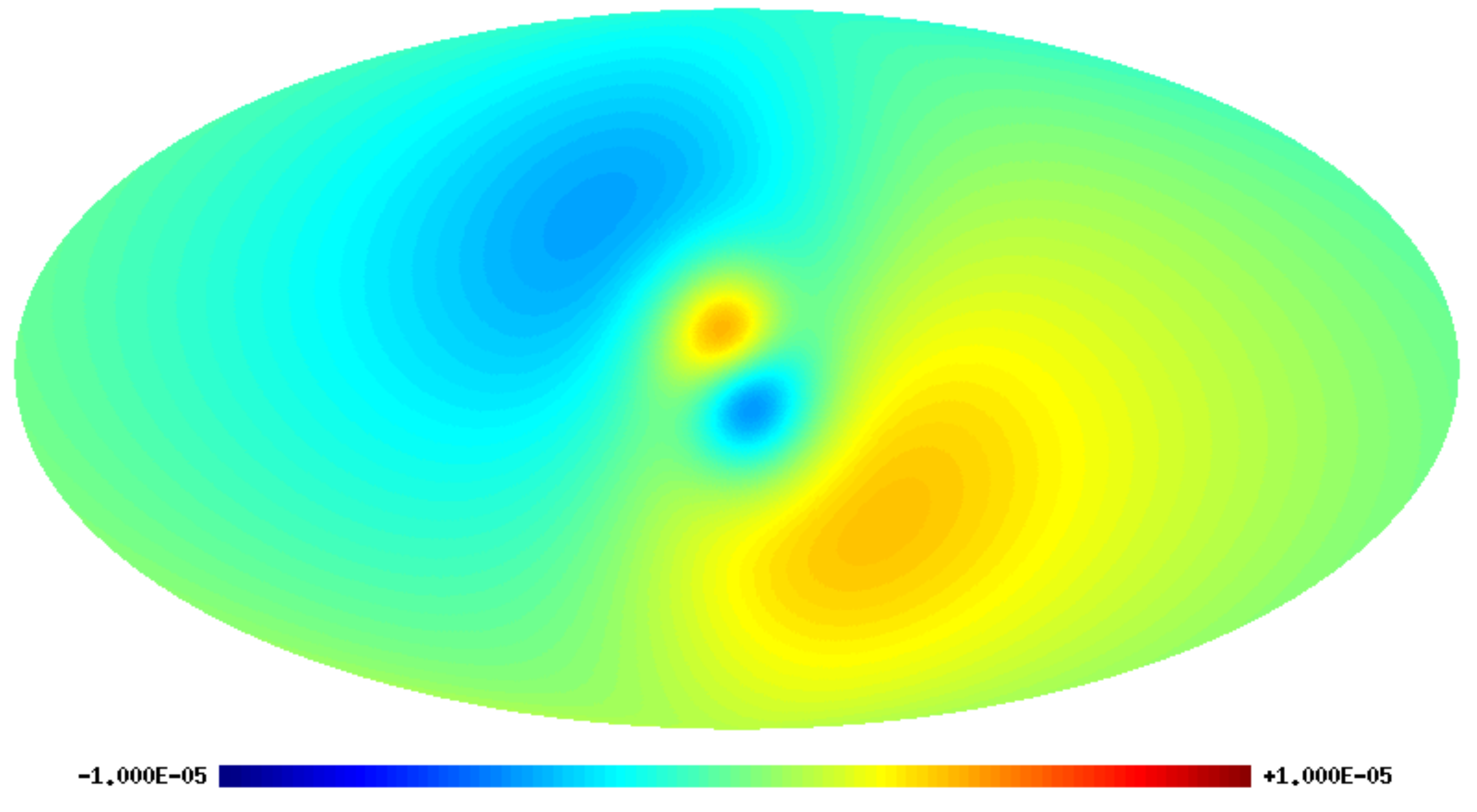} 
  \includegraphics[bb= 0 40 800 440,clip=,width=\plotwidth]{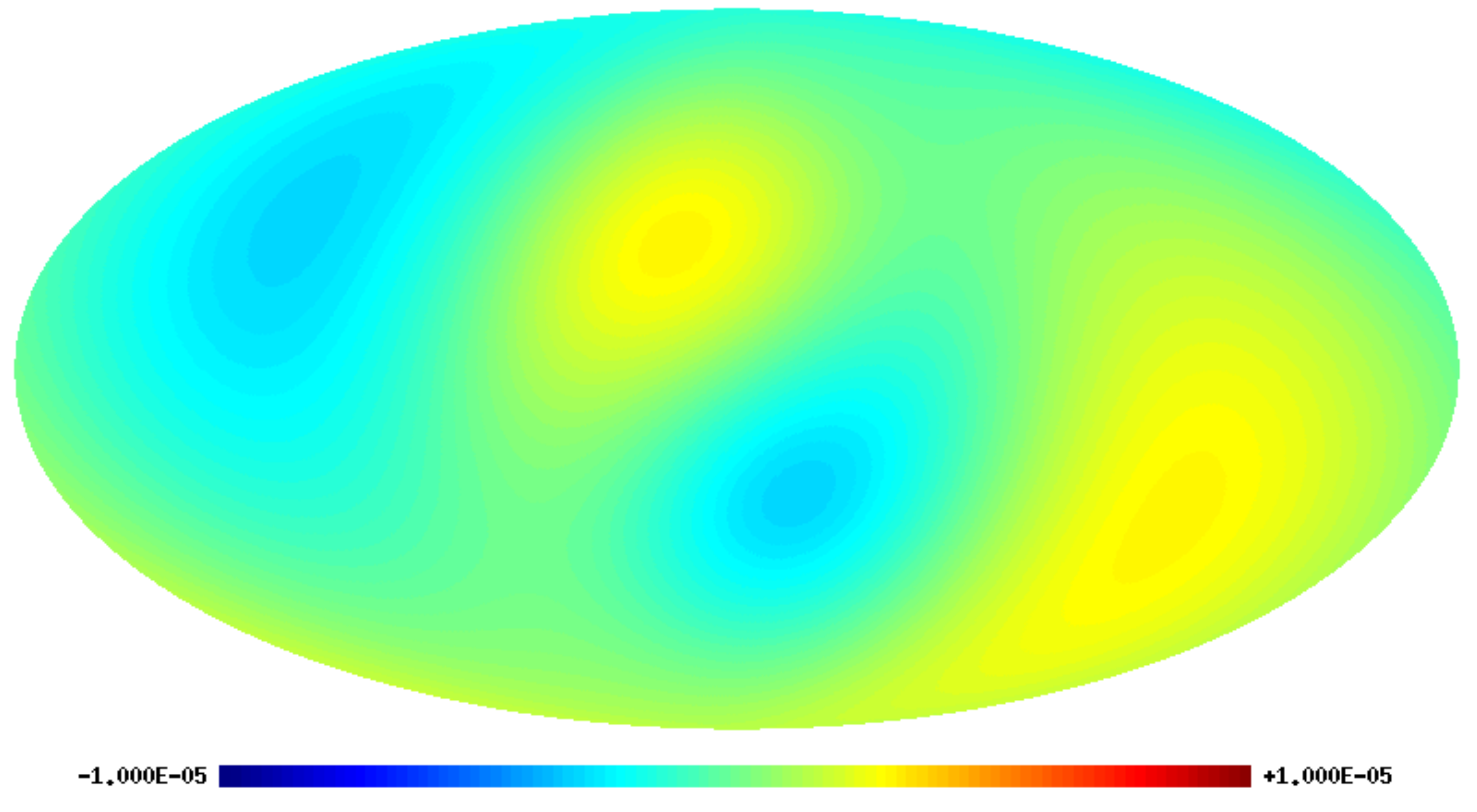}
}
\caption[Simulated \bianchiviih\ maps]{Simulated deterministic CMB
  temperature contributions in 
  \bianchiviih\ cosmologies for varying $\bx$ and $\Dentot$ (left-to-right
  $\Dentot \in \{0.10, 0.50, 0.95\}$; top-to-bottom $\bx\in\{0.1, 0.3,
  0.7, 1.5, 6.0 \}$).  In these maps the swirl pattern typical of
  Bianchi-induced temperature fluctuations is rotated from the South
  pole to the Galactic centre for illustrational purposes. }
\label{fig:bianchi_sims}
\end{figure}





\section{Data description} 
\label{sec:Data}

We use \Planck\ maps that have been processed by the various component-separation pipelines described in \cite{planck2013-p06}. The methods produce largely consistent maps of the sky, with detailed differences in pixel intensity, noise properties, and masks. Here, we consider maps produced by the \commanderruler , \nilc, \smica\ and \sevem\ methods. Each provides its own mask and we also consider the conservative common mask.

We note that because our methods rely on rather intensive pixel- or
harmonic-space calculations, in particular considering a full set of
three-dimensional orientations and, for the likelihood methods,
manipulation of an anisotropic correlation matrix, computational
efficiency requires the use of data degraded from the native
\texttt{HEALPix} \citep{gorski:2005} $N_\mathrm{side}=2048$ resolution
of the \Planck\ maps. Because the signatures of either a
multiply-connected topology or a Bianchi model are most prominent on
large angular scales, this does not result in a significant loss of
ability to detect and discriminate amongst the models (see
Sect.~\ref{sub:Simulations}). 
However, it is worth pointing out that the likelihood and matched circles
methods are sensitive to different angular scales as applied to \Planck\ data here.
The likelihood method explicitly retains only low-$\ell$ (large-scale) information in its correlation matrix, whereas the matched-circles method considers anisotropies at angular scales down to
tens of arcminutes (still  large
in comparison to the native resolution of the
\Planck\ maps). Of course, the matched-circles method exploits the correlation of the small-scale patterns along matched circles potentially separated by large angles; this effect is not generated by
intrinsically large angular scale anisotropies but by the boundary conditions of
the fundamental domain imposed by the multiply connected topology. As
described in Sect.~\ref{sec:method_circles}, the matched circles statistic used here damps the anisotropies at the
largest angular scales relative to those at smaller scales so
sensitivity of the method does not rely on the former.

The topology analyses both rely on degraded maps and masks. The
matched-circles method smooths with a 30\arcm\ Gaussian filter and
degrades the maps to $N_\mathrm{side}=512$, and uses a mask derived
from the \sevem\ component separation method.
Because the performance of the matched-circles
statistic depends on anisotropies on smaller angular scales, it
can be significantly degraded by the point source cut. As there are
   more point sources detected in the \Planck\ maps than in the 
    \textit{WMAP} maps, the problem of point source masking is
  more severe in the present case. We
mask only those point sources from the full-resolution
$f_\mathrm{sky}=0.73$ \sevem\ mask with amplitude, after smoothing and
extrapolation to the 143 or 217\,GHz channels, greater than the
faintest source originally detected at those
frequencies. The mask derived in this way retains
$f_\mathrm{sky}=0.76$ of the sky.

The likelihood method smooths the maps and masks with an 11\deg\
Gaussian filter and then degrades them to $N_\mathrm{side}=16$ and
conservatively masks out any pixel with more than $10\,\%$ of its
original subpixels masked. At full resolution, the common mask retains
a fraction $f_\mathrm{sky}=0.73$ of the sky, and $f_\mathrm{sky}=0.78$
when degraded to $N_\mathrm{side}=16$ (the high-resolution
point-source masks are largely filled in the degraded masks). The
Bianchi analysis is performed in harmonic space, and so does not
require explicit degradation in pixel space. Rather, a
  \emph{noisy mask} is added in pixel space to effectively marginalise
  the pixel values in the masked region (as described in more detail
  below; see also \citealt{mcewen:bianchi}), before the data are
transformed at full resolution into harmonic space and considered only
up to a specified maximum harmonic $\ell$, where correlations due to
the mask are taken into account.

Different combinations of these maps and masks are used to discriminate between the topological and anisotropic models described in Sect.~\ref{sec:Correlations}.


\section{Methods} 
\label{sec:Methods}

\subsection{Topology: circles in the sky} 
\label{sec:method_circles}

The first set of methods, exemplified by the circles-in-the-sky of
\citet{cornish1998}, involves a frequentist analysis using a
  statistic which is expected to differ
between the models examined. For the circles, this uses the fact that
the intersection of the topological fundamental domain with the
surface of last scattering is a circle, which one potentially views
from two different directions in a multiply-connected universe. Of
course, the matches are not exact due to noise, foregrounds, the
integrated Sachs-Wolfe (ISW) and Doppler effects along the different
lines of sight.

By creating a statistic based on the matching of different such
circles, we can compare Monte Carlo simulations of both a
simply-connected, isotropic null model with specific anisotropic or
topological models.  We may then calibrate detections and
non-detections using Monte Carlo simulations. In principle, these
simulations should take into account the complications of noise,
foreground contributions, systematics, the ISW and Doppler
effects. However, they do not include gravitational lensing of the CMB
as the lensing deflection angle is small compared to the minimal
angular scale taken into account in our analysis. Note that the null
test is generic (i.e., not tied to a 
specific topology) but any detection must be calibrated with specific
simulations for a chosen topology or anisotropic model.
A very similar technique can be used for polarisation by taking into
account the fact that the polarisation pattern itself is now not directly
repeated, but rather that the underlying quadrupole radiation field
around each point on the sky is now seen from different directions
\citep{bielewicz2012}.
These methods have been applied successfully to \textit{COBE} DMR
and \textit{WMAP} data, and have recently been shown to be feasible
for application to \Planck\ data \citep{bielewicz2012}.

The idea of using the matched circles to study topology
is due to \citet{cornish1998}. In that work, a
statistical tool was developed to detect correlated circles in all sky maps of the CMB
anisotropy --- the circle comparison statistic. In our studies we will
use version of this statistic optimised for the small-scale
anisotropies as defined by \citet{cornish2004}:
\begin{equation} \label{eqn:s_statistic_fft}
S_{i,j}^{+}(\alpha, \phi_\ast)=\frac{2 \sum_m |m| \Delta T_{i,m}^{}
  \Delta T_{j,m}^\ast
e^{-{\rm i} m \phi_\ast}}{\sum_n |n| \left( |\Delta T_{i,n}|^2+|\Delta
  T_{j,n}|^2\right)}\ ,
\end{equation}
where $\Delta T_{i,m}$ and $\Delta T_{j,m}$ denote the Fourier
coefficients of the temperature fluctuations around two circles of angular
radius $\alpha$ centered at different points on the sky, $i$ and $j$,
respectively, with relative phase $\phi_\ast$. The $m$th harmonic of
the temperature anisotropies around the circle is weighted by the factor $|m|$,
taking into account the number of degrees of freedom per mode. Such
weighting enhances the contribution of small-scale structure relative to large-scale
fluctuations and is especially important since the large-scale
fluctuations are dominated by the ISW effect. This can obscure
the image of the last scattering surface and reduce the ability to
recognise possible matched patterns on it.

The above $S^{+}$ statistic corresponds to pair of circles with the points
ordered in a clockwise direction (phased). For
alternative ordering, when along one of the circles the points are
ordered in an anti-clockwise direction (anti-phased), the
Fourier coefficients $\Delta T_{i,m}$ are complex conjugated, defining the $S^{-}$ statistic. This
allows the detection of both orientable and non-orientable
topologies. For orientable topologies the matched circles have
anti-phased correlations while for non-orientable topologies they have
a mixture of anti-phased and phased correlations.

The statistic has a range over the interval
$[-1,1]$. Circles that are perfectly matched have $S=1$, while
uncorrelated circles will have a mean value of $S=0$. Although the
statistic can also take negative values for the temperature anisotropy
generated by the Doppler term \citep{bielewicz2012}, anticorrelated
circles are not expected for the total temperature anisotropy
considered in this work. To find matched circles for each radius
$\alpha$, the maximum value $S_{\rm max}^{\pm}(\alpha) =
{\rm max}_{i,j,\phi_\ast} \, S_{i,j}^{\pm}(\alpha,\phi_\ast)$
is determined.

Because general searches for matched circles are computationally very
intensive, we restrict our analysis to a search for pairs of
circles centered around antipodal points, so called back-to-back
circles. As described above, the maps were also downgraded to $N_{\rm
  side}=512$, which greatly speeds up the computations required, but
with no significant loss of discriminatory power, as seen in
Sect.~\ref{sec:applic_simulations}. More details on the numerical
implementation of the algorithm can be found in the paper by
\cite{bielewicz2011}. 

As mentioned in Sect.~\ref{sub:Topology}, the constraints we will
  derive concern topologies that predict matching pairs of
  back-to-back circles. However, the constraints do not apply to those universes
for which the orientation of the matched circles is impossible to
detect due to partial masking on the sky. Because of the larger sky
fraction removed by the \Planck\ common mask than for  
\emph{WMAP} this probability is larger for the analysis of the
\Planck\ maps. Moreover, the smaller fraction of the sky used in the
search of matched circles results in a false detection level
larger with our $f_\mathrm{sky}=0.76$ mask than for the
$f_\mathrm{sky}=0.78$ 7-year KQ85 \emph{WMAP} mask. As a result we obtain weaker 
--- but more conservative --- constraints on topology than for similar
analyses of \emph{WMAP} data \citep{bielewicz2011}.
 
To draw any conclusions from an analysis based on the statistic $S_{\rm
  max}^{\pm}(\alpha)$, it is very important to correctly estimate the
threshold for a statistically significant match of circle pairs.
We used 300 Monte Carlo simulations of CMB maps, described in detail
in Section \ref{sec:applic_simulations}, to establish the threshold such that fewer than
1\,\% of simulations would yield a false event.


\subsection{Bayesian analyses} 
\label{sub:Bayesian analyses}

The second set of methods take advantage of the fact that the
underlying small-scale physics is unchanged in both anisotropic and
topological models compared to the standard cosmology, and thus a
Gaussian likelihood function will still describe the statistics of
temperature and polarization on the sky, albeit no longer with
isotropic correlations. When considering specific topologies,
these likelihood methods instead calculate the pixel-pixel correlation
matrix. This has been done for various torus
topologies (which are a continuous family of possibilities) in the
flat Universe as well as for locally hyperbolic and spherical
geometries (which have a discrete set of possibilities for a given
value of the curvature). More general likelihood-based techniques have
been developed for generic mild anisotropies in the initial power
spectrum \citep{Hanson2009}, which may have extension to other
models. For the Bianchi setting, an isotropic zero-mean Gaussian likelihood is
  recovered by subtracting a deterministic Bianchi component from the
  data, where the cosmological covariance matrix remains diagonal in
  harmonic space but masking introduces non-diagonal structure that
  must be taken into account.

Because these methods use the likelihood function directly, they can
take advantage of any detailed noise correlation information that is
available, including any correlations induced by the
foreground-removal process. We denote the data by the vector
$\fitdata$, which may be in the form of harmonic coefficients $d_{\ell
  m}$ or pixel temperatures $d_p$ or, in general, coefficients of the
temperature expansion in any set of basis functions. We denote the
model under examination by the discrete parameter $M$, which can take
on the appropriate value denoting the usual isotropic case, or the
Bianchi case, or one of the possible multiply-connected universes. The
continuous parameters of model $M$ are given by the vector $\Theta$,
which for this case we can partition into $\Theta_\mathrm{C}$ for the
cosmological parameters shared with the usual isotropic and
simply-connected case, and $\Theta_\mathrm{A}$ which denotes the parameters for
the appropriate anisotropic case, be it a topologically non-trivial
universe or a Bianchi model. Note that all of the anisotropic cases
contain ``nuisance parameters'' which give the orientation of either
the fundamental domain or the Bianchi template which we can
marginalize over as appropriate.

Given this notation, the posterior distribution for the parameters of
a particular model, $M$, is given by Bayes' theorem:
\begin{equation}\label{eq:bayestheorem}
	P(\Theta|\fitdata,M) = \frac{P(\Theta|M)P(\fitdata|\Theta,M)}{P(\fitdata|M)}\;.
\end{equation} 
Here, $P(\Theta|M)=P(\Theta_\mathrm{C},\Theta_\mathrm{A}|M)$ is the joint prior
probability of the standard cosmological parameters $\Theta_\mathrm{C}$ and those
describing the anisotropic universe $\Theta_\mathrm{A}$, $P(\fitdata|\Theta,M)\equiv{\cal L}$ is the
likelihood, and the normalizing constant $P(\fitdata|M)$ is the
Bayesian evidence, which can be used to compare the models to one
another.

We will usually take the priors to be simple ``non-informative''
distributions (e.g., uniform over the sphere for orientations, uniform
in length for topology scales, etc.) as appropriate. The form of the
likelihood function will depend on the anisotropic model: for
multiply-connected models, the topology induces anisotropic
correlations, whereas for the Bianchi model, there is a deterministic
template, which depends on the Bianchi parameters, in addition to the standard
isotropic cosmological perturbations. We will assume that any other
non-Gaussian signal (either from noise or cosmology) is negligible \citep{planck2013-p09,planck2013-p09a} and
use an appropriate multivariate Gaussian likelihood.

Given the signal and noise correlations, and a possible Bianchi
template, the procedure is similar to that used in standard
cosmological-parameter estimation, with a few complications.
Firstly, the evaluation of the likelihood function is computationally
  expensive and usually limited to large angular scales. This means
  that in practice the effect of the topology on the likelihood is
  usually only calculated on those large scales.
Secondly, the orientation of the fundamental domain or Bianchi template
  requires searching (or marginalizing) over three additional
  parameters, the Euler angles.


\subsubsection{Topology}

In topological studies, the parameters of the model consist of
$\Theta_\mathrm{C}$, the set of cosmological parameters for the fiducial
best-fit flat cosmological model, and $\Theta_\mathrm{T}$, the topological
parameters which include the set of compactification lengths $L_x,
L_y, L_z$ for flat toroidal model or the curvature parameter
$\Omega_{K}$ for curved spaces, and a choice of compactification $T$. In
our studies we keep $\Theta_\mathrm{C}$ fixed, and
vary $\Theta_\mathrm{T}$ for a select choice of compactifications listed in
Sect.~\ref{sub:Topology}. These parameters define the predicted
two-point signal correlation matrix $C_{pp^\prime}$ for each model,
which are precomputed. Additional internal parameters, including the
amplitude of the signal $A$ and the angles of orientation of the
fundamental domain of the compact space relative to the sky $\varphi$ 
(e.g., parameterized by a vector of the three Euler angles),
are maximized and/or marginalized over during likelihood evaluation.

The likelihood, i.e., the probability to find a temperature data map
$\vec{ d}$ with associated noise matrix $\mtrx{ N}$ given a certain
topological model is then given by
\begin{eqnarray}\label{eq:fullskylike}
\lefteqn{P(\vec{d}| \mtrx{C}[ \Theta_\mathrm{C},\Theta_\mathrm{T},T],A,\varphi)}\nonumber \\
&& \propto \frac{1}{\sqrt{|A \mtrx{C} + \mtrx{ N}|}} 
\exp\left\{ - \frac{1}{2} \vec{d}^* (A \mtrx{ C} + \mtrx{ N} )^{-1} \vec{d} \right\} \, .
\end{eqnarray}

Working with a cut-sky, it is often easier to start the analysis with data
and a correlation matrix given in pixel space. However, especially
in the realistic case of negligible noise on large scales, the matrix
$\mtrx{ C} + \mtrx{ N}$ is poorly conditioned in pixel space, and pixel space
evaluation of the likelihood is, as a rule, not robust. Indeed, there
are typically more pixels than independent modes that carry
information about the signal (e.g., even in the standard isotropic case, sub-arcminute pixels would not be useful due to beam-smoothing; with anisotropic correlations and masked regions of the sky, more complicated linear combinations of pixels even on large scales may have very little signal content). Therefore in general we expand the
temperature map $d_p$, the theoretical correlation matrix
$C_{pp^\prime}$ and the noise covariance matrix $N_{pp^\prime}$ in a
discrete set of mode functions $\psi_n(p)$, orthonormal over the
pixelized sphere, possibly with weights $w(p)$, $\sum_p w(p)
\psi_n(p)\psi^*_{n^\prime}(p) = \delta_{nn^\prime}$, obtaining the
coefficients of expansion
\begin{eqnarray}
d_n &=& \sum_p d_p \psi^*_n(p) w(p); \nonumber \\
C_{nn^\prime} &=& \sum_p \sum_{p^\prime} C_{pp^\prime} \psi_n(p) \psi_{n^\prime}^*(p^\prime) w(p) w(p^\prime); \nonumber \\
N_{nn^\prime} &=& \sum_p \sum_{p^\prime} C_{pp^\prime}\psi_n(p) \psi_{n^\prime}^*(p^\prime) w(p) w(p^\prime) \, .
\end{eqnarray}
Next we select $N_\mathrm{m}$ such modes for comparison and consider the likelihood marginalized over the remainder of the modes
\begin{eqnarray}
\lefteqn{p( \fitdata | \mtrx{C}[ \Theta_\mathrm{C},\Theta_\mathrm{T},T],\varphi, A) \propto} \nonumber \\
&&\frac{1}{\sqrt{|A \mtrx{ C} + \mtrx{ N} |_M}} 
\exp\left\{ - \frac{1}{2} \sum_{n=1}^{N_\mathrm{m}} d_n^* (A \mtrx{ C} + \mtrx{ N})^{-1}_{nn^\prime} d_{n^\prime} \right\} \, ,
\label{eq:L_modes}
\end{eqnarray}
where $\mtrx{ C}$ and $\mtrx{ N}$ are restricted to the $N_\mathrm{m} \times N_\mathrm{m}$
block of chosen
modes. Flexibility in choosing mode functions and their number $N_\mathrm{m}$ is
used to achieve the compromise between the robust invertibility of the projected
$\mtrx{ C} + \mtrx{ N}$ matrix on the one hand, and the amount of discriminating
information retained in the data on the other. The weights $w(p)$ can
be used to improve the accuracy of transforms on a pixelized sky.

For full-sky analysis the natural choice of the mode functions is the
set of ordinary spherical harmonics $Y_{lm}(p)$ which leads to standard harmonic analysis 
with modes limited to a suitably chosen $\ell_\mathrm{max}$.
Here, where we focus on masked data, we have made a somewhat different
choice. As a mode set for comparison we use the $N_\mathrm{m}=837$ largest
eigenvectors of the $C_{pp^\prime}$ matrix, restricted to the masked sky,
for the fiducial flat isotropic model with best-fit parameters
$\Theta_\mathrm{C}$. We emphasize that the correlation matrix computed for this reduced dataset has fewer modes, but contains no additional assumptions beyond those of the original $C_{pp'}$.

Since computation of $C_{pp^\prime}$ matrices for a range of
topological models is expensive, we do not aim to determine the full
Bayesian evidence $P(\vec{ d} | T)$ which would require marginalization
over all parameters $\Theta_\mathrm{C}$, $\Theta_\mathrm{T}$,
an overall amplitude of the correlation matrix $A$ (proportional to the physical amplitude $\sigma_8$ or the scalar amplitude $A_{\rm s}$) , and
orientation (Euler angles) $\varphi$, and would in addition be sensitive to the prior probabilities 
assumed for the size of the fundamental domain. Instead we directly compare the likelihood along the
changing set of $\Theta_\mathrm{T}$ that has as its limit the flat
fiducial model defined by $\Theta_\mathrm{C}$.  In case of toroidal
topology such a limit is achieved by taking compactification lengths
to infinity, while for curved models we vary $\Omega_{K}$ in comparison
to the flat limit $\Omega_{K}=0$. In the latter case, for the spherical
spaces we change $\Omega_\Lambda$ and $H_0$ together with $\Omega_{K}$ to
track the CMB geometrical degeneracy line in which the recombination sound speed, initial fluctuations,
and comoving distance to the last scattering surface are kept constant \citep[e.g.,][]{BET97,ZalSel97,Stompor1999}, and
for hyperbolic spaces we vary $\Omega_{K}$ while keeping $H_0$ and
$\Omega_\Lambda-\Omega_{\rm m}$ fixed to fiducial values.   Note that hyperbolic multi-connected spaces, in contrast to tori and the single-action
positive curvature manifolds considered in this paper, are not only anisotropic but also inhomogeneous. Therefore,
the likelihood is expected to be dependent on the position of the observer. We do not study this dependence here.

For each parameter choice, we find the likelihood at the best orientation
$\varphi$ of the topology with respect to the sky after marginalizing
over the amplitude $A$ of the signal (hence, this can be considered a
\emph{profile likelihood}
with respect to the orientation parameters).
This likelihood is compared both
with the fiducial model applied to the observed temperature map and
with the likelihood of the topological model applied to the simulated
realization of the isotropic map drawn from the fiducial model. Such a
strategy is optimized for the detection of topological signatures. For non-detections, the marginalized likelihood can be a better probe of the overall power of the data to reject a non-trivial topology, and so 
for real data below, we also show the likelihood marginalized over the orientations $\varphi$.
We estimate the marginalized likelihood from the random sample of 10,000 orientations,
drawn statistically uniformly on the $S^3$ sphere of unit quaternions representing
rotations of the fundamental domain relative to the observed sky.

\subsubsection{Bianchi} 

For the Bianchi analysis the posterior distribution of the
parameters of model $\fitmodelsel$ is given by Bayes' Theorem,
specified in Eq.~\ref{eq:bayestheorem}, similar to the topological setting.
The approach of \cite{mcewen:bianchi} is followed, where the likelihood is
made explicit in the context of fitting a deterministic Bianchi template embedded in
a stochastic \cmb\ background, defined by the power spectrum $C_\el(\cosmoparam)$
for a given cosmological model with parameters \cosmoparam. The
\bianchiviih\ parameters are denoted $\bparam$. The corresponding
likelihood is given by \begin{equation}
  \label{eqn:bianchi_likelihood}
  \prob( \fitdata  | \bparam, \cosmoparam) \propto
  \frac{1}{\sqrt{|\mtrx{X}(\cosmoparam) |} }{\exp\bigl[-\chi^2(\cosmoparam, \bparam) / 2\bigr]}
  \spcend, 
\end{equation}
where
\begin{equation}
    \chi^2(\cosmoparam, \bparam) = 
    \bigl[\fitdata-\fittmpl(\bparam)\bigr]^\dagger \mtrx{X}^{-1}(\cosmoparam) \bigl[\fitdata-\fittmpl(\bparam)\bigr]
\end{equation}
and $\fitdata = \{ \fitdataalm \}$ and $\fittmpl(\bparam) = \{
\fittmplalm(\bparam) \}$ are the spherical harmonic coefficients of
the data and Bianchi template, respectively, considered up to the
harmonic band-limit $\elmax$.  A band-limit of $\elmax=32$ is
considered in the subsequent analysis for computational
  tractibility and since this is sufficient to capture the structure
of the \cmb\ temperature fluctuations induced in \bianchiviih\ models
in the vacinity of the best-fit model found in \textit{WMAP}
data (see, e.g., \citealt{mcewen:2006:bianchi}).  The likelihood is
computed in harmonic space where rotations of the Bianchi template can
be performed efficiently \citep{mcewen:2006:bianchi}.

The covariance matrix $\mtrx{X}(\cosmoparam)$ depends on whether the
full-sky or partial-sky masked setting is considered. In the full-sky
setting $\mtrx{X}(\cosmoparam) = \mtrx{C}(\cosmoparam)$ as first
considered by \cite{bridges:2006b}, where $\mtrx{C}(\cosmoparam)$ is
the diagonal CMB covariance matrix with entries $C_\el(\cosmoparam)$
on the diagonal.  In the case of a zero Bianchi component,
Eq.~\ref{eqn:bianchi_likelihood} then reduces to the likelihood
function used commonly to compute parameter estimates from the power
spectrum estimated from \cmb\ data (\eg\ \citealt{verde:2003}).  In
the masked setting considered subsequently, the situation is a little
more involved.

In order to handle a mask in the harmonic space analysis of Bianchi
models we follow the approach of \cite{mcewen:bianchi}, where
\textit{masking noise} is added to the data to effectively marginalise
over the pixel values of the data in the masked region.  The masking
noise $m$ is chosen to be zero-mean and large in the masked region of the
data, and zero elsewhere.  Consequently, the masking noise is
anisotropic over the sky but may be chosen to be uncorrelated, and may
thus be defined by its covariance
\begin{equation}
\label{eqn:mnoise_covariance_real}
\langle 
m(\omega_i) \, m^\ast(\omega_j)
\rangle
= \delta_{ij}\,
\sigma_m^2(\omega_i)
\spcend ,
\end{equation}
where $\delta_{ij}$ is Kronecker delta symbol, $\omega_i$ denotes the
angular coordinate of pixel $i$, and the variance of the noise for
pixel $i$ is given by a constant value in the masked regions
$\sigma_m^2(\omega_i)=\Sigma_m^2$ and zero elsewhere.  By
synthetically adding masking noise that is much larger than the
original data in the masked region of the sky, we effectively
marginalise over the pixel values of the data in this region.
The noisy mask introduces coupling in harmonic space that
must be accounted for in the analysis.  The covariance matrix of the
resultant data is given by  $\mtrx{X}(\cosmoparam) =
\mtrx{C}(\cosmoparam) + \mtrx{M}$, where $\mtrx{M}$ is the
non-diagonal mask covariance matrix:
\begin{equation}
\label{eqn:mnoise_covariance_harmonic}
\mtrx{M}_{\ell m}^{\ell^\prime m^\prime} 
= 
\langle
m_{\ell m} \,
m_{\ell^\prime m^\prime}
\rangle 
\simeq
\sum_{\omega_i} 
\sigma_m^2(\omega_i)
Y_{\ell m}^\ast(\omega_i) \,
Y_{\ell^\prime m^\prime}(\omega_i) \,
\Omega_{i}^2
\spcend ,
\end{equation}
and $\Omega_{i}$ is the area of pixel $i$ (see \cite{mcewen:bianchi}
for further details).

The $\chi^2$ of the likelihood for the Bianchi
case hence differs from the topology case by the nonzero Bianchi template $\fittmpl$
and the use of a correlation matrix $\mtrx{M}$ to account for the
presence of the mask.

In the most physically motivated scenario, the Bianchi and
cosmological parameters are coupled (\eg\ the total density of the
Bianchi and standard cosmological model are identical). However, it is
also interesting to consider Bianchi templates as phenomenological
models with parameters decoupled from the standard cosmological
parameters, particularly for comparison with previous studies.  Both
scenarios are considered in the subsequent analysis.  In the decoupled
scenario a flat cosmological model is considered, whereas in the
coupled scenario an open cosmological model is considered to be
consistent with the \bianchiviih\ model; we label these models the
flat-decoupled-Bianchi model and the the open-coupled-Bianchi model,
respectively.

To determine whether the inclusion of a Bianchi component better
describes the data the Bayesian evidence is examined, as given by
\begin{equation}
  E = 
  \prob(\fitdata | \fitmodelsel) =
  \int \dx \Theta \,
  \prob(\fitdata | \Theta, \fitmodelsel) \,
  \prob(\Theta | \fitmodelsel) 
  \spcend .
\end{equation}
Using the Bayesian evidence to distinguish between models naturally
incorporates Occam's razor, trading off model simplicity and accuracy.
In the absence of any prior information on the preferred model, the
Bayes factor given by the ratio of Bayesian evidences (i.e.,
$E_1/E_2$) is identical to the ratio of the model probabilities given
the data. The Bayes factor is thus used to distinguish models.  The
Jeffreys scale \citep{jeffreys:1961} is often used as a rule-of-thumb
when comparing models via their Bayes factor. The log-Bayes factor
$\Delta {\rm ln} E = {\rm ln} (E_1/E_2)$ (also called the log-evidence
difference) represents the degree by which the model corresponding to
$E_1$ is favoured over the model corresponding to $E_2$, where: $0
\leq \Delta {\rm ln} E < 1$ is regarded as inconclusive; $1 \leq
\Delta {\rm ln} E < 2.5$ as significant; $2.5 \leq \Delta {\rm ln} E <
5$ as strong; and $\Delta {\rm ln} E \geq 5 $ as conclusive (without
loss of generality we have assumed $E_1 \geq E_2$).  For reference, a
log-Bayes factor of 2.5 corresponds to odds of 1 in 12, approximately,
while a factor of 5 corresponds to odds of 1 in 150, approximately.

The {\tt ANICOSMO}\footnote{\url{http://www.jasonmcewen.org/}} code
\citep{mcewen:bianchi} is used to perform a Bayesian analysis of
\bianchiviih\ models, which in turn uses the public {\tt
  MultiNest}\footnote{\url{http://www.mrao.cam.ac.uk/software/multinest/}}
code \citep{feroz:multinest1,feroz:multinest2} to sample the posterior
distribution and compute evidence values by nested sampling
\citep{skilling:2004}. We sample the parameters describing the
\bianchiviih\ model and those describing the standard cosmology
simultaneously.

\subsection{Simulations and Validation} 
\label{sub:Simulations}

\subsubsection{Topology} 
\label{sec:applic_simulations}

\begin{figure}[htbp]
	\centering
		\includegraphics[width=0.9\columnwidth,angle=180]{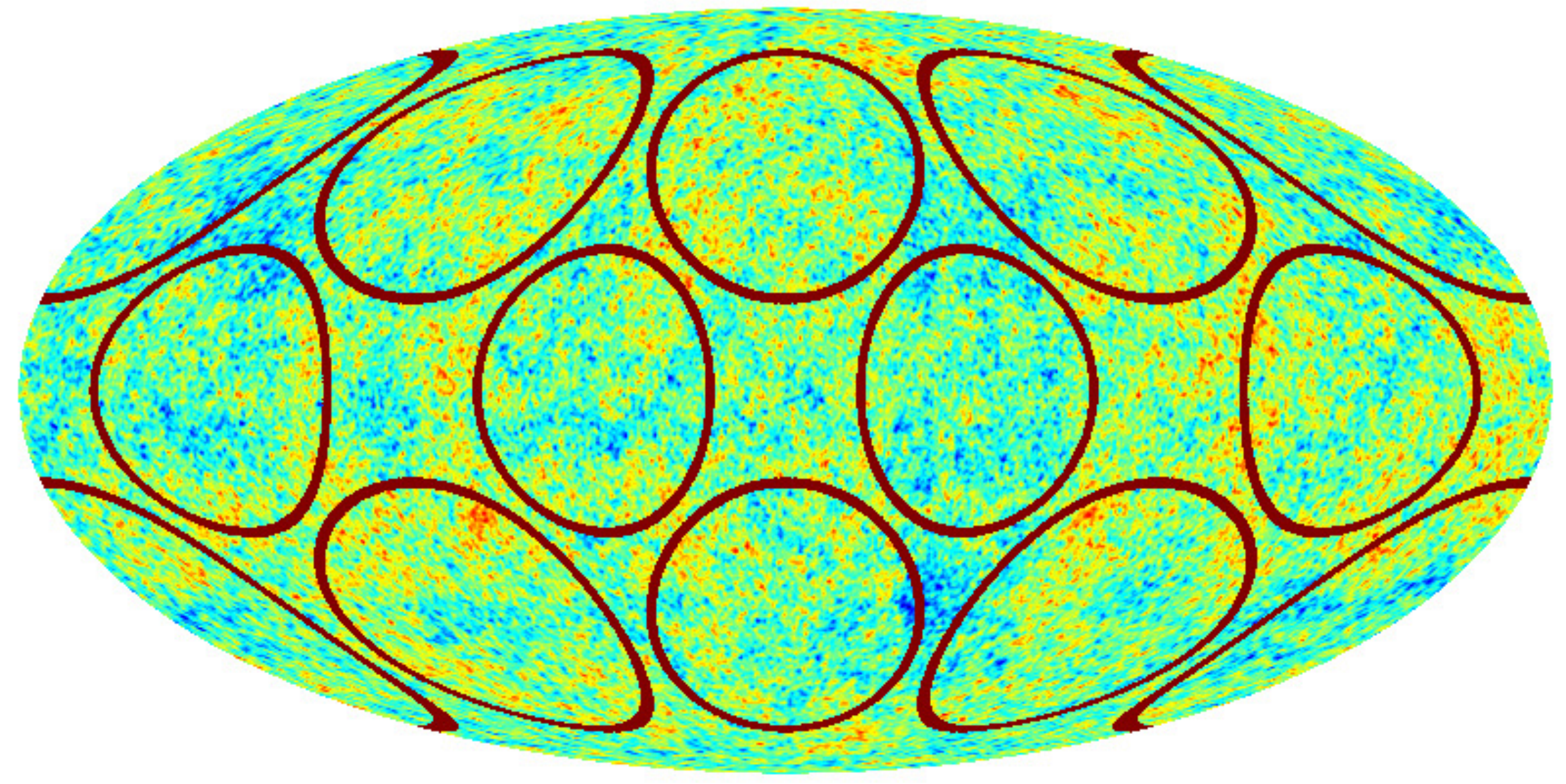}
	\caption{A simulated map of the CMB sky in a universe with a $T[2,2,2]$ toroidal topology. The dark circles show the locations of the same 
          slice through the last scattering surface seen on opposite sides of the sky. They correspond to matched circles with radius $\alpha\simeq 24^\circ$.}
	\label{fig:map_cmb_t222}
\end{figure}

\begin{figure}
\centering
   \includegraphics[width=1.0\columnwidth]{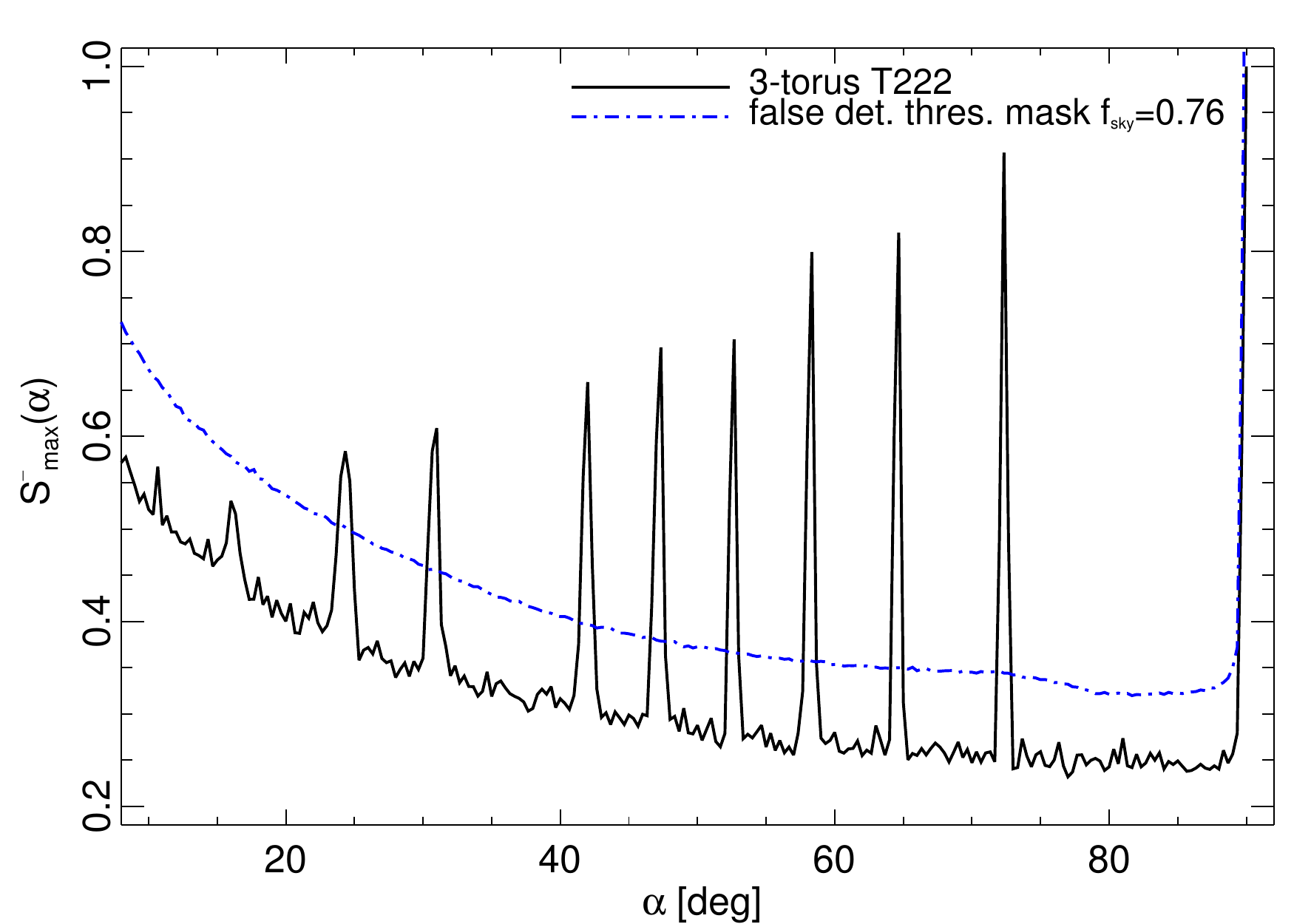}
\caption{An example of the $S_{\rm max}^{-}$ statistic as a function
  of circle radius $\alpha$ for a simulated
  CMB map (shown in Fig.~\ref{fig:map_cmb_t222}) of a
  universe with the topology of a cubic 3-torus with dimensions $L = 2H_0^{-1}$ 
  (solid line). The dash-dotted line show the false detection
  level established such that fewer than 1\,\% out of 300 Monte Carlo simulations of the CMB map, smoothed
  and masked in the same way as the data, would yield a false event.
  }
\label{fig:smax_t222}
\end{figure}

\paragraph{Circles-in-the-Sky}
Before beginning the search for pairs of matched circles in the
\emph{Planck} data, we validate our algorithm using simulations of
the CMB sky for a universe with 3-torus topology for which the
dimension of the cubic fundamental domain is $L=2H_0^{-1}$, and with
cosmological parameters corresponding to the $\Lambda CDM$ model
\citep[see][Table 1]{komatsu2010} determined from the 7-year
\emph{WMAP} results combined with the measurements of the distance from the
baryon acoustic oscillations and the Hubble constant. We performed simulations computing directly the
$a_{\ell m}$ coefficients up to the multipole of order $\ell_{\rm max}
= 500$ as described in \cite{bielewicz2011} and convolving them with
the same smoothing beam profile as used for the data, i.e., a Gaussian
beam with 30\arcm\ FWHM. In particular,
we verified that our code is able to find all pairs of matched circles
in such a map. The map with marked pairs of matched circles with radius
$\alpha\simeq24^\circ$ and the statistic $S_{\rm max}^{-}(\alpha)$ for the
map are shown in \fig\ref{fig:map_cmb_t222} and
\fig\ref{fig:smax_t222}, respectively. Note that the peak amplitudes
in the statistic, corresponding to the temperature correlation for
matched circles, decrease with radius of the circles.
\citet{cornish2004} noted that this is primarily caused by the
Doppler term, which becomes increasingly anticorrelated for circles
with radius smaller than $45^\circ$.

The intersection of the peaks in the matching statistic with the false
detection level estimated for the CMB map corresponding to the
simply-connected universe defines the minimum radius of the correlated circles which
can be detected for this map. The height of the peak with the smallest radius seen in
\fig\ref{fig:smax_t222} indicates that the minimum radius is about
$\alpha_{\rm min} \approx 20^\circ$. 

For the Monte Carlo simulations of the CMB maps for the simply-connected
universe we used the same cosmological parameters as for the
multi-connected universe, i.e., corresponding to the $\Lambda$CDM model
determined from the 7-year \emph{WMAP} results. The maps were 
also convolved with the same beam profile as for the simulated map for the
3-torus universe and data, as well as masked with the same cut 
used for the analysis of data.
The false detection threshold was established such that fewer than
1\,\% of 300 Monte Carlo simulations would yield a false event.

\paragraph{Bayesian Analysis}

\begin{figure}
	\centering
	\includegraphics[width=1.0\columnwidth]{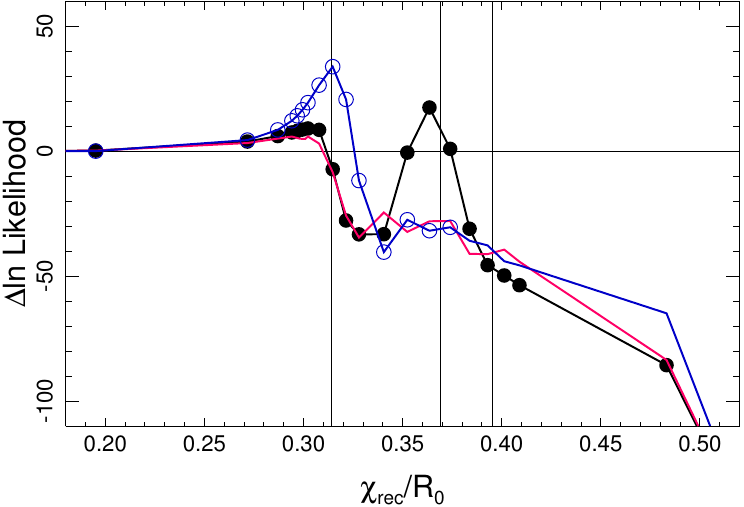}
	\includegraphics[width=1.0\columnwidth]{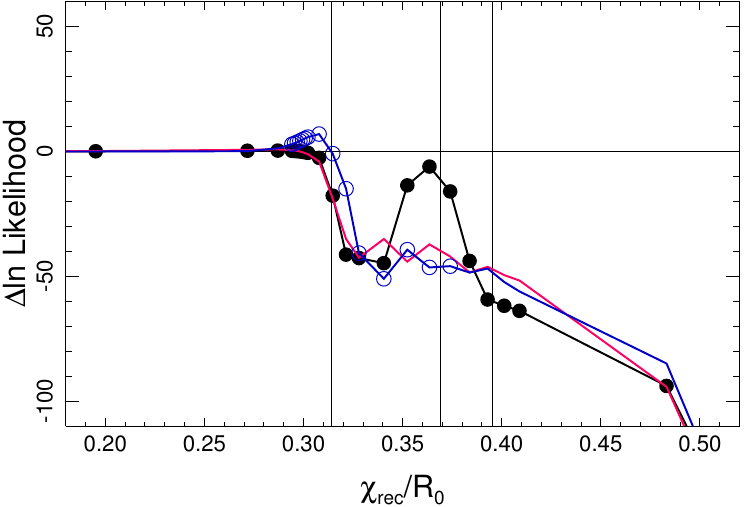}
	\caption{Test for likelihood detectability of compactified space for the 
	example of a dodecahedral ($I^*$)
	closed universe.  
	The vertical axis shows the log-likelihood relative to the largest model considered.
	Values are given for the orientations of the models which maximize the likelihood (top) and  marginalized
	over the orientations (bottom).
	Different size models are tested against two {\tt HEALPix} $N_\mathrm{side}=16$ temperature
	realizations drawn from the model with $\chi_\mathrm{rec}/R_0=0.314={\cal R}_\mathrm{i}$ (blue) 
	and $\chi_\mathrm{rec}/R_0=0.361$ (black). 
	No noise is added and the
	common mask has been applied. 
	Dots mark the positions of the models for which the likelihoods were computed.
	 The vertical lines show characteristic
	scales of the fundamental domain of the models in the units of
	curvature, from smaller to larger, ${\cal R}_\mathrm{i}/R_0$, ${\cal R}_\mathrm{m}/R_0$ and
	${\cal R}_\mathrm{u}/R_0$.  The variable
	$\chi_\mathrm{rec}/R_0$ gives the size of the last scattering surface in the
	same units. The $R_0 \to \infty$ limit corresponds
	to the flat simply-connected space.
	Both maximized and marginalized likelihoods show a detection relative
	to  the isotropic sky realization drawn 
	from the fiducial flat infinite universe (red) with the detection stronger for smaller spaces.
	However only the maximized likelihood 
	unambigously distinguishes the correct compact model from spaces that exceed the last-scattering diameter,
	which shows that
	the likelihood for small models is narrowly peaked at the correct orientation and suppressed
	otherwise.}
\label{fig:Spherical-validation}
\end{figure}

Because of the expense of the calculation of the correlation matrix,
we wish to limit the number of three-dimensional wavevectors
$\vec{k}$ we consider, as well as the number of spherical harmonic modes
$\ell$, and finally the number of different correlation matrices as a
whole. We need to ensure that the full set of matrices $C_{\ell
  \ell'}^{m m'}$ that we calculate contains all of the available
information on the correlations induced by the topology in a
sufficiently fine-grained grid. For this purpose, we consider the
Kullback-Leibler (KL) divergence as a diagnostic \citep[see, e.g.,][for applications of the KL
divergence to topology]{Kunz:2005wh,Kunz:2008fq}. The KL divergence
between two probability distributions $p_1(x)$ and $p_2(x)$ is given
by
\begin{equation}
	d_\mathrm{KL} = \int p_1(x)\ln\frac{p_1(x)}{p_2(x)}\;dx\;.
\end{equation}
If the two distributions are Gaussian with correlation matrices $\mtrx{C}_1$
and $\mtrx{C}_2$, this expression simplifies to
\begin{equation}
	d_\mathrm{KL} = -\frac12 \left[ \ln\left|\mtrx{C}_1 \mtrx{C}_2^{-1}\right| + \mathrm{Tr}\left(\mtrx{I} - \mtrx{C}_1 \mtrx{C}_2^{-1}\right)\right]\;,
\end{equation}
and is thus a measure of the discrepancy between the correlation
matrices. The KL divergence can be interpreted as the ensemble 
average of the log-likelihood-ratio $\Delta\textrm{ln}{\cal L}$ between realizations of the two distributions. 
Hence, they enable us to probe the ability to tell if, on average, we can distinguish 
realizations of $p_1$ from a fixed $p_2$ without having to perform a brute-force Monte Carlo integration. 
Thus, the KL divergence is related to ensemble averages of the likelihood-ratio plots that we present for simulations (Fig.~\ref{fig:Spherical-validation}) and real data (Sect.~\ref{sec:Results}), but does not depend on simulated or real data.

\begin{figure}[htbp]
	\centering
		\includegraphics[width=1\columnwidth]{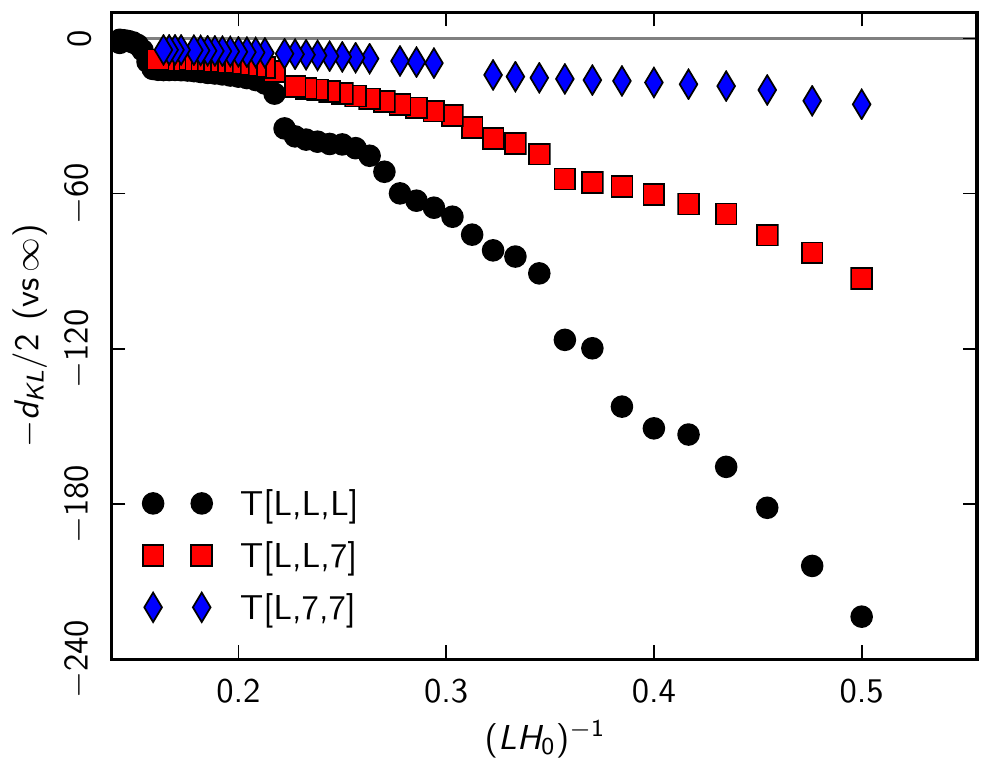}
	\caption{
	The KL divergence computed for torus models as a function of the (inverse) length of a side of the cube. $T[L_1,L_2,L_3]$ refers to a torus with edge lengths $L_i$.} 
	\label{fig:KL}
\end{figure}

We first use the KL divergence to determine the size of the fundamental
domain which we can consider to be equivalent to the simply-connected case (i.e., the limit in which all dimensions of the fundamental domain go to infinity). 
We note that in our
standard $\Lambda$CDM model, the distance to the surface of last
scattering is $\chi_\mathrm{rec} \approx 3.1416(H_0)^{-1}$. We would
naively expect that as long as the sphere enclosing the last
scattering surface can be enclosed by the fundamental domain
($L=2\chi_\mathrm{rec}$), we would no longer see the effects of
non-trivial topology. However, because the correlation matrix includes
the full three-dimensional correlation information (not merely the
purely geometrical effects of completely correlated points) we would
see some long-scale correlation effects even for larger fundamental
domains. In Fig.~\ref{fig:KL} we show the KL
divergence (as a function of $(LH_0)^{-1}$ so that the simply-connected limit $L\to\infty$ is at a finite position) for the $T[L,L,L]$ (cubic), $T[L,L,7]$ (chimney) and
$T[L,7,7]$ (slab) spaces and show that it begins to level off for
$(LH_0)^{-1}\lsim1/5$, although these topologies are still distinguishable from the $T[7,7,7]$ torus which is yet closer to the value for a simply-connected universe $d_\mathrm{KL}[7,7,7]\simeq1.1$. 
These figures indicate that a length of $L=7H_0^{-1}$ is an acceptable proxy for the simply-connected infinite Universe. The figures, as well as the likelihoods computed on simulations and
data, show steps and other structures on a variety of scales
corresponding to the crossing of the different length scales of the
fundamental domain ${\cal R}_\mathrm{u}$, ${\cal R}_\mathrm{m}$, and
${\cal R}_\mathrm{i}$ crossing the last scattering surface; smaller
fundamental domains with longer intersections with the last scattering surface are easier to detect.

\begin{figure}[htbp]
		\includegraphics[width=\hsize]{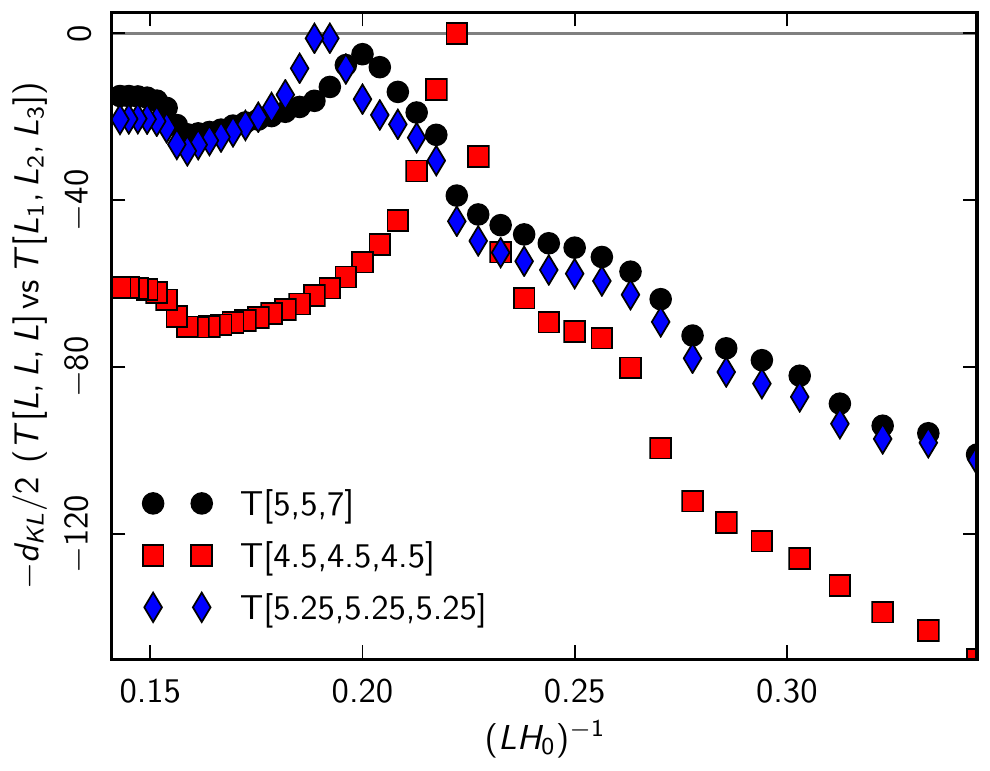}
	\caption{The KL divergence between a supposed correct model and other models. We show differences of cubic tori with respect to  models with $(LH_0)^{-1}=1/4.5\simeq0.22$ (aligned with our grid of models), $(LH_0)^{-1}=1/5.25\simeq0.19$ (in between the gridpoints) and and a $T[5,5,7]$ chimney model with $(LH_0)^{-1}=1/5$ in two directions and $(LH_0)^{-1}=1/7\simeq0.14$ in the third.}
	\label{fig:KLgrid}
\end{figure}

Computational limitations further prevent us from calculating the
likelihood at arbitrary values of the fundamental domain size
parameters. We must therefore ensure that our coarse-grained
correlation matrices are sufficient to detect a topology even if it
lies between our gridpoints. In Fig.~\ref{fig:KLgrid} we show the KL
divergence as a function of the size of the fundamental domain,
relative to various models, both aligned with our grid ($LH_0=4.5$) and in between our grid points ($LH_0=5.25$). We see that the peak
is wide enough that we can detect a peak within $\delta LH_0\sim0.1$
of the correct value. We also show that we can detect anisotropic
fundamental domains even when scanning through cubic tori: we show a case which approximates a ``chimney''
universe with one direction much larger than the distance to the last
scattering surface.

\begin{figure}[htbp]
	\centering
		\includegraphics[width=\hsize]{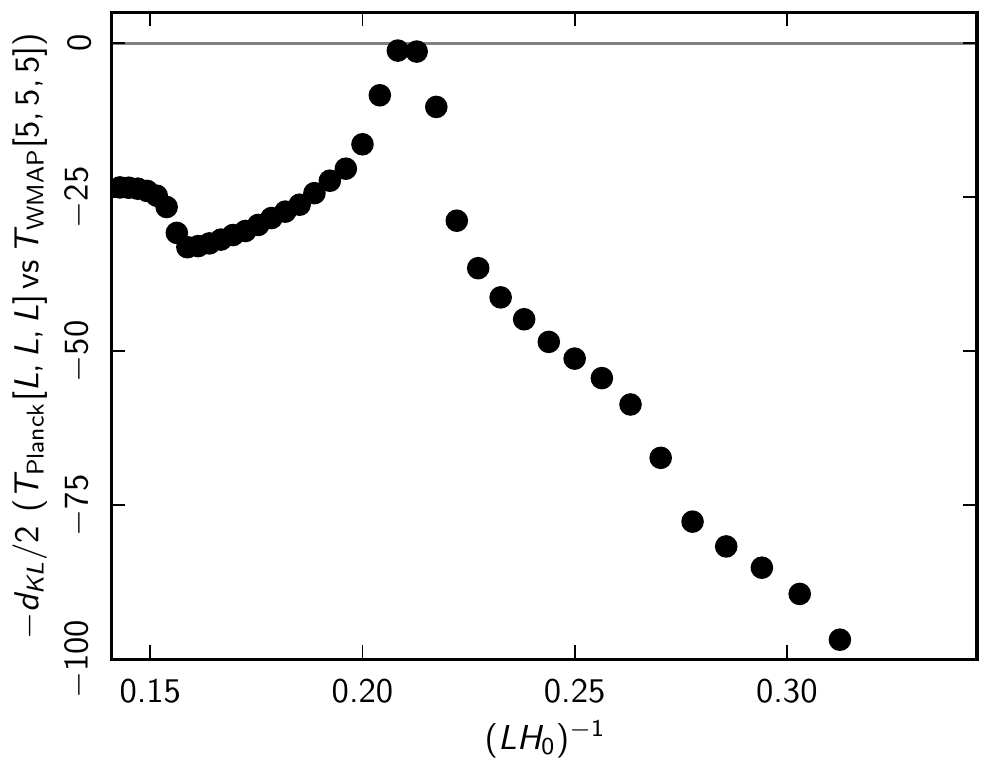}
	\caption{The KL divergence between a model generated with the \textit{WMAP} best-fit cosmological parameters as a background cosmology and a $T[5,5,5]$ cubic torus topology with respect to a \Planck\ best-fit cosmology and a varying cubic topology.}
	\label{fig:KLparams}
\end{figure}

Because our topological analyses do not simultaneously vary the background
cosmological parameters along with those describing the topology, we
also probe the sensitivity to the cosmology. In Fig.~\ref{fig:KLparams}
we show the effect of varying the fiducial cosmology from the
\cite{planck2013-p11} best-fit values to those reported by \textit{WMAP}
\citep{komatsu2010}.\footnote{We use the \texttt{wmap7+bao+h0} results from 
\url{http://lambda.gsfc.nasa.gov}.}
We see that this induces a small bias of
$\delta LH_0\simeq0.2$ but does not hinder the ability to detect a non-trivial
topology. This indicates that small deviations from the correct background cosmology do not hinder our ability to detect (or rule out) topological signals.

\begin{figure}[htbp]
	\centering
		\includegraphics[width=0.9\columnwidth]{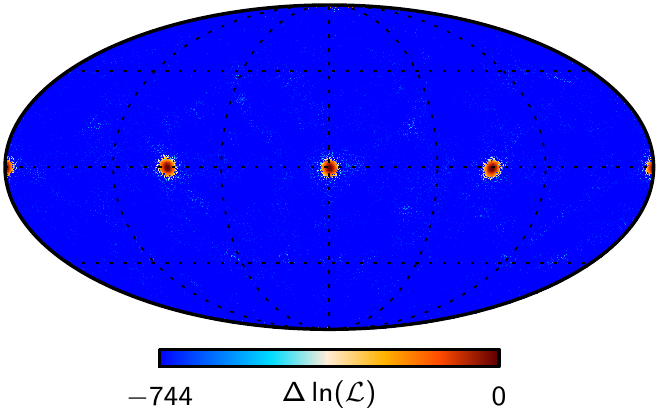}
	\caption{The log-likelihood with respect to the peak as a function of the orientation of the fundamental $T[2,2,2]$ torus domain for the simulations. The third Euler angle is marginalized over. We see peaks at the orientations corresponding to the six faces of the cubic fundamental domain (there are peaks at the North and South poles, which are difficult to see in this projection).}
	\label{fig:pawel_heat_plot}
\end{figure}

We have also directly validated the topological Bayesian techniques with simulations. In
Fig.~\ref{fig:pawel_heat_plot} we show the log-likelihood for the above $T[2,2,2]$
simulations as a function of two of the Euler angles, maximized over the third. We find a strong peak
at the correct orientation, with a multiplicity due to the degenerate orientations corresponding to the faces of the cube (there are peaks at the North and South poles, which are difficult to see in this projection). Note that the peaks correspond to ratios of more than $\exp(700)$ compared to the relatively smooth minima elsewhere.

In Fig.~\ref{fig:Spherical-validation} we also test the ability of the Bayesian
likelihood technique to detect the compactification of the space in 
the simulated temperature realizations drawn from the dodecahedral closed
model. For curved geometries, the size of the fundamental domain is fixed with respect to the varying curvature scale ($R_0$), whereas the distance to the last scattering $\chi_\mathrm{rec}$ is constant.
Hence we plot the likelihood as a function of $\chi_\mathrm{rec}/R_0$, inversely proportional to the scale of the fundamental domain.

Two mulitply-connected realizations of the sky were tested: one corresponding to the space
in which the last scattering sphere can be just inscribed into the fundamental 
domain, $\chi_\mathrm{rec}={\cal R}_\mathrm{i}$, when just the first large angle correlations appear,
and the second drawn from a somewhat smaller space for which $\chi_\mathrm{rec}=R_\mathrm{e}$.
We see detections in both cases, stronger as the fundamental domain shrinks relative to $\chi_\mathrm{rec}$. We also calculate the likelihood for a model known to be simply-connected. 
Note that the likelihood taken at the best orientations of the compact models generically  shows a
slight increase relative to that for the limiting simply
connected space as one brings the size of the fundamental
domain down to the size of the last scattering surface ($\chi_\mathrm{rec}\approx{\cal R}_\mathrm{i}$), followed, in
the absence of signal in the map, by a rapid drop as soon as the
models smaller than $\chi_\mathrm{rec}$ are applied. This small increase is
also present in the fiducial exactly isotropic sky, a single realization of which is shown in the figure, but is a generic feature irrespective of the topology being tested (occurring also in models with $R<{\cal R}_\mathrm{i}$), and thus should not 
be taken as an indication for compact topology. The reason for the
increase is the possibility of aligning the model with a weak anisotropic
correlation feature with chance patterns of a single sky realization.
However the fit drastically worsens as soon as the correlation
features in a model become pronounced. Moreover, the feature becomes considerably less significant when the likelihood is marginalized over the orientation (Euler angles) of the fundamental domain.

All of these results (KL divergences and likelihoods) were computed with $\ell_\mathrm{max}=40$, corresponding approximately to $N_\mathrm{side}=16$, indicating that this is more than adequate for detecting even relatively small fundamental domains such as the $T[2,2,2]$ case simulated above. We also calculate $d_\mathrm{KL}$ between the correlation matrices for the $T[7,7,7]$ torus (as a proxy for the simply-connected case) and the $T[5,5,5]$ torus, as a function of the maximum multipole $\ell_\mathrm{max}$ used in the calculation of the correlation matrix: we find that $d_\mathrm{KL}$ continues to increase beyond $\ell_\mathrm{max}=60$. Thus, higher-resolution maps (as used by the matched-circles methods) contain more information, but with the very low level of noise in the \Planck\ CMB maps, $\ell_\mathrm{max}=40$ would nonetheless give a robust detection of a multiply-connected topology, even with the conservative foreground masking we apply.

We note that it is difficult to compress the content of these likelihood figures down to limits upon the size of the fundamental domain. This arises because it is difficult to provide a physically-motivated prior distribution for quantities related to the size of the fundamental domain. Most naive priors would diverge toward arbitrarily large fundamental domain sizes or would otherwise depend on arbitrary limits to the topological parameters.


\subsubsection{Bianchi} 
\label{ssub:Bianchi}

The {\tt ANICOSMO} code \citep{mcewen:bianchi} is used to perform a Bayesian analysis of
\bianchiviih\ models, which has been extensively validated by \cite{mcewen:bianchi}
already; we briefly summarise the validation performed for the masked
analysis.  In \cite{mcewen:bianchi} a CMB map is simulated, in which a
simulated Bianchi temperature map with a large vorticity (\ie,
amplitude) is embedded, before applying a beam, adding isotropic noise
and applying a mask.
Both the underlying cosmological and Bianchi parameters used to
generate the simulations are well recovered.  For this
simulation the coupled Bianchi model is favoured over $\Lambda$CDM,
with a log-Bayes factor of $\Delta \ln E \sim 50$. As expected, one
finds that the log-Bayes factor favours $\Lambda$CDM in simulations
where no Bianchi component is added.  For further details see
\cite{mcewen:bianchi}.






\section{Results} 
\label{sec:Results}

We now discuss the results of applying the
circles-in-the-sky and likelihood 
 methods to \Planck\ data to study topology and  \bianchiviih\ cosmologies.
 
\subsection{Topology} 
\label{sub:topology}

\begin{figure}
\centering
\includegraphics[width=\columnwidth]{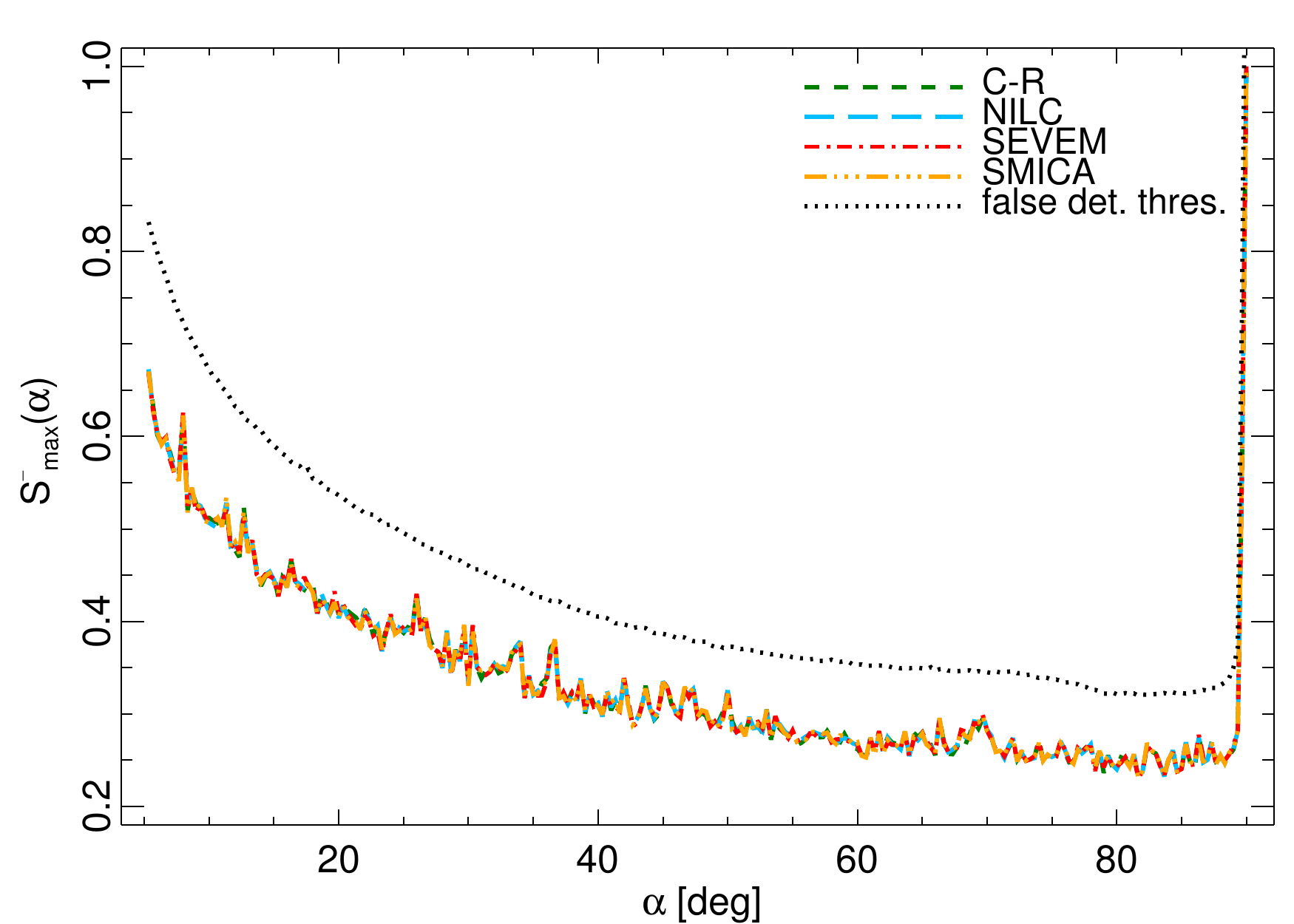}
\includegraphics[width=\columnwidth]{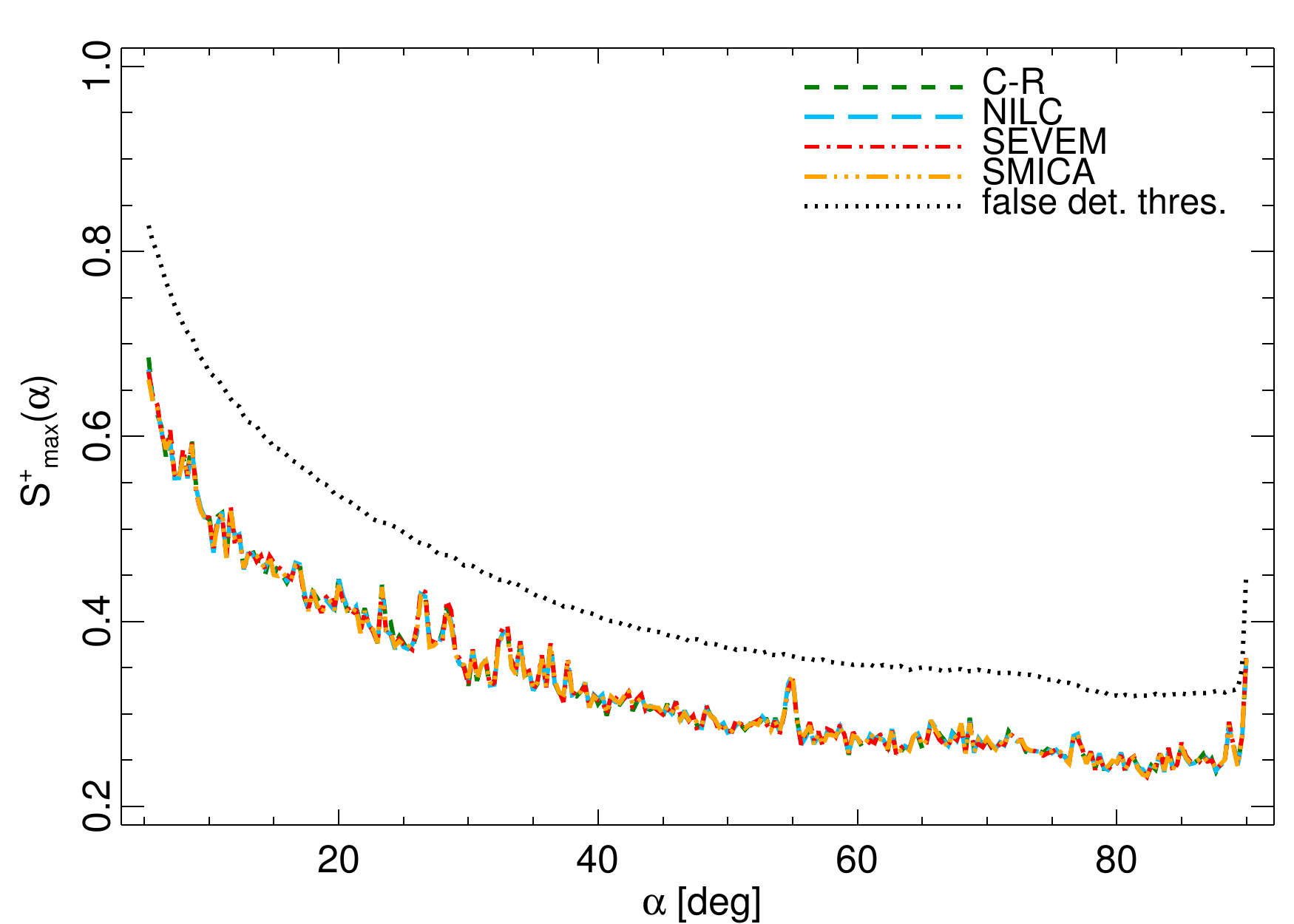}
\caption{The $S_{\rm max}^-$ (upper) and $S_{\rm max}^+$ (lower)
  statistics as a function of circle radius $\alpha$ for the \emph{Planck} CMB
  maps estimated using \commanderruler\ (red short dashed line),
  \nilc\ (blue long dashed line), \sevem\ (green dot-dashed line) and
  \smica\ (orange three dots-dashed line). Dotted line shows the false detection
  level established such that fewer than 1\% out of 300 Monte Carlo simulations of the CMB map, smoothed
  and masked in the same way as the data, would yield a false event. The peak
  at $90^\circ$ corresponds to a match between two copies of the same
  circle of radius $90^\circ$ centered around two antipodal points.}
\label{fig:smax_dx9}
\end{figure}

Neither the circles-in-the-sky search nor the likelihood method find
evidence for a multiply-connected topology.
We show the matched circle statistic in \fig\ref{fig:smax_dx9}.
We do not find any statistically significant correlation of circle
pairs in any map. As seen in \fig\ref{fig:smax_t222}, the minimum
radius at which the peaks expected for the matching statistic are
larger than the false detection level is $\alpha_{\rm min}
\approx 20^\circ$. Thus, we can exclude at the confidence level of
99\,\% any topology that predicts
matching pairs of back-to-back circles larger than this radius,
assuming that relative orientation of the fundamental domain and mask
allows its detection. This implies that in a flat universe described otherwise by the \Planck\ 
fiducial $\Lambda$CDM model, a $99\,\%$ confidence-limit lower bound on the size 
of the fundamental domain is
$L/2 \gsim \chi_{\rm rec}\cos(\alpha_{\rm min}) = 0.94\chi_\mathrm{rec} = 13.2\,\mathrm{Gpc}$. 
This is better than the limits from the marginalized likelihood
  ratios below for the tori and octahedron topologies and
  slightly worse than the limits for the dodecahedron and truncated
  cube. However, this constraint is not limited only to these few
  topologies. The frequentist analysis provides constraints upon a much wider class
  of topologies than those explicitly considered in the Bayesian
  likelihood approach; it concerns all
  topologies listed in Sect.~\ref{sub:Topology}.

\begin{figure}[htbp]
	\centering
		\includegraphics[width=1.0\columnwidth]{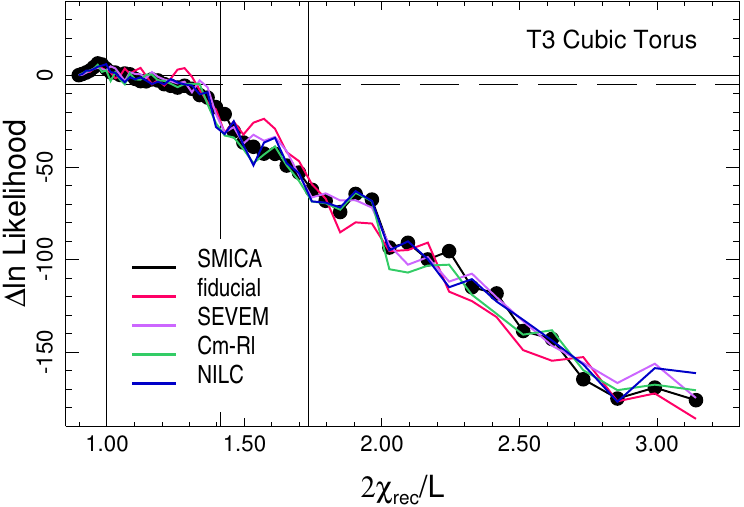}
		\includegraphics[width=1.0\columnwidth]{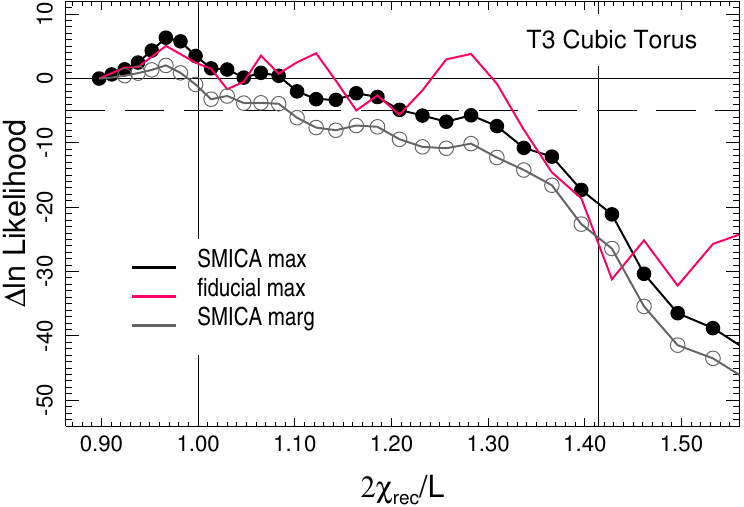}
	\caption{\emph{Top}: the likelihood as a function of the length of an edge of the fundamental domain $L$ for a cubic-torus topology. In this figure, $\chi_\mathrm{rec}$ gives the distance to the surface of recombination. The data are 
	component-separated CMB temperature 
	maps degraded to {\tt HEALPix} $N_\mathrm{side}=16$
	resolution and smoothed with an $\mathrm{FWHM}=660\,\arcm$
	Gaussian filter. The common mask of $f_\mathrm{sky}=0.78$ is used.
	The likelihood is marginalized over the amplitude
	of fluctuations, but maximized over the orientation of the fundamental domain.
	Lines for different estimates of the CMB temperature from \Planck\ data are black:
	\smica; magenta: \sevem; green: \commanderruler; blue: \nilc.  The red line is for a simulated isotropic sky
	from a fiducial flat simply-connected model. 
	Noise has been accounted for but is negligible at $N_\mathrm{side}=16$.
	The likelihoods are normalized to match the likelihood  obtained
	with the common mask in the $R_0 \to \infty$ isotropic flat limit.
	The vertical lines mark the positions where $\chi_{rec}$ is equal to the characteristic sizes
	of the fundamental domain, from left to right, 
	${\cal R}_\mathrm{i}=L/2$, ${\cal R}_\mathrm{m}=\sqrt{2} L/2$ and ${\cal R}_\mathrm{u}=\sqrt{3} L/2$.
	Dots, superimposed onto the \smica\ curve, designate the discrete set of
	models studied.
	\emph{Bottom}: zoom into the transitional region near $\chi_\mathrm{rec} \approx {\cal} R_\mathrm{i}$.  
	Black \Planck\ \smica\ and red fiducial curves are the same as in the top panel.
	The grey curve (open circles) is the likelihood marginalized over the orientations for the \Planck\ \smica\ map.
	Only ${\cal R}_\mathrm{i}$ and ${\cal R}_\mathrm{m}$ are within the scale range shown.
	}
	\label{fig:lnlikelihoodTorus}
\end{figure}

The likelihood method also show no evidence of a multiply connected universe.
We present the likelihood for various models. In Fig.~\ref{fig:lnlikelihoodTorus} we show the likelihood
(marginalized over amplitude and maximized over orientation of the fundamental domain) for the cubic torus,
fixing the background cosmology to the best-fit flat Universe \Planck\  model \citep{planck2013-p11}. We see that this is maximized for $L>2\chi_\mathrm{rec}$, i.e., showing no evidence for non-trivial topology.
Note that the likelihood shows mild features as the size goes through the other scales associated with the topology,
in particular a small increase in the likelihood when the scale of the
inscribed sphere ${\cal R}_\mathrm{i}$ is crossed.
However, the same increase is found when the toroidal
model is compared to a single realization of a strictly isotropic fiducial sky, and thus, should not be interpreted as a
detection of multi-connected topology. The origin of this likelihood behaviour at best fit angles is that
the freedom of orientation can be used to align
small enhancements in large-angle correlations in the anisotropic $L \approx 2 {\cal R}_\mathrm{i} $ model
with random features in 
the given single realization of the sky. When marginalized over all possible orientations the effect is significantly
reduced; the slight rise is $\Delta\textrm{ln}{\cal L}\simeq 1.9$ from a likelihood of $P=650$, which is comparable to the numerical noise inherent in our stochastic integration.
For even smaller spaces, more extensive correlations of the temperature
can no longer be accommodated and for $L < 2 {\cal R}_\mathrm{m} $ the likelihood of the  T3 cubic toroidal model drops 
quickly, although not strictly monotonically. 

\begin{figure}[htbp]
	\centering
		\includegraphics[width=1.0\columnwidth]{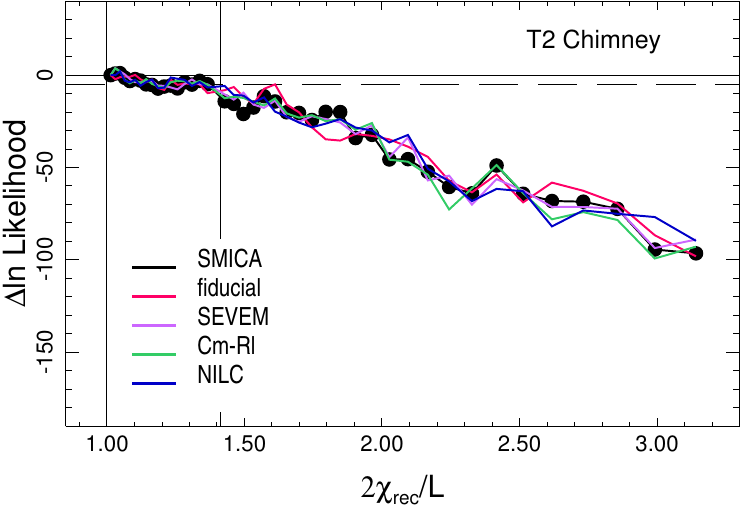}
	\caption{Same as Fig.~\ref{fig:lnlikelihoodTorus}, but for a toroidal space with one large dimension fixed
	at $7H_0^{-1}$ and two short dimensions of equal size $L$ (approximating the ``chimney'' space). 
	${\cal R}_\mathrm{i}$ and ${\cal R}_\mathrm{m}$ are marked while ${\cal R}_\mathrm{u}=\infty$
	}
	\label{fig:lnlikelihoodTorusRod}
\end{figure}

\begin{figure}[htbp]
	\centering
		\includegraphics[width=1.0\columnwidth]{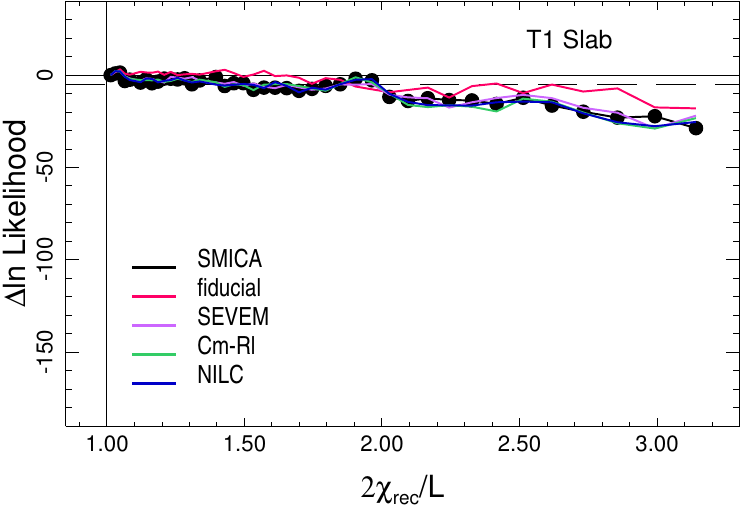}
	\caption{Same as Fig.~\ref{fig:lnlikelihoodTorus}, but for a toroidal space with two large dimensions
	fixed at $7H_0^{-1}$ and one short dimension of variable $L$ (approximating the ``slab'' space).
	${\cal R}_\mathrm{i}$ is marked while ${\cal R}_\mathrm{m}={\cal R}_\mathrm{u}=\infty$.}
	\label{fig:lnlikelihoodTorusSlab}
\end{figure}

In Figs.~\ref{fig:lnlikelihoodTorusRod} and 
\ref{fig:lnlikelihoodTorusSlab} we show the likelihood for the $T[L,L,7]$ chimney and $T[L,7,7]$ slab topologies,
which are also maximized in the simply-connected limit. The T2 chimney, with only two compact dimensions, is less constrained than the T3 cube, and the T1 slab, with one compact dimension, even less so.

\begin{figure}[htbp]
	\centering
		\includegraphics[width=1.0\columnwidth]{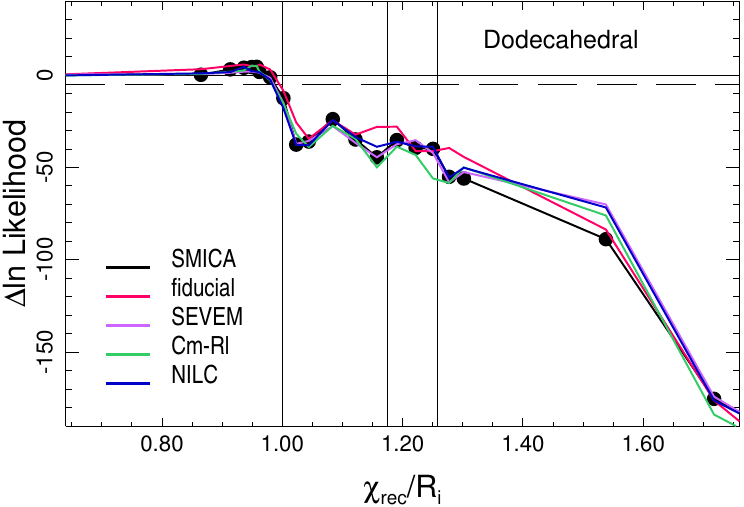} 
		\includegraphics[width=1.0\columnwidth]{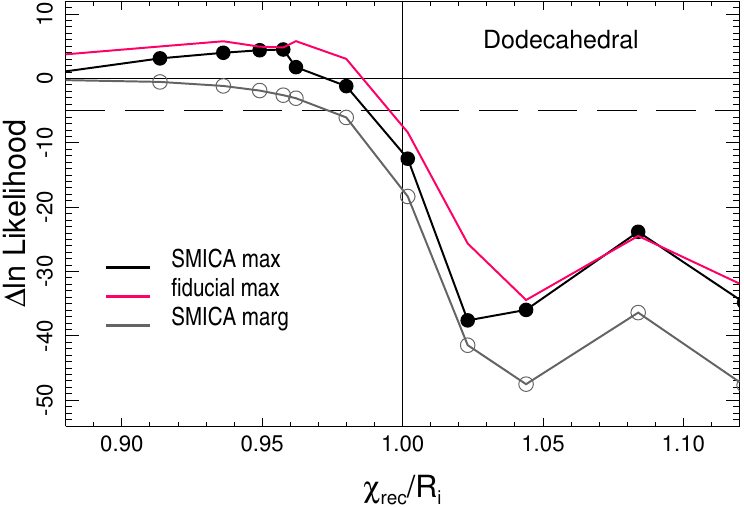}
	\caption{\emph{Top}: the likelihood as a function of the distance to last scattering
	surface in curvature units for 
	a locally spherical multiply-connected universe 
	with a dodecahedral ($I^*$) fundamental domain with ${\cal R}_\mathrm{i}=0.31 R_0$.
	Lines are for different estimates of the CMB temperature
	from \Planck\ data as in Fig.~\ref{fig:lnlikelihoodTorus}.
	In this figure, the $\chi_\mathrm{rec}/R_0$ parameterizes the
	position of the model on the geometrical degeneracy line
	which links $H_0$ and $\Omega_\Lambda$ with $\Omega_{K}$.
	The degeneracy relations are approximated as 
	$\Omega_\Lambda = 0.691 + 2.705 \Omega_{K}$ and
	$H_0 = 67.8 + 388 \Omega_{K} + 1200 \Omega_{K}^2$.
	The red reference curve is for the random isotropic realization
	from a fiducial flat model. Vertical lines mark when $\chi_\mathrm{rec}$ equals each of ${\cal R}_\mathrm{i}, {\cal R}_\mathrm{m}$,
	and $ {\cal R}_\mathrm{u}$, the characteristic scales of the fundamental domain. 
	\emph{Bottom}: zoom into the transitional region near $\chi_\mathrm{rec} \approx {\cal} R_\mathrm{i}$.  
	Both the likelihood at the best orientation of
	the domain versus the sky (black for the \Planck\ \smica\ CMB map and red for the fiducial realization, as in the top panel)
	and the likelihood marginalized over the orientations for \Planck\ \smica\ map
	(gray curve, open circles) are shown. 
	} 
	\label{fig:lnlikelihoodI}
\end{figure}

\begin{figure}[htbp]
	\centering
		\includegraphics[width=1.0\columnwidth]{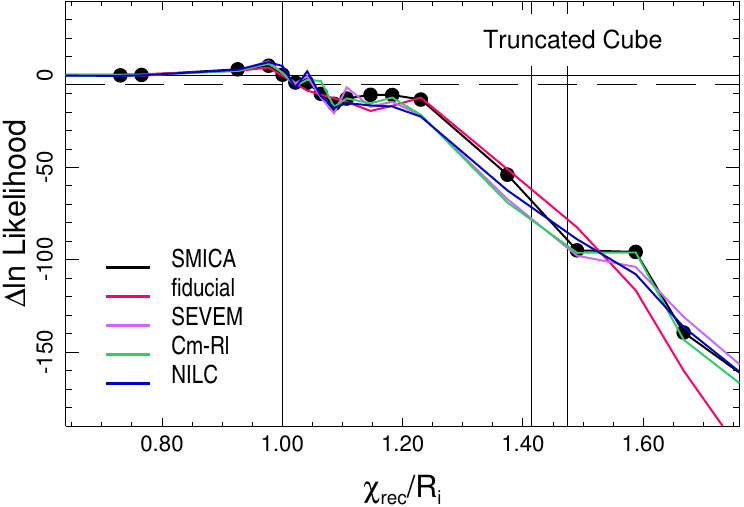}
		\includegraphics[width=1.0\columnwidth]{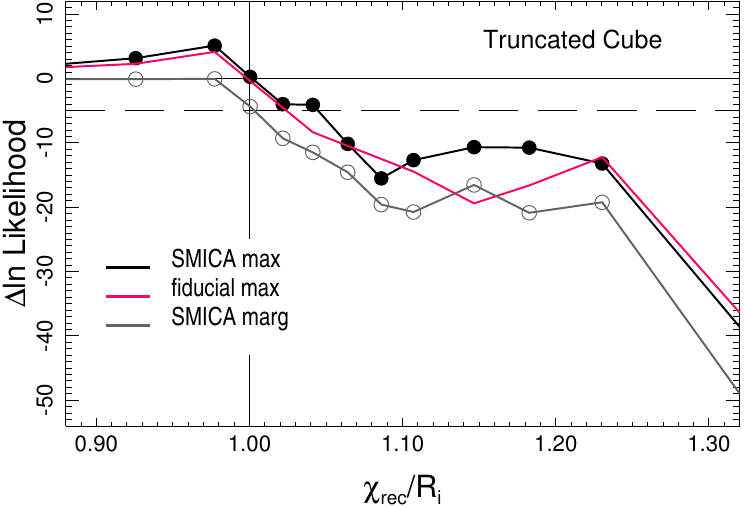}
	\caption{
	Likelihood for a constant positive curvature
	multiply-connected universe with a truncated cube ($O^*$)
        fundamental domain with $R_{\rm i}=0.39 R_0$.
	Notation is the same as in Fig.~\ref{fig:lnlikelihoodI}.
	}
	\label{fig:lnlikelihoodO}
\end{figure}

\begin{figure}[htbp]
	\centering
		\includegraphics[width=1.0\columnwidth]{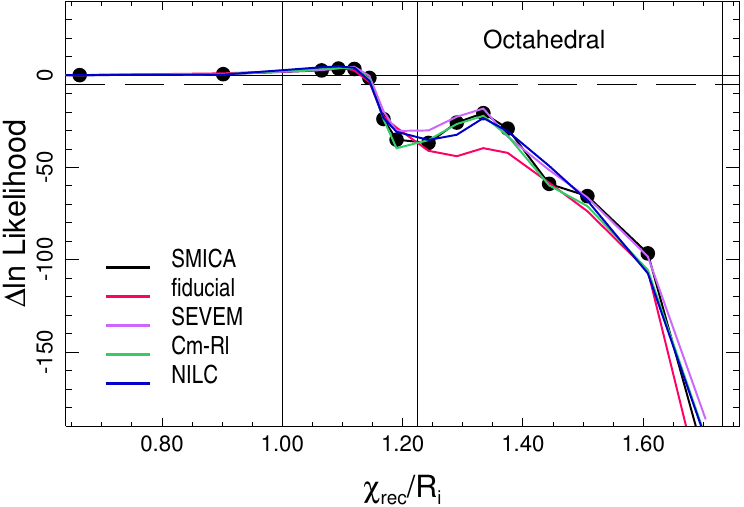}
		\includegraphics[width=1.0\columnwidth]{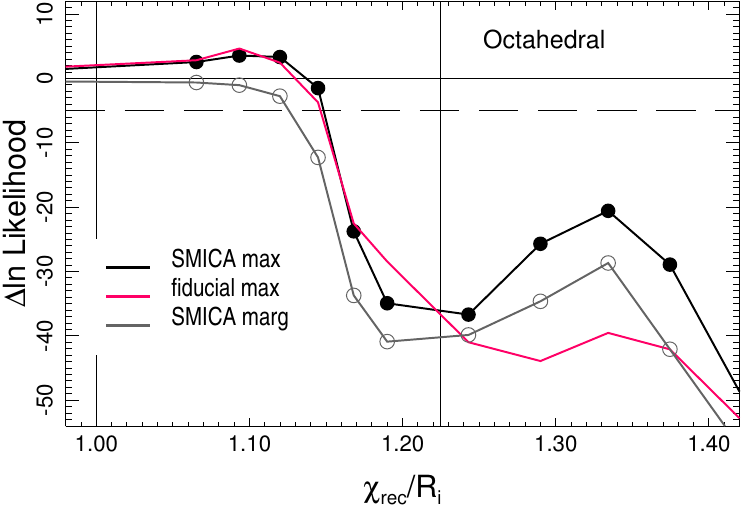}
	\caption{
	Likelihood for a constant positive curvature 
	multiply-connected universe with an octahedral ($T^*$)
        fundamental domain with $R_{\rm i}=0.45 R_0$.
	Notation is the same as in Fig.~\ref{fig:lnlikelihoodI}.}
	\label{fig:lnlikelihoodT}
\end{figure}

We find similar limits for the topologies allowed in a closed universe with a locally spherical geometry. In Fig.~\ref{fig:lnlikelihoodI} we show the likelihood for the dodecahedral fundamental domain, in
Fig.~\ref{fig:lnlikelihoodO} for the truncated cube, and in Fig.~\ref{fig:lnlikelihoodT} for the octahedron. In this case, we do not fix the background cosmological model, but rather account for the geometrical degeneracy
line which links $H_0$ and $\Omega_\Lambda$ with $\Omega_{K}$. The degeneracy relations are approximated as 
$\Omega_\Lambda = 0.691 + 2.705 \Omega_{K}$ and $H_0 = 67.8 + 388 \Omega_{K} + 1200 \Omega_{K}^2$. 
As in the toroidal case, there is no detection of a small space at the level expected from the simulations
of Sect.~\ref{sec:Methods}. Fundamental domains larger than the last scattering diameter are preferred for
the dodecahedral and truncated cube spaces with somewhat weaker restriction for
the octahedral case. Note that an observationally motivated prior on $H_0$ or $\Omega_{K}$ would be yet
more restrictive on the fundamental domain size.
For all three topologies, again as in the toroidal case, the maximum of the likelihood at best fit orientation
is detected for the finite volume spaces with $\chi_\mathrm{rec}\approx{\cal R}_\mathrm{i}$ at the level
$\Delta\textrm{ln}{\cal L}\approx +4$ relative to the fiducial flat simply-connected  model.
Since this feature is seen in the isotropic fiducial sky as well, we cannot take it as an indication
of a detection of a multi-connected space. In the case of curved spaces we see that this mild increase
disappears when we consider the likelihood marginalized over orientations. 
	
We present numerical limits for these flat and positively curved spaces in Table~\ref{tab:topolims}.
Because of the one-sided nature of these limits, we characterize the shape of the likelihood by the
steepness of its fall from the value as the scale of the fundamental domain goes to infinity
(i.e., the simply-connected limit). Hence, we show limits for $\Delta\textrm{ln}{\cal L}<-5$, (roughly equivalent to a $3\sigma$ --- 99\,\% confidence limit --- fall for a Gaussian; because of the very steep gradient,
the $2\sigma$ limits are very similar) and $\Delta\textrm{ln}{\cal L}<-12.5$ ($5\sigma$). Note that the limits differ depending on whether we marginalize or maximize the likelihood over the orientation angles.
We show lower limits on the quantity ${\cal R}_{\rm i}$ ($L/2$ for a torus with edge length $L$) in units of the last scattering distance $\chi_\mathrm{rec}$ (in conventional units, $\chi_\mathrm{rec}\approx14\,\mathrm{Gpc}$
for the fiducial \Planck\ parameters; \citealt{planck2013-p11}). In most cases, the limits are roughly
${\cal R}_\mathrm{i}\gsim\chi_\mathrm{rec}$ --- the scale of the fundamental domain must be greater than that of
the last scattering surface.  We place the most restrictive limits on the dodecahedron with
${\cal R}_\mathrm{i}>1.03\chi_\mathrm{rec}$ using marginalized values for  $\Delta\textrm{ln}{\cal L}<-5$. Conversely, the chimney and slab spaces
are less constrained as the expected correlations are weaker in one or two directions; for the slab space, we only
constrain ${\cal R}_\mathrm{i}=L/2\gsim0.5\chi_\mathrm{rec}$.


\begin{table}[tmb]                 
\begingroup
\newdimen\tblskip \tblskip=5pt
\caption{Lower limits on the size of the fundamental domain for different multiply-connected spaces, in units of the distance to the last scattering surface, $\chi_\mathrm{rec}$. 
For the torus, slab, and chimney, we present limits on the quantity $L/2$; in curved spaces,
limits are on the inscribed-sphere topology scale ${\cal R}_i$. For the columns labelled ``max'', 
we maximize the probability over the orientation of the fundamental domain; for ``marg'', we marginalize  over orientation. 
}
\label{tab:topolims}                            
\nointerlineskip
\vskip -3mm
\footnotesize
\setbox\tablebox=\vbox{
   \newdimen\digitwidth 
   \setbox0=\hbox{\rm 0} 
   \digitwidth=\wd0 
   \catcode`*=\active 
   \def*{\kern\digitwidth}
   \newdimen\signwidth 
   \setbox0=\hbox{+} 
   \signwidth=\wd0 
   \catcode`!=\active 
   \def!{\kern\signwidth}
\halign{#\hfil\tabskip=1.5em&                            
\hfil#\hfil&
\hfil#\hfil&
\hfil#\hfil&
\hfil#\hfil&
\hfil#\hfil&
\hfil#\hfil&
\hfil#\hfil\/\tabskip=0pt\cr
\noalign{\doubleline}
\hfil Space\hfil& Quantity&& \multispan2\hfil$\Delta\textrm{ln}{\cal L}<-5$\hfil&&   \multispan2\hfil$\Delta\textrm{ln}{\cal L}<-12.5$\hfil\cr  
& && max& marg&& max& marg\cr                                
\noalign{\vskip 3pt\hrule\vskip 5pt}
T3 Cubic Torus&                 $L/(2\chi_\mathrm{rec})$&& 0.83& 0.92&& 0.76& 0.83\cr
    T2 Chimney&                 $L/(2\chi_\mathrm{rec})$&& 0.71& 0.71&& 0.63& 0.67\cr
       T1 Slab&                 $L/(2\chi_\mathrm{rec})$&& 0.50& 0.50&&   \ldots &   \ldots \cr
  Dodecahedron&  ${\cal R}_\mathrm{i}/\chi_\mathrm{rec}$&& 1.01& 1.03&& 1.00& 1.01\cr
Truncated Cube&  ${\cal R}_\mathrm{i}/\chi_\mathrm{rec}$&& 0.95& 1.00&& 0.81& 0.97\cr
    Octahedron&  ${\cal R}_\mathrm{i}/\chi_\mathrm{rec}$&& 0.87& 0.89&& 0.87& 0.88\cr
\noalign{\vskip 5pt\hrule\vskip 3pt}}}
\endPlancktable                    
\endgroup
\end{table}                        

\begin{figure}[htbp]
	\centering
		\includegraphics[width=1.0\columnwidth]{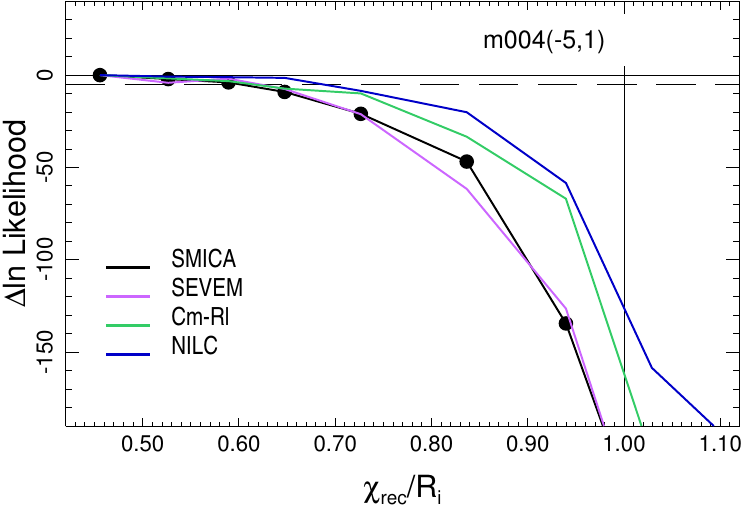}
		\includegraphics[width=1.0\columnwidth]{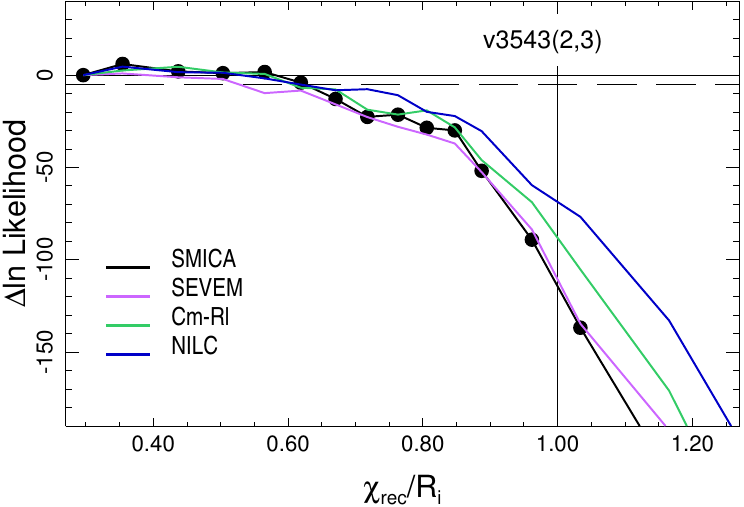}
	\caption{
	Likelihood for two constant negative curvature 
	multiply-connected universe, \emph{top}: m004($-5$,1);  \emph{bottom}: v3543(2,3).
	Notation is as in Fig.~\ref{fig:lnlikelihoodI} except that only ${\cal R}_\mathrm{i}/R_0$ is shown by vertical lines.}
	\label{fig:lnlikelihoodm004}
\end{figure}

In Fig.~\ref{fig:lnlikelihoodm004} we show the likelihood for the two  
hyperbolic models listed in Table~\ref{tbl:spaces}, which also show no detection of the multi-connected topology.
In the hyperbolic case we space the range of space sizes by varying
$\Omega_{K}$ while keeping $\Omega_{\Lambda}-\Omega_{\rm m}$
as well as $H_0$ constant at fiducial values.

All of these results show at least some increase in the likelihood for certain orientations when one of the characteristic scales of the fundamental domain (${\cal R}_\mathrm{u}$, ${\cal R}_\mathrm{m}$, or ${\cal R}_\mathrm{i}$) just exceed the surface of last scattering, and so no longer produces matched patterns, but induces extra correlations at large angular separations. Chance patterns can then mimic these correlations, and this is exacerbated by our conservative sky masks, which allow arbitrary patterns in the masked regions. 

\subsection{Bianchi}

Masked \Planck\ data are analysed for evidence of a \bianchiviih\
component, where the prior parameter ranges adopted are the same as
those specified by \cite{mcewen:bianchi}. The analysis is performed on
the \smica\ component-separated map, using the mask defined for this
method, and is repeated on the \sevem\ component-separated map for
validation purposes (using the mask defined for the \sevem\ method). The
Bayes factors for the various \bianchiviih\ models and the equivalent
standard cosmological models are shown in
Table~\ref{tbl:bianchi_evidences_dx9_delta}.

\begin{table}
\caption{Log-Bayes factor relative to equivalent
  $\Lambda$CDM model (positive favours Bianchi model).}
\label{tbl:bianchi_evidences_dx9_delta}
\centering
\begin{tabular}{lcc} 
\noalign{\doubleline}
{\hfil Model}           & \smica & \sevem \\ 
\noalign{\vskip 3pt\hrule\vskip 5pt}
\hfil Flat-decoupled-Bianchi (left-handed)& $\phantom{-}2.8 \pm 0.1$ & $\phantom{-}1.5 \pm 0.1$ \\
\hfil Flat-decoupled-Bianchi (right-handed)& $\phantom{-}0.5 \pm 0.1$ & $\phantom{-}0.5 \pm 0.1$ \\
\hfil Open-coupled-Bianchi (left-handed)            & $\phantom{-}0.0 \pm 0.1$ & $\phantom{-}0.0 \pm 0.1$ \\
\hfil Open-coupled-Bianchi (right-handed)           & $-0.4 \pm 0.1$ &           $-0.4 \pm 0.1$ \\
\noalign{\vskip 5pt\hrule\vskip 3pt}
\end{tabular}
\end{table}

\begin{figure*}
\centering
\begin{subfigure}[b]{170mm}
\includegraphics[viewport=0 0 1 1,clip=,width=0.23\textwidth]{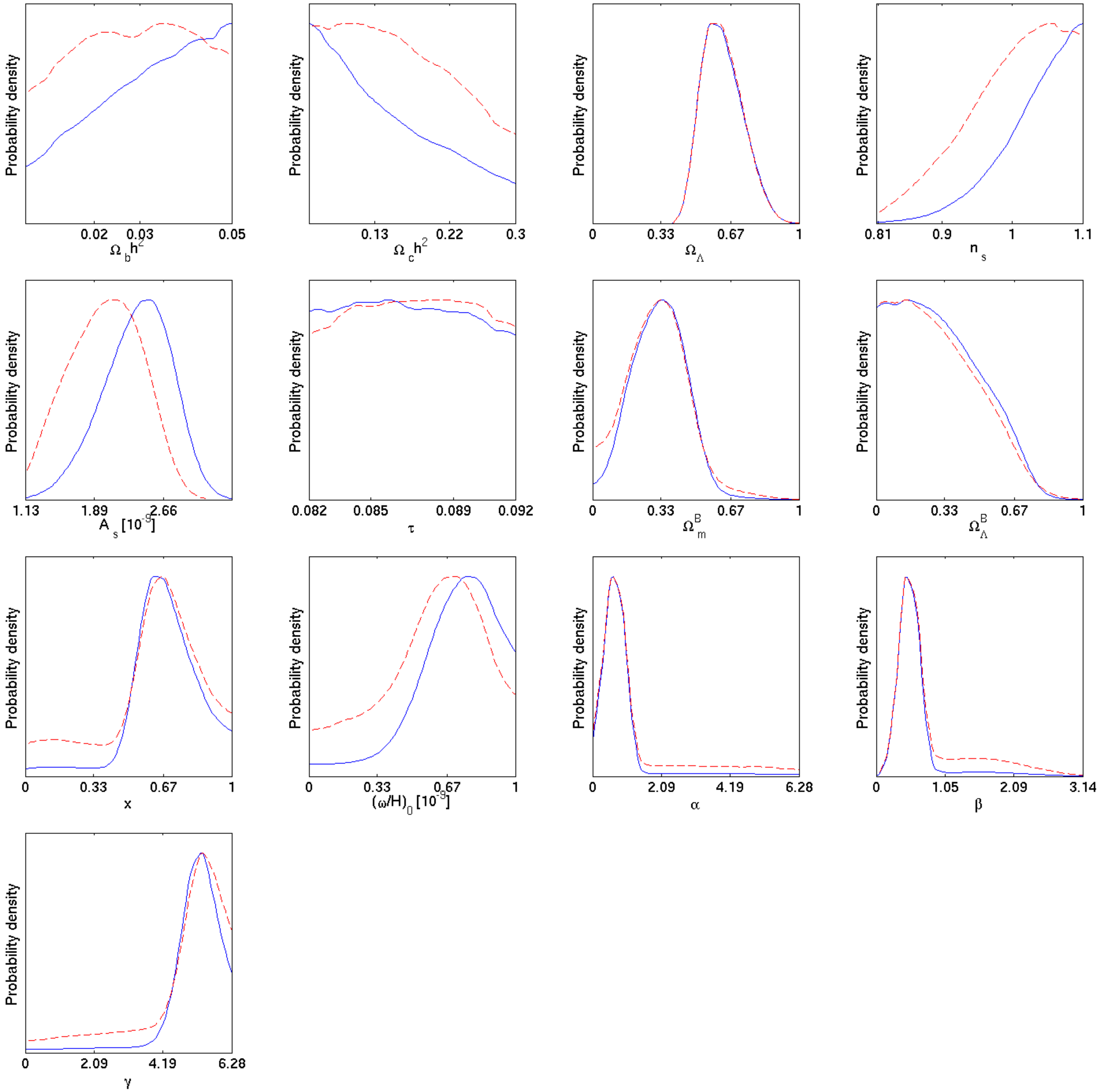}
\includegraphics[viewport=560 550 830 830,clip=,width=0.23\textwidth]{figures_bianchi/dx9_delta/posterior_flat_bianchi_decoupled_lmax032_smica_sevem}
\includegraphics[viewport=840 550 1110 830,clip=,width=0.23\textwidth]{figures_bianchi/dx9_delta/posterior_flat_bianchi_decoupled_lmax032_smica_sevem}
\includegraphics[viewport=0 270 270 550,clip=,width=0.23\textwidth]{figures_bianchi/dx9_delta/posterior_flat_bianchi_decoupled_lmax032_smica_sevem}\\
\includegraphics[viewport=280 270 550 550,clip=,width=0.23\textwidth]{figures_bianchi/dx9_delta/posterior_flat_bianchi_decoupled_lmax032_smica_sevem}
\includegraphics[viewport=560 270 830 550,clip=,width=0.23\textwidth]{figures_bianchi/dx9_delta/posterior_flat_bianchi_decoupled_lmax032_smica_sevem}
\includegraphics[viewport=840 280 1110 550,clip=,width=0.23\textwidth]{figures_bianchi/dx9_delta/posterior_flat_bianchi_decoupled_lmax032_smica_sevem}
\includegraphics[viewport=0 0 270 270,clip=,width=0.23\textwidth]{figures_bianchi/dx9_delta/posterior_flat_bianchi_decoupled_lmax032_smica_sevem}
\caption{Flat-decoupled-Bianchi model.}
\label{fig:posteriors_planck_decoupled}
\end{subfigure} \\*[3mm]
\begin{subfigure}[b]{170mm}
\includegraphics[viewport=840 830 1110 1110,clip=,width=0.23\textwidth]{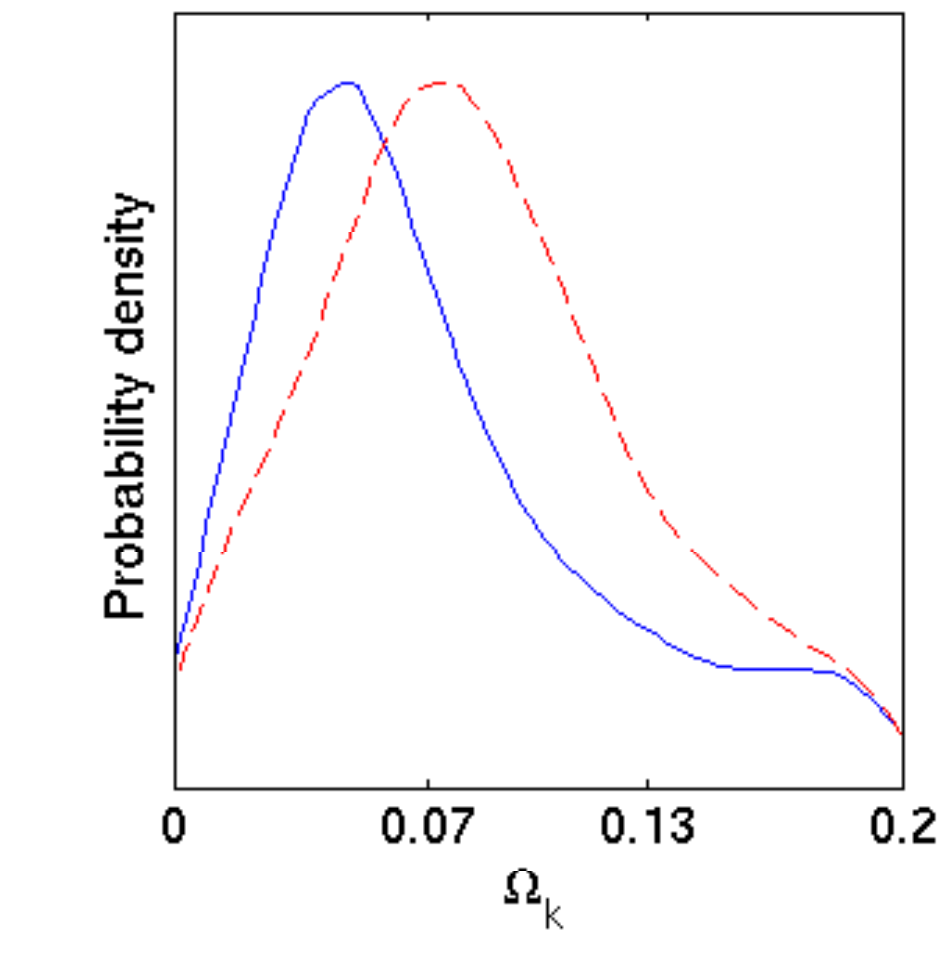}
\includegraphics[viewport=840 550 1110 830,clip=,width=0.23\textwidth]{figures_bianchi/dx9_delta/posterior_open_bianchi_coupled_lmax032_smica_sevem}
\includegraphics[viewport=0 270 270 550,clip=,width=0.23\textwidth]{figures_bianchi/dx9_delta/posterior_open_bianchi_coupled_lmax032_smica_sevem}
\includegraphics[viewport=280 270 550 550,clip=,width=0.23\textwidth]{figures_bianchi/dx9_delta/posterior_open_bianchi_coupled_lmax032_smica_sevem}\\
\includegraphics[viewport=560 270 830 550,clip=,width=0.23\textwidth]{figures_bianchi/dx9_delta/posterior_open_bianchi_coupled_lmax032_smica_sevem}
\includegraphics[viewport=840 280 1110 550,clip=,width=0.23\textwidth]{figures_bianchi/dx9_delta/posterior_open_bianchi_coupled_lmax032_smica_sevem}
\includegraphics[viewport=0 0 270 270,clip=,width=0.23\textwidth]{figures_bianchi/dx9_delta/posterior_open_bianchi_coupled_lmax032_smica_sevem}
\includegraphics[viewport=280 0 550 270,clip=,width=0.23\textwidth]{figures_bianchi/dx9_delta/posterior_open_bianchi_coupled_lmax032_smica_sevem}
\caption{Open-coupled-Bianchi model.}
\label{fig:posteriors_planck_coupled}
\end{subfigure}
\caption{Posterior distributions of Bianchi parameters recovered from
  \Planck\ \smica\ (solid curves) and \sevem\ (dashed curves)
  component-separated data for left-handed models.  \Planck\ data
  provide evidence in support of a Bianchi component in the phenomenological
  flat-decoupled-Bianchi model (panel a) but not in the physical
  open-coupled-Bianchi model (panel b).}
\label{fig:posteriors_planck}
\end{figure*}

\begin{figure*}[htbp]
\centering
\begin{subfigure}[b]{110mm}
	\includegraphics[width=110mm]{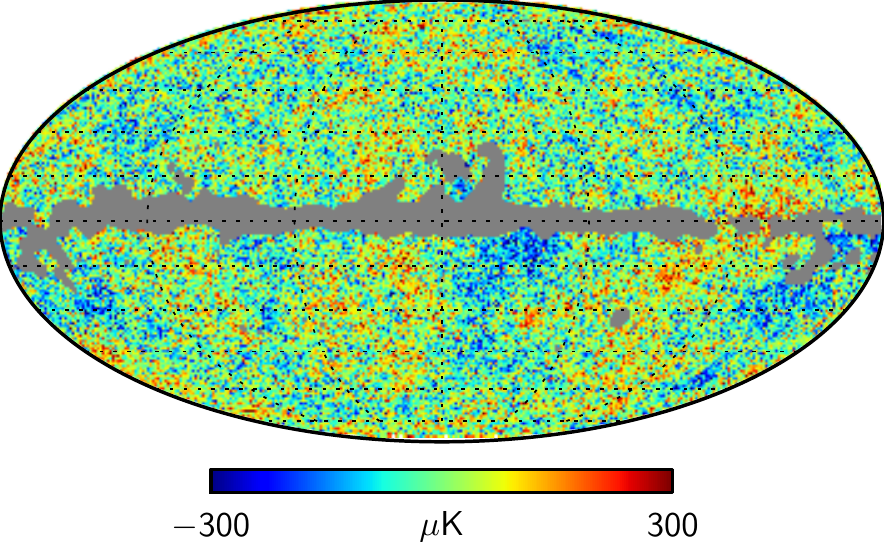}
	\caption{\smica\ CMB map.}
\end{subfigure}
\begin{subfigure}[b]{110mm}
	\includegraphics[width=110mm]{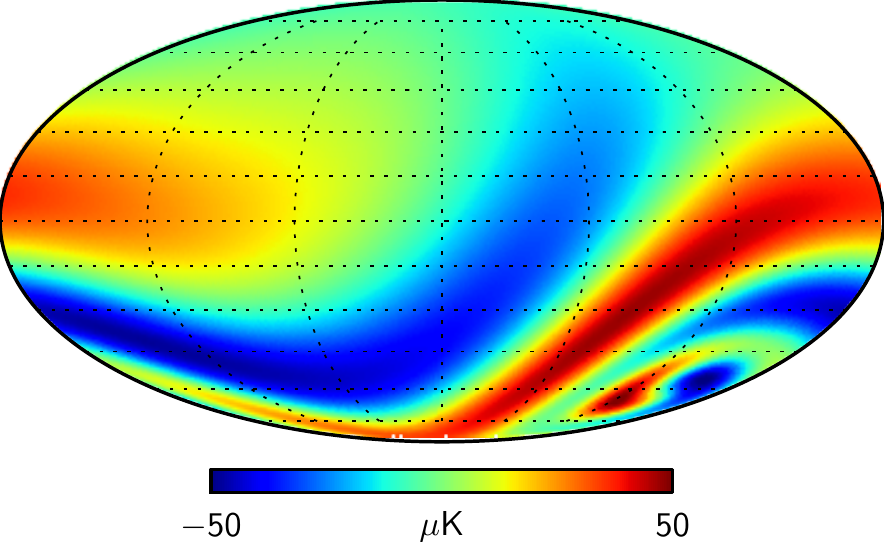}
	\caption{Best-fit \bianchiviih\ map. \label{fig:bianchi_bestfit_template}}
\end{subfigure}
\begin{subfigure}[b]{110mm}
	\includegraphics[width=110mm]{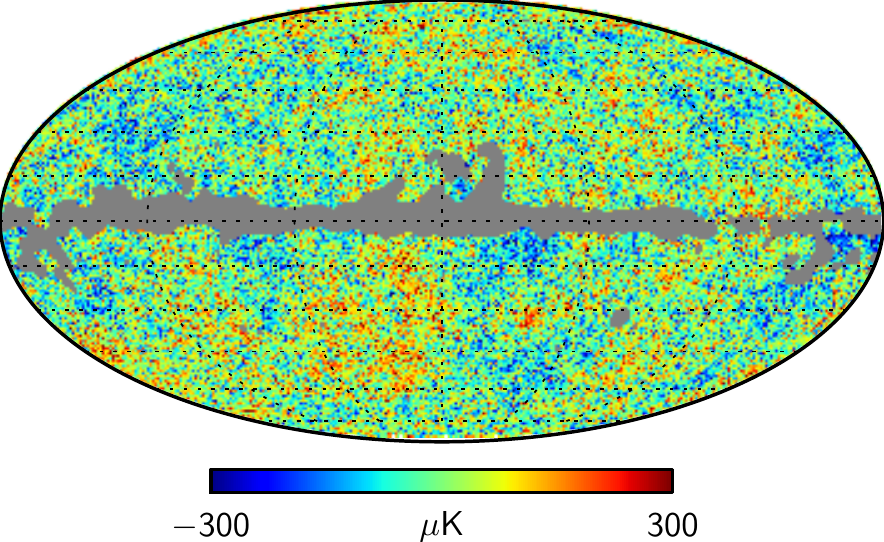}
	\caption{\smica\ CMB map with best-fit Bianchi component removed.}
\end{subfigure}
\caption{Best-fit template of left-handed flat-decoupled-\bianchiviih\ model subtracted
  from \Planck\ \smica\ component-separated data. Before subtraction,
  the peak-to-peak variation is $\pm594 \, \mu$K, reduced to $\pm564\,
  \mu$K after subtraction.}
\label{fig:bianchi_corrected_cmb}
\end{figure*}

\begin{figure*}[htbp]
\centering
	\includegraphics[width=110mm]{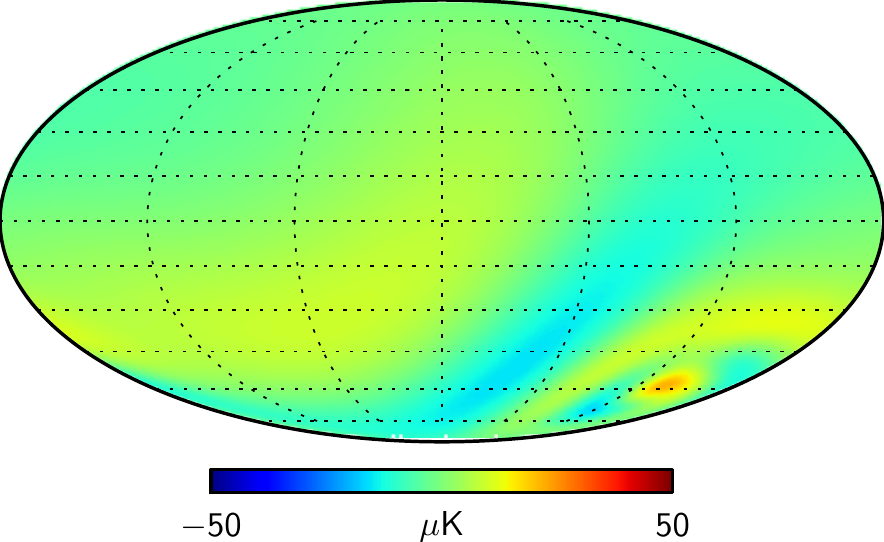}
\caption{Difference between best-fit template of flat-decoupled-\bianchiviih\ model recovered
from \textit{WMAP} 9-year data and from \Planck\ \smica\ component-separated data.}
\label{fig:bianchi_wmap_diff}
\end{figure*}

For the phenomenological flat-decoupled-Bianchi model, evidence in support of a
left-handed Bianchi template is found. On the Jeffreys
scale \citep{jeffreys:1961}, evidence for this model would be referred to as strong for the
\smica\ map and significant for the \sevem\ map. For both \smica\ and \sevem\
component-separated data, recovered posterior distributions for the
flat-decoupled-Bianchi model are shown in
Fig.~\ref{fig:posteriors_planck_decoupled}, where similar posterior
distributions are recovered for both component separation
methods. Recall that the Bianchi parameters are decoupled from the
standard cosmology in the flat-decoupled-Bianchi model, hence for this
model $\Omega_{\rm m}^{\rm B}$ and $\Omega_{\Lambda}^{\rm B}$ are
specific to the Bianchi model and should not be compared with standard
values. The maximum a posteriori (MAP) best-fit template found
for \smica\ component-separated data is shown in
Fig.~\ref{fig:bianchi_bestfit_template}, with the difference between
this template and the template found in \textit{WMAP} 9-year data
\citep{mcewen:bianchi} shown in Fig.~\ref{fig:bianchi_wmap_diff}. Note
that the template found in \Planck\ data is very similar to the
template found in \textit{WMAP} 9-year data \citep{mcewen:bianchi},
which in turn is similar to the template first found by
\cite{jaffe:2005}. However, the template found in \textit{WMAP} 9-year
data \citep{mcewen:bianchi} is only significant in full-sky data, but
not when the 9-year KQ75 \textit{WMAP} mask \citep{bennett2012} is
applied.  Since the \Planck\ \smica\ and \sevem\ masks are less
conservative than the KQ75 mask, these findings suggest data near the
Galactic plane may be playing a considerable role in supporting a
Bianchi component in \Planck\ data.  The \smica\ CMB map and a
Bianchi-subtracted version of this map are also shown in
Fig.~\ref{fig:bianchi_corrected_cmb}.  The best-fit parameters of the
templates found in \Planck\ \smica\ and \sevem\ component-separated data
are displayed in Table~\ref{tbl:bianchi_bestfit_parameters}, for both
the MAP and mean-posterior estimates. The analysis was also performed
on a \smica\ component-separated Gaussian simulation, yielding a null
detection (i.e., no evidence for a Bianchi component), as expected.

\newcommand{\degdot}{\ensuremath{\!\stackrel{\circ}{\textstyle.}\!}}
\begin{table*}
\caption{Parameters recovered for left-handed flat-decoupled-Bianchi
  model. \Planck\ data favour the inclusion of a Bianchi component in
  this phenomenological model.}
\label{tbl:bianchi_bestfit_parameters}
\centering
\begin{tabular}{lccccccc} 
\noalign{\doubleline}
Bianchi Parameter     && \multicolumn{2}{c}{\smica} && \multicolumn{2}{c}{\sevem} \\
                      && MAP & Mean && MAP & Mean \\
\noalign{\vskip 3pt\hrule\vskip 5pt}
\hfil$\Omega_{\rm m}^{\rm B}$   && $0.38$ & $0.32\pm0.12$ && $0.35$ & $0.31 \pm 0.15$ \\
\hfil$\Omega_{\Lambda}^{\rm B}$ && $0.20$ & $0.31\pm0.20$ && $0.22$& $0.30 \pm 0.20$  \\
\hfil\bx                     && $0.63$ & $0.67\pm0.16$ && $0.66$&  $0.62 \pm 0.23$  \\
\hfil$(\omega/H)_0$         && $8.8 \times 10^{-10}$ & $(7.1 \pm 1.9) \times 10^{-10}$ && $9.4 \times 10^{-10}$& $(5.9 \pm 2.4)  \times 10^{-10}$  \\
\hfil\eula                  && \phantom{3}$38\degdot8$ & \phantom{2}$51\degdot3 \pm 47\degdot9$ && \phantom{3}$40\degdot5$& \phantom{2}$77\degdot4 \pm 80\degdot3$  \\
\hfil\eulb                  && \phantom{3}$28\degdot2$ & \phantom{2}$33\degdot7 \pm 19\degdot7$ && \phantom{3}$28\degdot4$& \phantom{2}$45\degdot6 \pm 32\degdot7$  \\
\hfil\eulc                  && $309\degdot2$ & $292\degdot2 \pm 51\degdot9$ && $317\degdot0$& $271\degdot5 \pm 80\degdot7$  \\
\noalign{\vskip 5pt\hrule\vskip 3pt}
\end{tabular}
\end{table*}

\begin{figure*}[htbp]
\begin{subfigure}[b]{88mm}
	\includegraphics[width=88mm]{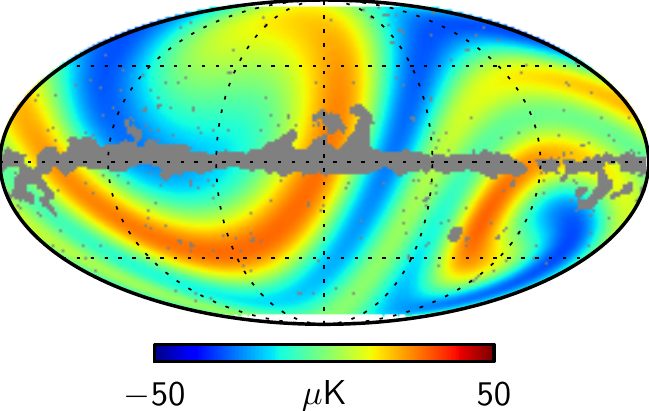}
	\caption{\smica}
\end{subfigure}
\begin{subfigure}[b]{88mm}
	\includegraphics[width=88mm]{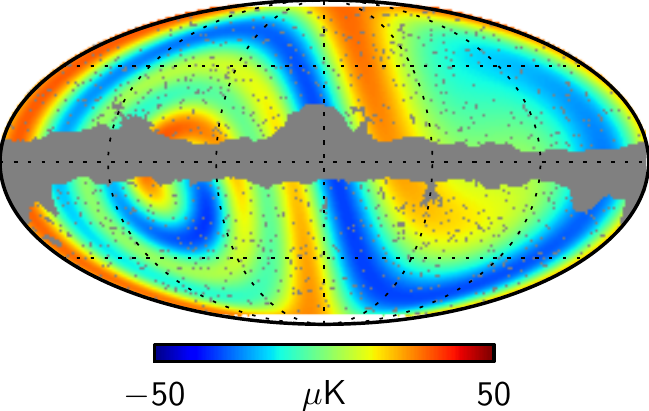}
	\caption{\sevem}
\end{subfigure}
\caption{Best-fit templates of left-handed open-coupled-\bianchiviih\
  model recovered from \Planck\ \smica\ and \sevem\ component-separated
  data.  The Bayes factors for this model indicate that \Planck\ data
  do not favour the inclusion of these Bianchi maps.}
\label{fig:bianchi_bestfit_template_coupled}
\end{figure*}

For the most physically motivated open-coupled-Bianchi model where
the \bianchiviih\ model is coupled to the standard cosmology, there is
no evidence in support of a Bianchi contribution. 
Recovered posterior distributions for the open-coupled-Bianchi model
are shown in Fig.~\ref{fig:posteriors_planck_coupled} for both \smica\
and \sevem\ component-separated data.  Although the cosmological Bianchi
parameters agree reasonably well between these different
component-separated data, the posterior distributions recovered for
the Euler angles differ.  For \sevem\ data, an additional mode of the
posterior distribution is found; the mode found with \smica\ data is
still present in \sevem\ data but is not dominant.  Consequently, the
best-fit estimates for the Euler angles differ between the \smica\
and \sevem\ component-separated data.  Note that the additional mode
found in \sevem\ data is also present in \textit{WMAP} 9-year data
\citep{mcewen:bianchi}.  The resulting best-fit parameters for the the
open-coupled-Bianchi model are displayed in
Table~\ref{tbl:bianchi_bestfit_parameters_coupled}, while the
corresponding MAP best-fit maps are shown in
Fig.~\ref{fig:bianchi_bestfit_template_coupled}.  Nevertheless, for
both \smica\ and \sevem\ data the Bayes factors computed 
(Table~\ref{tbl:bianchi_evidences_dx9_delta}) do not favour the
inclusion of any Bianchi component for the open-coupled-Bianchi model.
\Planck\ data thus do not provide evidence in support of \bianchiviih\
cosmologies. However, neither is it possible to conclusively discount
\bianchiviih\ cosmologies in favour of $\Lambda$CDM cosmologies. The
constraints \mbox{$(\omega/H)_0 < 7.6 \times 10^{-10}$} (95\%
confidence level) on the vorticity of the physical coupled
\bianchiviih\ left-handed models and \mbox{$(\omega/H)_0 < 8.1 \times
  10^{-10}$} (95\% confidence level) for right-handed models are
recovered from \smica\ component-separated data. 


\begin{table*}
\caption{Parameters recovered for left-handed open-coupled-Bianchi model. \Planck\ data do not favour the inclusion of a Bianchi component in this model and some parameters are not well constrained.}
\label{tbl:bianchi_bestfit_parameters_coupled}
\centering
\begin{tabular}{lccccccc}
\noalign{\doubleline}
Bianchi Parameter     && \multicolumn{2}{c}{\smica} && \multicolumn{2}{c}{\sevem} \\
                      && MAP & Mean && MAP & Mean \\ 
\noalign{\vskip 3pt\hrule\vskip 5pt}
\hfil$\Omega_{k}$   && $0.05$ & $0.07\pm0.05$ && $0.09$ & $0.08 \pm 0.04$ \\
\hfil$\Omega_{\rm m}^{\rm B}$   && $0.41$ & $0.33\pm0.07$ && $0.41$ & $0.32 \pm 0.07$ \\
\hfil$\Omega_{\Lambda}^{\rm B}$ && $0.55$ & $0.60\pm0.07$ && $0.50$& $0.59 \pm 0.07$  \\
\hfil\bx                     && $0.46$ & $0.44\pm0.24$ && $0.38$&  $0.39 \pm 0.22$  \\
\hfil$(\omega/H)_0$         && $5.9 \times 10^{-10}$ & $(4.0 \pm 2.4) \times 10^{-10}$ && $9.3 \times 10^{-10}$& $(4.5 \pm 2.8)  \times 10^{-10}$  \\
\hfil\eula                  && \phantom{2}$57\degdot4$ & $122\degdot5 \pm 960\degdot0$ && $264\degdot1$& $188\degdot6 \pm 98\degdot7$  \\
\hfil\eulb                  && \phantom{2}$54\degdot1$ & \phantom{1}$70\degdot8 \pm 35\degdot5$ && \phantom{2}$79\degdot6$& \phantom{1}$81\degdot1 \pm 31\degdot7$  \\
\hfil\eulc                  && $202\degdot6$ & $193\degdot5 \pm 77\degdot4$ && \phantom{2}$90\degdot6$& $160\degdot4 \pm 91\degdot1$  \\
\noalign{\vskip 5pt\hrule\vskip 3pt}
\end{tabular}
\end{table*}


\section{Discussion} 
\label{sec:Discussion}


We have used the \Planck\ temperature anisotropy maps to probe the
large-scale structure of spacetime. We have calculated the Bayesian likelihood for specific topological
models in universes with locally flat, hyperbolic and spherical geometries, all of which find no
evidence for a multiply-connected topology with a fundamental domain
within the last scattering surface. After calibration on simulations,
direct searches for matching circles resulting from the intersection
of the fundamental topological domain with the surface of last
scattering also give a null result at high confidence. These results use conservative masks of the sky, unlike previous \emph{WMAP} results, which used full-sky internal linear combination maps (not originally intended for cosmological studies) or less conservative foreground masks. Hence, the results presented here, while corroborating the previous non-detections, use a single, self-consistent, and conservative dataset. The masked sky also increases the possibility of chance patterns in the actual sky mimicking the correlations expected for topologies with a characteristic scale near that of the last scattering surface.  

Depending on the shape of the fundamental domain, we find ${\cal R}_\mathrm{i}\gsim\chi_\mathrm{rec}$ (Table~\ref{tab:topolims}) with detailed 99\,\% confidence limits (considering the likelihood marginalized over the orientation of the fundamental domain) varying from $0.9\chi_\mathrm{rec}$ for the cubic torus in a flat universe to $1.03\chi_\mathrm{rec}$ for the dodecahedron in a positively curved universe, with somewhat weaker constraints for poorly-proportioned spaces that are considerably larger along some directions. In the case of the torus and octahedron topologies, a tighter constraint of $0.94\chi_\mathrm{rec}$ comes from the matched circles method (albeit with a somewhat different interpretation of frequentist and Bayesian limits). The constraint derived using this method applies to a wide class of topologies, listed in Sect.~\ref{sub:Topology}, predicting matching pairs of back-to-back circles.

Note that the results derived using the likelihood method make use of the expected pixel-space correlations as a unique signal of non-trivial topology. Hence, although a small fundamental domain will suppress power on the largest scales of the CMB, observation of such low power on large scales as observed by \textit{COBE} \citep{bjk00}, and confirmed by \textit{WMAP} \citep{Luminet:2003bp}, is not sufficient for the detection of topology. Conversely, because our methods search directly for these correlations (and indeed marginalize over the amplitude of fluctuations), a slight modification of the background FRW cosmology by lowering power in some or all multipoles \citep{planck2013-p08} will not affect the ability to detect the correlations induced by such topologies.

Similarly, using a Bayesian analysis we find no evidence for a
physical, anisotropic \bianchiviih\ universe. However, \Planck\ data
do provide evidence supporting a phenomenological \bianchiviih\ component,
where the parameters of the Bianchi component are decoupled from
standard cosmology.  The resulting best-fit \bianchiviih\ template
found in \Planck\ data is similar to that found in \textit{WMAP} data
previously \citep{jaffe:2005,mcewen:bianchi}.  However, although this
Bianchi component can produce some of the (possibly anisotropic)
temperature patterns seen on the largest angular scales \citep[see
also][]{planck2013-p09}, there is no set of cosmological parameters
which can simultaneously produce these patterns and the observed
anisotropies on other scales.  Moreover, the parameters of the
best-fit \bianchiviih\ template in the decoupled setting are in strong disagreement 
with other measurements of the cosmological parameters.

These results are expected from previous measurements from
\textit{COBE} and \textit{WMAP}, but \Planck's higher sensitivity and
lower level of foreground contamination provides further
confirmation. We have shown that the results are insensitive to the
details of the preparation of the temperature maps (in particular, the
method by which the cosmological signal is separated from
astrophysical foreground contamination).  Future \Planck\ measurement
of CMB polarization will allow us to further test models of
anisotropic geometries and non-trivial topologies and may provide more
definitive conclusions, for example allowing us to moderately extend
the sensitivity to large-scale topology \citep{bielewicz2012}.



\begin{acknowledgements}
	
The development of \Planck\ has been supported by: ESA; CNES and CNRS/INSU-IN2P3-INP (France); ASI,
CNR, and INAF (Italy); NASA and DoE (USA); STFC and UKSA (UK); CSIC, MICINN, JA and RES (Spain);
Tekes, AoF and CSC (Finland); DLR and MPG (Germany); CSA (Canada); DTU Space (Denmark); SER/SSO
(Switzerland); RCN (Norway); SFI (Ireland); FCT/MCTES (Portugal); and PRACE (EU). A description of
the Planck Collaboration and a list of its members, including the technical or scientific activities
in which they have been involved, can be found at
\url{http://www.sciops.esa.int/index.php?project=planck&page=Planck_Collaboration}.
	
The authors thank the anonymous referee for helpful comments and acknowledge the use of 
the UCL Legion High Performance Computing Facility (Legion@UCL),
and associated support services, in the completion of this work. Part of the computations were
performed on the Andromeda cluster of the University of Geneve, the Hopper Cray XE6 at NERSC and on
the GPC supercomputer at the SciNet HPC Consortium. SciNet is funded by: the Canada Foundation for
Innovation under the auspices of Compute Canada; the Government of Ontario; Ontario Research Fund -
Research Excellence; and the University of Toronto. \end{acknowledgements}

\bibliography{topo,Planck_bib}

\raggedright 
\end{document}